\documentclass[showpacs,preprintnumbers,
superscriptaddress,amsmath]{revtex4}
\usepackage{epsfig}
\usepackage{graphicx}

\usepackage{bm} 

\renewcommand\a{\alpha}
\renewcommand\b{\beta}
\newcommand\g{\gamma}
\renewcommand\d{\delta}

\newcommand\m{\mu}

\newcommand{\non}{\nonumber\\}

\newcommand{\Diracslash}[1]{#1\llap{/\kern1pt}}
\newcommand{\diracslash}[1]{#1\llap{/\kern0pt}}

\newcommand{\be}{\begin{equation}}
\newcommand{\ee}{\end{equation}}
\newcommand{\bea}{\begin{eqnarray}}
\newcommand{\eea}{\end{eqnarray}}
\newcommand{\ba}[1]{\begin{array}{#1}}
\newcommand{\ea}{\end{array}}

\newcommand{\vg}{\bm{\gamma}}

\begin{document}

\title{A General Effective Action for High-Density Quark Matter} 

\author{Philipp T.\ Reuter}
\email{preuter@th.physik.uni-frankfurt.de}
\affiliation{
Institut f\"ur Theoretische Physik,
Johann Wolfgang Goethe-Universit\"at,
D-60054 Frankfurt, Germany
}
\author{Qun Wang}
\email{qwang@th.physik.uni-frankfurt.de}

\affiliation{
Institut f\"ur Theoretische Physik,
Johann Wolfgang Goethe-Universit\"at,
D-60054 Frankfurt, Germany
}

\affiliation{Physics Department, Shandong University, Jinan,
Shandong, 250100, P.R. China}

\author{Dirk H.\ Rischke}
\email{drischke@th.physik.uni-frankfurt.de}
\affiliation{
Institut f\"ur Theoretische Physik,
Johann Wolfgang Goethe-Universit\"at,
D-60054 Frankfurt, Germany
}

\date{\today} 

\begin{abstract}

We derive a general effective action for quark matter at nonzero 
temperature and/or nonzero density. For this purpose,
we distinguish irrelevant from relevant
quark modes, as well as hard from soft gluon modes by
introducing two separate cut-offs in momentum space, 
one for quarks, $\Lambda_q$, and one for gluons, $\Lambda_g$.
We exactly integrate out irrelevant quark modes
and hard gluon modes in the functional integral representation
of the QCD partition function. Depending on the specific
choice for $\Lambda_q$ and $\Lambda_g$, the resulting
effective action contains well-known effective actions
for hot and/or dense quark matter, for instance the ``Hard Thermal Loop'' or
the ``Hard Dense Loop'' action, as well as the high-density effective
theory proposed by Hong and others. We then apply our
effective action to review the calculation of the
color-superconducting gap parameter to subleading order 
in weak coupling, where the strong coupling constant $g \ll 1$. 
In this situation, relevant quark modes are
those within a layer of thickness $2 \Lambda_q$ around the
Fermi surface. The non-perturbative nature of the 
gap equation invalidates naive attempts to estimate the importance 
of the various contributions via
power counting on the level of the effective action.
Nevertheless, once the gap equation has been derived
within a particular many-body approximation scheme, 
the cut-offs $\Lambda_q, \, \Lambda_g$ 
provide the means to rigorously power count different contributions
to the gap equation. We recover the previous result for the
QCD gap parameter for the choice $\Lambda_q \alt g \mu \ll \Lambda_g
\alt \mu$, where $\mu$ is the quark chemical potential. 
We also point out how to improve this result beyond
subleading order in weak coupling.

\end{abstract}

\pacs{12.38.Mh, 24.85.+p} 

\maketitle

\section{Introduction}

Quark matter at small temperature $T$ and large quark chemical
potential $\mu$ is a color superconductor
\cite{RWreview,DHRreview}. While this discovery goes back
to the late 1970's \cite{bailinlove}, wider interest
in the phenomenon of color superconductivity has only recently been
generated by the observation that, within a simple
Nambu--Jona-Lasinio (NJL) -- type model \cite{nambu} for 
the quark interaction, the color-superconducting gap parameter 
assumes values of the order of 100 MeV \cite{ARWRSSV}.
Gap parameters of this magnitude would have important phenomenological
consequences for the physics of compact stellar objects, and possibly
even for heavy-ion collisions at laboratory energies of the order of
$\sim 10$ AGeV. It is therefore of paramount importance to
put the estimates from NJL-type models on solid ground
and obtain a more reliable result for the magnitude of the gap
parameter based on first principles. To this end, the
color-superconducting gap parameter was also computed in quantum
chromodynamics (QCD) \cite{son,schaferwilczek,rdpdhr,shovkovy,Hsu}.

At zero temperature, $T=0$, in weak coupling, $g \ll 1$, 
and in the mean-field approximation,
the gap equation for the color-superconducting gap parameter $\phi$ 
assumes the schematic form
\be \label{gapeq}
\phi = g^2\, \phi \left[ \zeta\, \ln^2 \left(\frac{\mu}{\phi}\right)
+ \beta\, \ln \left( \frac{\mu}{\phi}\right) + \alpha \right]\;.
\ee
The solution is
\be \label{gapsol}
\phi = 2 \, b\, \mu \, \exp \left( - \frac{c}{g} \right) \left[ 1 +
O(g) \right]\;.
\ee
The first term in Eq.\ (\ref{gapeq}) is of {\em leading\/} order
since, according to Eq.\ (\ref{gapsol}), $g^2 \ln^2 (\mu/\phi) \sim 1$.
It originates from the exchange of almost static, long-range, Landau-damped
magnetic gluons. One factor $\ln (\mu/\phi)$ is the standard BCS
logarithm which arises when integrating over quasiparticle modes 
from the bottom to the surface of the Fermi sea, 
$\int dq/\epsilon_q \sim \ln (\mu/\phi)$, where 
\be
\epsilon_q \equiv \sqrt{(q-\mu)^2 + \phi^2}
\ee 
is the quasiparticle energy in a superconductor. The second factor 
$\ln (\mu/\phi)$ comes from a collinear enhancement $\sim
\ln(\mu/\epsilon_q)$ in the exchange of almost
static magnetic gluons. The coefficient $\zeta$ determines the constant
$c$ in the exponent in Eq.\ (\ref{gapsol}). As was 
first shown by Son \cite{son},
\be
c  \equiv \frac{3 \pi^2}{\sqrt{2}} \;.
\ee
The second term in Eq.\ (\ref{gapeq}) is of {\em subleading\/} order,
$g^2 \ln (\mu/\phi) \sim g \ll 1$. It originates from two sources. 
The first is the exchange of
electric and non-static magnetic gluons 
\cite{schaferwilczek,rdpdhr,shovkovy,Hsu}. In this case, the
single factor $\ln (\mu/\phi)$ is the standard BCS logarithm. The
second source 
is the quark wave-function renormalization factor in dense quark matter
\cite{rockefeller,qwdhr}. Here, the BCS logarithm does not arise, but
the wave-function renormalization contains an additional $\ln
(\mu/\epsilon_q)$ which generates a $\ln(\mu/\phi)$. The coefficient
$\beta$ determines the prefactor $b$ of the exponent in Eq.\
(\ref{gapsol}).
For a two-flavor color superconductor,
\be
b \equiv 256\, \pi^4\, \left( \frac{2}{N_f \, g^2} \right)^{5/2} 
\exp\left( -\frac{\pi^2 + 4}{8} \right)\;,
\ee
where $N_f$ is the number of (massless) quark flavors participating in 
screening the gluon exchange interaction.
The third term in Eq.\ (\ref{gapeq}) is of {\em sub-subleading\/}
order, $\sim g^2$. The coefficient $\alpha$ determines the
$O(g)$ correction to the prefactor of the color-superconducting gap
parameter in Eq.\ (\ref{gapsol}). Since $\alpha$ has 
not yet been determined, the gap parameter can be reliably computed only
in weak coupling, i.e., when the $O(g)$ corrections to the prefactor
are small. 

Due to asymptotic freedom the QCD coupling
constant becomes small only at large momentum transfer. The
typical momentum scale in dense quark matter is given by the quark
Fermi momentum, $k_F \equiv  \sqrt{\mu^2 - m^2}$, where $m$ is the
quark mass. The Fermi momentum is equal
to $\mu$ up to terms of order $O(m^2/\mu)$. 
Thus, $g \ll 1$ only for asymptotically 
large $\mu \gg \Lambda_{\rm QCD}$, where $\Lambda_{\rm QCD}$ is the
QCD scale parameter.
The range of $\mu$ values of phenomenological
importance is, however, $\alt 1$ GeV.
Although the quark density $n$ is already quite large
at such values of $\mu$, $n \sim 10$ times
the nuclear matter ground state density, the coupling constant
is still not very small, $g \sim 1$. It is therefore of interest
to determine the coefficient of $g$ in the $O(g)$ corrections
to the prefactor in Eq.\ (\ref{gapsol}).
If it turns out to be small, one gains more confidence in
the extrapolation of the weak-coupling result
(\ref{gapsol}) to chemical potentials of order $\sim 1$ GeV.

Let us mention that an extrapolation of
the weak-coupling result (\ref{gapsol}) for a two-flavor color
superconductor, neglecting sub-subleading
terms altogether and assuming
the standard running of $g$ with the chemical potential $\mu$,
yields values of $\phi$ of the order of $\sim 10$ MeV at chemical
potentials of order $\sim 1$ GeV, cf.\ Ref.\ \cite{DHRreview}.
This is within one order of magnitude of the predictions based on 
NJL-type models and thus might lead one to conjecture
that the true value of $\phi$ will lie somewhere in the
range $\sim 10 - 100$ MeV.
However, in order to confirm this and to obtain a more
reliable estimate of $\phi$ at values of $\mu$ of relevance in
nature, one ultimately has to
compute all terms contributing to sub-subleading order.
 
Although possible in principle, this task is prohibitively 
difficult within the standard solution of the QCD gap
equation in weak coupling. So far, in the course of this solution
terms contributing at leading and subleading order have been identified.
However, up to date it remained unclear which terms one would have to
keep at sub-subleading order. Moreover, 
additional contributions could in principle 
arise at any order from diagrams neglected in the
mean-field approximation \cite{rockefeller,DHQWDHR}.
Therefore, it would be ideal to have 
a computational scheme which allows one
to determine {\em a priori\/}, i.e., at the outset of the
calculation, which terms contribute to the gap equation 
at a given order.

As a first step towards this goal, 
note that there are several scales in the problem. 
Besides the chemical potential $\mu$,
there is the inverse gluon screening length which is of
the order of the gluon mass parameter $m_g$. At zero temperature
and for $N_f$ massless quark flavors,
\be \label{gluonmass}
m_g^2 = N_f \, \frac{g^2 \mu^2}{6 \pi^2}\;,
\ee
i.e., $m_g \sim g \mu$.
Finally, there is the color-superconducting gap parameter $\phi$,
cf.\ Eq.\ (\ref{gapsol}). In weak coupling, $g \ll 1$, these three scales
are naturally ordered, $\phi \ll g \mu \ll \mu$.
This ordering of scales implies that the modes near the Fermi surface,
which participate in the formation of Cooper pairs and are therefore of
primary relevance in the gap equation, can be considered to be
independent of the detailed dynamics of the modes deep within the 
Fermi sea. This suggests that the most efficient way to compute 
properties such as the color-superconducting gap parameter is 
via an {\em effective theory for quark modes near the Fermi surface}.
Such an effective theory has been originally proposed by
Hong \cite{hong,hong2} and was subsequently refined by others
\cite{HLSLHDET,schaferefftheory,NFL,others}. 

At this point it is worth reviewing
the standard approach to derive an effective theory
\cite{polchinski,Kaplan}. In the most simple case, one has a single 
scalar field, $\phi$, and a single momentum scale, $\Lambda$, which 
separates relevant modes, $\varphi$, from irrelevant modes, $\psi$, $\phi =
\varphi + \psi$. The relevant modes live on spatial scales 
$L \gg 1/\Lambda$, while the irrelevant modes 
live on scales $l \lesssim 1/ \Lambda \ll L$.
In the derivation of the effective action, one is supposed to integrate out
the microscopic, irrelevant modes. 
Usually, however, this is not done explicitly. Instead, one
constructs all possible operators ${\cal O}_i$ composed of 
powers of the field $\varphi$ and its derivatives,
which are consistent with the symmetries of
the underlying theory, and writes the effective action as
\be \label{Seffstandard}
S_{\rm eff}[\varphi] = \int _X \sum_i g_i {\cal O}_i(\varphi)\;.
\ee
The coefficients, or {\em vertices\/}, $g_i$ determine the interactions of
the relevant modes $\varphi$. A priori, they are unknown 
functions of the single scale $\Lambda$, $g_i = g_i(\Lambda)$. 
All information about the microscopic scale $l$ is contained 
in these vertices. Since the microscopic scale 
$l \ll L$, the operators
${\cal O}_i$ are assumed to be {\em local\/} on the scale $L$.

The effective action (\ref{Seffstandard}) contains infinitely many terms.
In order to calculate physical observables within the effective
theory, one has to truncate the expansion after a finite number of
terms. One can determine the order of magnitude of various terms in the
expansion (\ref{Seffstandard}) via a dimensional scaling analysis which allows
to classify the operators as {\em relevant\/} (they become
increasingly more important as the scale $L$ increases),
{\em marginal\/} (they do not change under scale transformations),
and {\em irrelevant\/} (they become increasingly less 
important as the scale $L$ increases).
To this end, one determines the naive scaling dimension of the
fields, ${\rm dim}(\varphi) \equiv \delta$, from the free
term in the effective action. Then, if the operator ${\cal O}_i$ consists of 
$M$ fields $\varphi$ and $N$ derivatives, its scaling dimension 
is ${\rm dim} ({\cal O}_i) \equiv \delta_i = M \delta + N$.
The operator ${\cal O}_i$ is then of order $\sim L^{-\delta_i}$. 
For dimensional reasons the constant
coefficients $g_i$ must then be of order
$\sim \Lambda^{d-\delta_i}$, where $d$ denotes the dimensionality of
space-time. Including the integration over space-time,
the terms in the expansion (\ref{Seffstandard}) are then of 
order $\sim (L \Lambda)^{d-\delta_i}$.
Consequently, relevant operators must have $\delta_i < d$,
marginal operators $\delta_i = d$, and irrelevant operators $\delta_i > d$.
At a given scale $L$, one has to take into account only relevant, 
or relevant and marginal, or all three types of operators,
depending on the desired accuracy of the calculation.
The final result still depends on $\Lambda$ through the
coefficients $g_i(\Lambda)$. This dependence is 
eliminated by computing a physical observable in the
effective theory and in the underlying microscopic theory, and
matching the result at the scale $\Lambda$.

There are, however, cases where this naive dimensional scaling analysis fails
to identify the correct order of magnitude, and thus the relevance, 
of terms contributing to the effective action.
Let us mention three examples. For the first example, consider effective
theories where, in contrast to the above assumption, 
the vertices $g_i$ are in fact {\em non-local\/} functions. Such 
theories are, for instance, given by
the ``Hard Thermal Loop'' (HTL) or ``Hard Dense Loop'' (HDL) 
effective actions \cite{braatenpisarski,LeBellac}. 
In these effective theories, valid at length scales $L \sim 1/(gT)$ or
$\sim 1/(g \mu)$, respectively,
there are terms $g_n\, A^n$ in the effective action, which are
constructed from a quark or gluon (or ghost) loop
with $n$ external gluon legs;
$A$ is the external gluon field with $\delta = 1$. The 
coefficients $g_n$ are non-local and 
do not only depend on the scale $\Lambda \alt T$, or $\alt \mu$, 
but also on the relevant momentum scale $1/L \sim g T$, or $\sim g \mu$. 
Naively, one would expect
$g_n$ to belong to a local $n$-gluon operator and to
scale like $\Lambda^{4-n}$. Instead,
it scales like $L^{n-4}$ \cite{braatenpisarski}. 
For arbitrary $n$, the corresponding term $g_n\, A^n$
in the effective action then scales like $L^4$, independent
of the number $n$ of external gluon legs.

The second example pertains to the situation when there is
more than one single momentum scale $\Lambda$.
As explained above, for a single scale $\Lambda$ and a given length scale
$L$, the naive dimensional scaling analysis unambiguously determines
the order of magnitude of the terms in the expansion (\ref{Seffstandard}).
Now suppose that there are two scales, $\Lambda_1$ and $\Lambda_2$.
Then, the vertices $g_i$ may no longer be functions of a single scale,
say $\Lambda_1$, but could also depend on the ratio of
$\Lambda_2/\Lambda_1$. 
Two scenarios are possible: (a) two terms in the expansion
(\ref{Seffstandard}), say $g_n {\cal O}_n$ and
$g_m {\cal O}_m$, with the {\em same\/} scaling behavior may still be
of a {\em different\/} order of magnitude, or (b) the two terms can have
a {\em different\/} scaling behavior, but may still be of the {\em
same\/} order of magnitude.
In case (a), all that is required is that the operators ${\cal O}_n$
and ${\cal O}_m$ scale in the same manner, say $L^{-k}$, and that
$g_n \sim \Lambda_2^{d-k}$, but $g_m \sim \Lambda_1^{d-k}$.
If $\Lambda_1 \ll \Lambda_2$, $g_m \gg g_n$, and thus the two
terms are of different order of magnitude.
In case (b), let us assume $1/L \ll \Lambda_1 \ll \Lambda_2$,
with $\Lambda_1/ \Lambda_2 \sim 1/(\Lambda_1 L) \sim \epsilon \ll 1$ and let
us take the fields $\varphi$ to have naive scaling dimension $\delta =1$.
Then, at a given length scale $L$,
a term $g_n \varphi^n$, with a coefficient $g_n$ of order 
$\Lambda_2^{d-n}$, can be of the same order of magnitude 
as a term  $g_m \varphi^m$, $m \neq n$, if the coefficient
$g_m \sim \Lambda_1^{d-m} (\Lambda_2/ \Lambda_1)^k$ with $k= d+ m - 2n$.
Although the scaling behavior of the two terms is quite different as
$L$ increases, they can be of the same order of magnitude,
if the interesting scale $L$ happens to be
$\sim \Lambda_2/\Lambda_1^2$. In both cases (a) and (b) 
the naive dimensional scaling analysis fails to correctly sort the operators
${\cal O}_i$ with respect to their order of magnitude.

The third example where the naive dimensional scaling analysis fails
concerns quantities which have to be calculated self-consistently.
Such a quantity is, for instance,
the color-superconducting gap parameter which is computed
from a Dyson-Schwinger equation within a given many-body
approximation scheme. In this case, the self-consistent solution scheme
leads to large logarithms, like the BCS logarithm in Eq.\
(\ref{gapeq}). These logarithms cannot be identified {\em a priori\/}
on the level of the effective action, but only emerge {\em in the
course\/} of the calculation \cite{rdpdhr}.

In order to avoid these failures of the standard approach,
in this paper we pursue a different venue to 
construct an effective theory. We introduce cut-offs in momentum space
for quarks, $\Lambda_q$, and gluons, $\Lambda_g$. These cut-offs
separate relevant from irrelevant quark modes and soft from hard
gluon modes. We then explicitly integrate out irrelevant quark and
hard gluon modes and derive a general effective action for
hot and/or dense quark-gluon matter. One advantage of this approach is
that we do not have to guess the form of the possible operators
${\cal O}_i$ consistent with the symmetries of the underlying theory.
Instead, they are exactly derived from first principles. 
Simultaneously, the vertices $g_i$ are no longer unknown,
but are completely determined. 
Moreover, in this way we construct {\em all\/} possible operators and thus do
not run into the danger of missing a potentially important one.

We shall show that the standard HTL and HDL effective actions are
contained in our general effective action for a certain choice of the
quark and gluon cut-offs $\Lambda_q,\, \Lambda_g$. Therefore,
our approach naturally generates non-local terms in the
effective action, including their correct scaling behavior which, as
mentioned above, does not follow the rules of the naive 
dimensional scaling analysis. We also show that the action of
the high-density effective theory
derived by Hong and others 
\cite{hong,hong2,HLSLHDET,schaferefftheory,NFL,others}
is a special case of our general effective action. In this case, relevant
quark modes are located within a layer of width $2 \Lambda_q$ around the
Fermi surface.

The two cut-offs, $\Lambda_q$ and $\Lambda_g$, introduced in our approach
are in principle different, $\Lambda_q \neq \Lambda_g$. 
The situation is then as in the second example mentioned above, where the 
naive dimensional scaling analysis fails to
unambiguously estimate the order of magnitude of the various terms in the
effective action. Within the present approach, this problem does not
occur, since all terms, which may occur in the effective action, are 
automatically generated and can be explicitly kept in the further
consideration.
We shall show that in order to produce the correct result for the 
color-superconducting gap parameter to subleading order in weak
coupling, we have to demand $\Lambda_q \alt g \mu \ll \Lambda_g
\alt \mu$, so that $\Lambda_q/\Lambda_g \sim g \ll 1$.
Only in this case, the dominant contribution to the QCD gap equation
arises from almost static magnetic gluon exchange, while subleading
contributions are due to electric and non-static magnetic
gluon exchange.

The color-superconducting gap parameter is computed 
from a Dyson-Schwinger equation for the quark propagator.
In general, this equation corresponds to a self-consistent resummation of
all one-particle irreducible (1PI) diagrams for the quark self-energy.
A particularly convenient way to derive Dyson-Schwinger
equations is via the Cornwall-Jackiw-Tomboulis (CJT) formalism \cite{CJT}. 
In this formalism, one constructs the set
of all two-particle irreducible (2PI) vacuum diagrams from the
vertices of a given tree-level action. The functional derivative of
this set with respect to the full propagator then defines the 
1PI self-energy entering the Dyson-Schwinger equation.
Since it is technically not feasible to include all possible
diagrams, and thus to solve the Dyson-Schwinger equation exactly,
one has to resort to a many-body approximation scheme, which
takes into account only particular classes of diagrams. 
The advantage of the CJT formalism is that such an approximation
scheme is simply defined by a truncation of the set of 2PI diagrams.
However, in principle there is no parameter which controls the
accuracy of this truncation procedure.

The standard QCD gap equation in mean-field approximation studied in Refs.\
\cite{schaferwilczek,rdpdhr,shovkovy} follows from this approach
by including just the sunset-type diagram which is constructed
from two quark-gluon vertices of the QCD tree-level action
(see, for instance, Fig.\ \ref{Gamma2eff} below).
We also employ the CJT formalism to derive the gap equation for
the color-superconducting gap parameter. However, we construct
all diagrams of sunset topology from the vertices of
the general {\em effective\/} action derived in this work.
The resulting gap equation is equivalent to the gap equation in 
QCD, and the result for the gap parameter to subleading order
in weak coupling is identical to that in QCD, provided 
$\Lambda_q \alt g \mu \ll \Lambda_g \alt \mu$.
The advantage of using the effective
theory is that the appearance of the two scales $\Lambda_q$ and
$\Lambda_g$ considerably facilitates the power counting
of various contributions to the gap equation as compared to full QCD.
We explicitly demonstrate this in the course of the calculation 
and suggest that, within this approach, 
it should be possible to 
identify the terms which contribute beyond subleading order
to the gap equation. Of course, for a complete sub-subleading
order result one cannot restrict oneself to the sunset diagram,
but would have to investigate other 2PI diagrams as well.
This again shows that an {\em a priori\/} estimate of the relevance
of different contributions on the level of the effective action
does not appear to be feasible for quantities which have to be computed
self-consistently.

This paper is organized as follows. In Sec.\ \ref{II} we
derive the general effective action by explicitly integrating out
irrelevant quark and hard gluon modes. In Sec.\ \ref{III} we
show that the well-known HTL/HDL effective action, as well as the
high-density effective theory proposed by Hong and others,
are special cases of this general effective action for particular
choices of the quark and gluon cut-offs $\Lambda_q$ and $\Lambda_g$,
respectively. Section \ref{IV} contains the application of the
general effective action to the computation of the
color-superconducting gap parameter. In Sec.\ \ref{V} we conclude
this work with a summary of the results and an outlook.

Our units are $\hbar=c=k_B=1$. 4-vectors are denoted by
capital letters, $K^\mu = (k_0, {\bf k})$, with ${\bf k}$ being a
3-vector of modulus $|{\bf k}| \equiv k$ and direction
$\hat{\bf k}\equiv {\bf k}/k$. For the summation over Lorentz
indices, we use a notation familiar from Minkowski space, with metric
$g^{\mu \nu} = {\rm diag}(+,-,-,-)$, although 
we exclusively work in compact Euclidean space-time with 
volume $V/T$, where $V$
is the 3-volume and $T$ the temperature of the system. 
Space-time integrals 
are denoted as $\int_0^{1/T} d \tau \int_V d^3{\bf x} \equiv
\int_X$. Since space-time is compact, energy-momentum space is
discretized, with sums $(T/V)\sum_{K} \equiv T\sum_n (1/V) \sum_{\bf
k}$. For a large 3-volume $V$, the sum over 3-momenta
can be approximated by an integral, $(1/V)\sum_{\bf k} \simeq
\int d^3 {\bf k}/(2 \pi)^3$. For bosons, the sum over $n$ runs over
the bosonic Matsubara frequencies $\omega_n^{\rm b} = 2n \pi T$, while 
for fermions, it runs over the fermionic Matsubara frequencies
$\omega_n^{\rm f} = (2 n+1)\pi T$. In our Minkowski-like notation 
for four-vectors, $x_0 \equiv t \equiv -i \tau$, 
$k_0 \equiv -i \omega_n^{\rm b/f}$. The 4-dimensional delta-function
is conveniently defined as $\delta^{(4)}(X) \equiv \delta(\tau)\,
\delta^{(3)}({\bf x}) = -i \, \delta(x^0)\, \delta^{(3)}({\bf x})$.

\section{Deriving the effective action} \label{II}

In this section, we derive a general effective action for
hot and/or dense quark matter. We start from the QCD
partition function in the functional integral representation 
(Sec.\ \ref{IIa}). We first
integrate out irrelevant fermion degrees of freedom
(Sec.\ \ref{Intquarks}) and then hard gluon degrees of freedom
(Sec.\ \ref{Intgluons}). The final result is Eq.\ (\ref{Seff}) in
Sec.\ \ref{IId}. We remark that the same result could have been
obtained by first integrating out hard gluon modes, and then
irrelevant fermion modes, but the intermediate steps leading to the 
final result are less transparent.

\subsection{Setting the stage} \label{IIa}

The partition function for QCD in the absence of external
sources reads
\be \label{ZQCD}
{\cal Z} = \int {\cal D} A \, 
\exp \left\{ S_A [A]\right\}\, {\cal Z}_q[A] \,\, .
\ee
Here the (gauge-fixed) gluon action is
\be \label{SA}
S_A[A] = \int_X \left[ - \frac{1}{4} F^{\mu \nu}_a (X) \, F_{\mu \nu}^a
(X) \right] + S_{\rm gf}[A] + S_{\rm ghost}[A] \,\,,
\ee
where $F_{\mu \nu}^a = \partial_\mu A_\nu^a - \partial_\nu A_\mu^a
+ g f^{abc} A_\mu^b A_\nu^c$ is the gluon field strength tensor,
$S_{\rm gf}$ is the gauge-fixing part, and $S_{\rm ghost}$ the
ghost part of the action.

The partition function for quarks in the presence of gluon fields is
\be \label{Zq}
{\cal Z}_q[A] = \int {\cal D} \bar{\psi} \, {\cal D}\psi\,
\exp \left\{ S_q[A,\bar{\psi},\psi] \right\}\,\, ,
\ee
where the quark action is
\be
S_q[A,\bar{\psi},\psi] = \int_{X} \bar{\psi}(X) \,
 \left( i \Diracslash{D}_X + \mu \gamma_0 - m \right) 
 \, \psi(X) \,\,,
\ee
with the covariant derivative $D^\mu_X = \partial_X^\mu - ig
A^\mu_a(X) T_a$; $T_a$ are the generators of the $SU(N_c)_c$
gauge group.
In fermionic systems at nonzero density, it is advantageous to
additionally introduce charge-conjugate fermionic degrees of freedom,
\be \label{cc}
\psi_C(X) \equiv C \, \bar{\psi}^T(X) \;\;, \;\;\;\;
\bar{\psi}_C (X) \equiv \psi^T (X)\, C \;\; ,\;\;\;\;
\psi(X) \equiv C \, \bar{\psi}^T_C(X) \;\;, \;\;\;\;
\bar{\psi} (X) \equiv \psi^T_C (X)\, C \;\; ,
\ee
where $C \equiv i \gamma^2 \gamma_0$ is the charge-conjugation matrix,
$C^{-1} = C^\dagger = C^T = -C$, $C^{-1} \gamma_\mu^T C = - \gamma_\mu$;
a superscript $T$ denotes transposition. We may then rewrite the 
quark action in the form
\be \label{quarkaction}
S_q[A, \bar{\Psi}, \Psi] = 
\frac{1}{2} \int_{X,Y} \bar{\Psi}(X) \, {\cal G}_0^{-1} (X,Y)\,
\Psi(Y) + \frac{g}{2} \int_X \bar{\Psi}(X) \, \hat{\Gamma}^\mu_a
A_\mu^a(X) \, \Psi(X) \,\, ,
\ee
where we defined the Nambu-Gor'kov quark spinors
\be
\Psi \equiv \left( \begin{array}{c}
                    \psi \\
                    \psi_C \end{array} \right) \;\; , \;\;\;\;
\bar{\Psi} \equiv ( \bar{\psi} , \bar{\psi}_C )\,\, ,
\ee
and the free inverse quark propagator in the Nambu-Gor'kov basis
\be
{\cal G}_0^{-1}(X,Y)  \equiv \left( \begin{array}{cc} 
                          [G_0^+]^{-1}(X,Y) & 0 \\
                           0 & [G_0^-]^{-1}(X,Y) \end{array} \right)\,\, ,
\ee
with the free inverse propagator for quarks and charge-conjugate
quarks
\be
[G_0^\pm]^{-1}(X,Y) \equiv (i \Diracslash{\partial}_X \pm \mu
\gamma_0 - m )\, \delta^{(4)}(X-Y)\,\, .
\ee
The quark-gluon vertex in the Nambu-Gor'kov basis is defined as
\be \label{NGvertex}
\hat{\Gamma}^\mu_a \equiv \left( \begin{array}{cc}
                               \gamma^\mu T_a & 0 \\
                               0 & -\gamma^\mu T_a^T 
                              \end{array} \right) \,\, .
\ee
As we shall derive the effective action in momentum space,
we Fourier-transform all fields, as well as the free inverse
quark propagator,
\begin{subequations} \label{FT}
\bea
\Psi(X) & = & \frac{1}{\sqrt{V}} \sum _K e^{-i K \cdot X} \, \Psi(K)
\,\, , \\
\bar{\Psi}(X) & = & \frac{1}{\sqrt{V}} \sum_K e^{ i K \cdot X} \,
\bar{\Psi}(K)
\,\, , \\
{\cal G}_0^{-1}(X,Y) & = & \frac{T^2}{V} \sum_{K,Q} e^{-i K \cdot X}\,
e^{i Q \cdot Y} \, {\cal G}_0^{-1}(K,Q)
\,\, , \\
A^\mu_a(X) & = & \frac{1}{\sqrt{TV}} \sum_P e^{-i P \cdot X} \,
A^\mu_a(P) 
\,\, . \label{FTA}
\eea
\end{subequations}
The normalization factors are chosen such that the Fourier-transformed
fields are dimensionless quantities.
The Fourier-transformed free inverse quark propagator is diagonal
in momentum space, too,
\be 
\label{G0FT}
{\cal G}_0^{-1}(K,Q) = \frac{1}{T} \left( \begin{array}{cc}
                             [G_0^+]^{-1}(K) & 0 \\
                              0 & [G_0^-]^{-1}(K) \end{array} \right)
\delta^{(4)}_{K,Q} \,\, ,
\ee
where $[G_0^\pm]^{-1}(K) \equiv \Diracslash{K} \pm \mu \gamma_0 - m$.

Due to the relations (\ref{cc}), the Fourier-transformed
charge-conjugate quark fields are related to the original fields
via $\psi_C(K) = C \bar{\psi}^T(-K)$, $\bar{\psi}_C(K) = \psi^T(-K)C$.
The measure of the functional integration over quark fields
can then be rewritten in the form
\bea
{\cal D} \bar{\psi} \, {\cal D} \psi & \equiv & \prod_K d\bar{\psi}(K) 
\, d \psi(K) = {\cal N} \prod_{(K,-K)} d\bar{\psi}(K) \,
d \psi(K)\, d\bar{\psi}(-K) \, d \psi(-K) \nonumber \\
& = & {\cal N}' \prod_{(K,-K)} d\bar{\psi}(K) \,
d \psi(K)\, d\bar{\psi}_C(K) \, d \psi_C(K)
= {\cal N}'' \prod_{(K,-K)} d \bar{\Psi}(K) \, d\Psi(K)
\equiv {\cal D} \bar{\Psi}\, {\cal D}{\Psi} \,\, , \label{measure}
\eea
with the constant normalization factors ${\cal N},\, {\cal N}',
\, {\cal N}''$.
The last identity has to be considered as a definition for
the expression on the right-hand side.

Inserting Eqs.\ (\ref{FT}) -- (\ref{measure}) into Eq.\ (\ref{Zq}),
the partition function for quarks becomes
\be \label{Zq2}
{\cal Z}_q [A]= \int {\cal D} \bar{\Psi}\, 
{\cal D} \Psi \exp \left[ \frac{1}{2} \, \bar{\Psi} \left( 
{\cal G}_0^{-1} + g {\cal A} \right) \Psi \right]
\,\, .
\ee
Here, we employ a compact matrix notation,
\be \label{compact}
\bar{\Psi} \, \left( {\cal G}_0^{-1} + g {\cal A} \right)\,   \Psi \equiv
\sum_{K,Q} \bar{\Psi}(K) \, \left[ {\cal G}_0^{-1}(K,Q) 
+ g {\cal A}(K,Q) \right] \, \Psi(Q)\;\; ,
\ee
with the definition
\be \label{calA}
{\cal A}(K,Q) \equiv \frac{1}{\sqrt{VT^3}}\, \hat{\Gamma}^\mu_a
A_\mu^a(K-Q) \,\, .
\ee
The next step is to integrate out irrelevant quark modes.

\subsection{Integrating out irrelevant quark modes} \label{Intquarks}

Since we work in a finite volume $V$, 
the 3-momentum ${\bf k}$ is discretized.
Let us for the moment also assume that there is an ultraviolet
cut-off (such as in a lattice regularization) on the
3-momentum, i.e., the space of modes labelled by
3-momentum has dimension $D < \infty$.
We define projection operators ${\cal P}_1,\, {\cal P}_2$ for 
relevant and irrelevant quark modes, respectively,
\be \label{project}
\Psi_1 \equiv {\cal P}_1 \, \Psi\;\; , \;\;\;\;
\Psi_2 \equiv {\cal P}_2 \, \Psi \;\; , \;\;\;\;
\bar{\Psi}_1 \equiv \bar{\Psi} \, \gamma_0 {\cal P}_1 \gamma_0 
\;\; , \;\;\;\;
\bar{\Psi}_2 \equiv \bar{\Psi} \, \gamma_0 {\cal P}_2 \gamma_0 
\;\;.
\ee
The subspace of relevant quark modes has dimension $N_1$ in
the space of 3-momentum modes, 
the one for irrelevant modes dimension $N_2$, with $N_1 + N_2 = D$.

At this point, it is instructive to give an explicit
example for the projectors ${\cal P}_{1,2}$. In the effective theory 
for cold, dense quark matter, which contains
the high-density effective theory
\cite{hong,hong2,HLSLHDET,schaferefftheory,NFL,others} 
discussed in Sec.\ \ref{IIIB} as special case
and which we shall apply in Sec.\ \ref{IV} to the computation of 
the gap parameter, the projectors are chosen as
\begin{subequations} \label{P12}
\bea
{\cal P}_1(K,Q) & \equiv & \left( \begin{array}{cc}
 \Lambda_{\bf k}^+ & 0 \\
 0 & \Lambda_{\bf k}^- \end{array} \right) \, 
\Theta(\Lambda_q - | k - k_F|) \, \delta^{(4)}_{K,Q} \;, \\
{\cal P}_2(K,Q) & \equiv & \left( \begin{array}{cc} 
\Lambda_{\bf k}^- + \Lambda_{\bf k}^+\, \Theta(| k - k_F| - \Lambda_q)
& 0 \\
0 &  \Lambda_{\bf k}^+ + \Lambda_{\bf k}^-\, \Theta(| k - k_F| -
\Lambda_q) \end{array} \right)\, 
\delta^{(4)}_{K,Q}\;.
\eea
\end{subequations}
Here, 
\be
\Lambda^e_{\bf k} \equiv \frac{1}{2 E_{\bf k}} \, 
\left[ E_{\bf k} + e \gamma_0 \left(\vg \cdot
{\bf k} + m \right) \right] \,\, , 
\ee 
are projection operators onto states with positive ($e = +$)
or negative ($e=-$) energy, where $E_{\bf k} = \sqrt{{\bf k}^2 +m^2}$
is the relativistic single-particle energy.
The momentum cut-off $\Lambda_q$ 
controls how many quark modes (with positive
energy) are integrated out. Thus, all quark modes within a layer of 
width $2 \Lambda_q$ around the Fermi surface are considered as 
relevant, while all antiquark modes and quark modes
outside this layer are considered as irrelevant.
Note that,
for the Nambu-Gor'kov components corresponding to charge-conjugate
particles, the role of the projectors onto positive and negative energy
states is reversed with respect to the Nambu-Gor'kov components
corresponding to particles.
The reason is that, loosely speaking, a particle is actually a 
charge-conjugate antiparticle. For a more rigorous proof
compute, for instance, $\psi_{C,1}(K) \equiv C \, \bar{\psi}_1^T(-K)$ using 
$\bar{\psi}_1(-K) = \bar{\psi}(-K) \,\gamma_0 \Lambda^+_{-{\bf
k}}\gamma_0$ (for $|k-k_F| \leq \Lambda_q$)
and $\gamma_0\, C [\Lambda_{-{\bf k}}^+]^T C^{-1}\,\gamma_0= 
\Lambda_{\bf k}^-$.
In Sec.\ \ref{III} we shall discuss other
choices for the projectors ${\cal P}_{1,2}$, pertaining to other
effective theories of hot and/or dense quark matter.
The following discussion in this section, however, will be completely
general and is not restricted to any particular choice for these projectors.

Employing Eq.\ (\ref{project}), the partition function (\ref{Zq2}) becomes
\be \label{Zq3}
{\cal Z}_q[A] = \int \prod_{n=1,2} {\cal D} \bar{\Psi}_n\, 
{\cal D} \Psi_n \, \exp \left( \frac{1}{2} \sum_{n,m=1,2} \bar{\Psi}_n \, 
{\cal G}^{-1}_{nm} \, \Psi_m \right)
\,\, .
\ee
{}From now on,  $\bar{\Psi}_{1,2}$, $\Psi_{1,2}$ are
considered as vectors restricted to the $N_{1,2}$-dimensional subspace of
relevant/irrelevant 3-momentum modes.
The matrices ${\cal G}_{nn}^{-1},\, n=1,2,$ are defined as
\be
{\cal G}_{nn}^{-1}(K,Q) = {\cal G}_{0,nn}^{-1}(K,Q) + g {\cal A}_{nn}(K,Q)\;,
\ee
where the indices indicate that, 
for a given pair of quark energies $k_0,\ q_0$, 
the 3-momenta ${\bf k}, \, {\bf q}$ belong to the subspace of
relevant ($n=1$) or irrelevant ($n=2$) quark modes, i.e.,
${\cal G}_{nn}^{-1}$ is an ($N_n \times N_n$)-dimensional matrix in
3-momentum space.
The matrices ${\cal G}_{nm}^{-1}$, $n \neq m$, reduce to
\be
{\cal G}_{nm}^{-1}(K,Q) = g \, {\cal A}_{nm}(K,Q)\;,
\ee
since ${\cal G}_0^{-1}$ is diagonal
in 3-momentum space, i.e. ${\cal G}_{0,nm}^{-1} \equiv 0$ for
$n \neq m$. For a given pair of quark energies
$k_0,\, q_0$, ${\cal G}_{nm}^{-1}$ is a $(N_n \times
N_m)$-dimensional matrix in 3-momentum space.

The Grassmann integration over the irrelevant quark fields $\bar{\Psi}_2,\,
\Psi_2$ can be done exactly, if one redefines them such that
the mixed terms $\sim {\cal G}_{nm}^{-1}$, $n \neq m$, are eliminated.
To this end, substitute 
\be
\Upsilon \equiv \Psi_2 + {\cal G}_{22}\, {\cal G}_{21}^{-1}\, \Psi_1\;\; ,
\;\;\;\; 
\bar{\Upsilon} \equiv \bar{\Psi}_2 + \bar{\Psi}_1\, 
{\cal G}_{12}^{-1}\, {\cal G}_{22}\; ,
\ee 
where ${\cal G}_{22}$ is the inverse of ${\cal G}_{22}^{-1}$, defined
on the subspace of irrelevant quark modes.
The result is
\be \label{Zq4}
{\cal Z}_q[A] 
= \int {\cal D} \bar{\Psi}_1\, 
{\cal D} \Psi_1 \, \exp \left[ \frac{1}{2} \, \bar{\Psi}_1 
\left( {\cal G}^{-1}_{11} - {\cal G}^{-1}_{12} \, {\cal G}_{22}
\, {\cal G}^{-1}_{21} \right) \Psi_1
+ \frac{1}{2}\, {\rm Tr}_q \ln {\cal G}_{22}^{-1} \right] \,\, .
\ee
The trace in the last term runs over all irrelevant quark momenta
$K$, and not only over pairs $(K,-K)$, as prescribed by 
the integration measure, Eq.\ (\ref{measure}).
This requires an additional factor $1/2$ in front of the trace. 
A more intuitive way of saying this is that this factor accounts 
for the doubling of the quark degrees of freedom in the Nambu-Gor'kov basis.
Of course, the trace runs not only over 4-momenta, but also 
over other quark indices, such as Nambu-Gor'kov, fundamental color, 
flavor, and Dirac indices. We indicated this by the subscript ``$q$''.

\begin{figure}[ht]
\includegraphics[width=13cm]{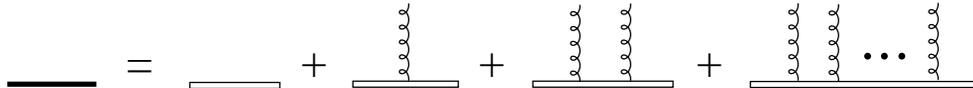}
\caption{The full propagator for irrelevant quarks. The right-hand
side symbolizes the expansion (\ref{expquark}). The free irrelevant
quark propagators ${\cal G}_{0,22}$ are denoted by double lines, the
gluon fields ${\cal A}_{22}$ by curly lines.}
\label{XXd}
\end{figure}

For a diagrammatic interpretation, it is advantageous to rewrite
\be
{\cal G}^{-1}_{11} - {\cal G}^{-1}_{12} \, {\cal G}_{22}
\, {\cal G}^{-1}_{21} 
\equiv {\cal G}_{0,11}^{-1}  + g {\cal B}\;,
\ee
where
\be \label{B}
g{\cal B} \equiv g {\cal A}_{11} - g {\cal A}_{12} \, 
{\cal G}_{22} \, g {\cal A}_{21}\;.
\ee
The propagator for irrelevant quark modes,
${\cal G}_{22}$, has an expansion in powers of $g$ times the gluon field,
\be  \label{expquark}
{\cal G}_{22} = {\cal G}_{0,22} \sum_{n=0}^\infty 
(-1)^n g^n \left[ {\cal A}_{22}\, {\cal G}_{0,22} \right]^n\;.
\ee
This expansion is graphically depicted in Fig.\ \ref{XXd}.

\begin{figure}[ht]
\includegraphics[width=11cm]{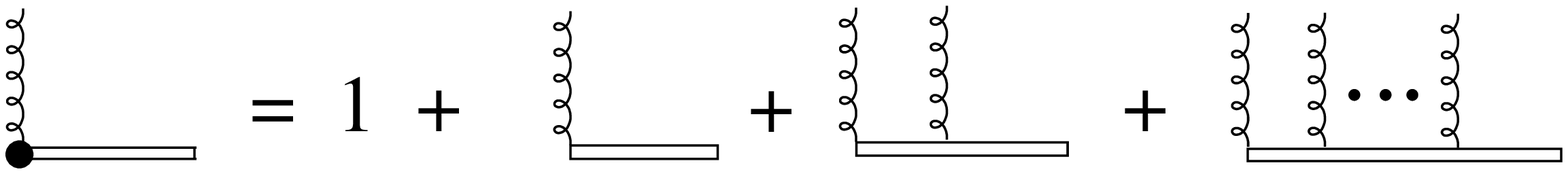}
\caption{The diagrammatic symbol for the factor 
$\left(1 + g {\cal A}\, {\cal G}_{0,22} \right)^{-1}$.}
\label{XXb}
\end{figure}

\begin{figure}[ht]
\includegraphics[width=13cm]{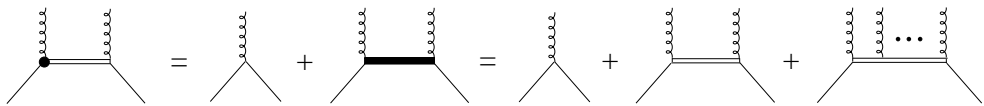}
\caption{The term $\bar{\Psi}_1\, g {\cal B} \, \Psi_1$. A relevant
quark field is denoted by a single solid line.}
\label{XXa}
\end{figure}

\begin{figure}[ht]
\includegraphics[width=12cm]{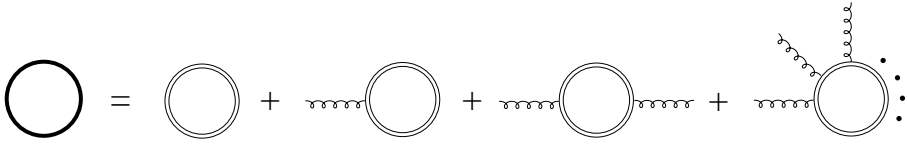}
\caption{The graphical representation of the term ${\rm Tr}_q \ln
{\cal G}_{22}^{-1}$ in Eq.\ (\ref{Zq4}).}
\label{XXc}
\end{figure}

Using this expansion, and suppressing the indices on ${\cal A}$,
Eq.\ (\ref{B}) can be symbolically written as
\be
g {\cal B} = \left( 1 + g {\cal A}\, {\cal G}_{0,22} \right)^{-1}
g {\cal A}\;,
\ee
which suggests the interpretation of the field ${\cal B}$ as
a ``modified'' (non-local) gluon field. In the diagrams to
be discussed below, the factor
$\left(1 + g {\cal A}\, {\cal G}_{0,22} \right)^{-1}$ will
be denoted by the diagrammatical symbol shown in Fig.\ \ref{XXb}.
With this symbol, the expression $\bar{\Psi}_1\, g {\cal B} \,
\Psi_1$ can be graphically depicted as shown in Fig.\ \ref{XXa}.

Since
\be \label{explnquark}
\ln {\cal G}_{22}^{-1} = \ln {\cal G}_{0,22}^{-1} 
- \sum_{n=1}^\infty \frac{(-1)^n}{n} \, g^n \, \left[ {\cal G}_{0,22}
\, {\cal A}_{22} \right]^n\; ,
\ee
the last term in the exponent in Eq.\ (\ref{Zq4}) also has a
graphical interpretation, shown in Fig.\ \ref{XXc}.

This concludes the integration over irrelevant quark modes. Note that
our treatment is (i) exact in the sense that no approximations have been
made and (ii) completely general, since it is independent of the
specific choice (\ref{P12}) for the projection operators.
The next step is to integrate out hard gluon modes.

\subsection{Integrating out hard gluon modes} \label{Intgluons}

Combining Eqs.\ (\ref{ZQCD}), (\ref{Zq4}), and (\ref{B}),
the partition function of QCD for relevant quark
modes and gluons reads
\begin{subequations} \label{ZQCD2}
\bea 
{\cal Z} & = &
\int {\cal D} \bar{\Psi}_1 \,  {\cal D} \Psi_1\, 
{\cal D} A \, \exp \left\{ S[A,\bar{\Psi}_1,\Psi_1] \right\}\,  \; \\
S[A,\bar{\Psi}_1,\Psi_1] & \equiv &
S_A[A] + \frac{1}{2} \, \bar{\Psi}_1 
\left\{ {\cal G}_{0,11}^{-1} +  g {\cal B}[A] \right\} \Psi_1
+ \frac{1}{2}\, {\rm Tr}_q \ln {\cal G}_{22}^{-1}[A] \;, \label{S}
\eea
\end{subequations}
where ${\cal D} A \equiv \prod_P d A(P)$. For
the sake of clarity, we restored
the functional dependence of the ``modified'' gluon field
${\cal B}$ and the inverse irrelevant quark propagator 
${\cal G}_{22}^{-1}$ on the gluon field $A$.

The gluon action in momentum space is 
\bea 
S_A [A] & = &
- \frac{1}{2} \sum_{P_1,P_2} A_\m^a(P_1)
\left[\Delta^{-1}_0\right]^{\mu\nu}_{ab}(P_1,P_2) A_\nu^b(P_2) \non
&  &
- \frac{1}{3!} \, \frac{g}{\sqrt{VT^3}} 
\sum_{P_1,P_2,P_3} \delta^{(4)}_{P_1+P_2+P_3,0} \,
{\cal V}_{\a \b \g}^{abc}(P_1,P_2,P_3) \,
A^{\a}_{a}(P_1) A^{\b}_{b}(P_2)A^{\g}_{c}(P_3)\non
&  & 
- \frac{1}{4!}\, \left(\frac{g}{\sqrt{VT^3}}\right)^2 
\sum_{P_1, \cdots ,P_4} \delta^{(4)}_{P_1+P_2+P_3+P_4,0}\, 
{\cal V}_{\a \b \g \d}^{abcd}\, 
A^{\a}_{a}(P_1) A^{\b}_{b}(P_2) A^{\g}_{c}(P_3) A^{\d}_{d}(P_4)  \non
&   &
+ {\rm Tr}_{gh} \ln {\cal W}^{-1} \;.
\label{Sgluon}
\eea
Here, $\Delta^{-1}_0(P_1,P_2)$ is the gauge-fixed inverse free gluon
propagator. 
To be specific, in general Coulomb gauge it reads
\begin{subequations}\label{D_0}
\bea 
\left[\Delta^{-1}_0 \right]^{\mu \nu}_{ab}(P_1,P_2) & \equiv& 
\frac{1}{T^2} \, \left[ \Delta^{-1}_0 \right]^{\mu \nu}_{ab}(P_1)\, 
\delta^{(4)}_{P_1, -P_2}\;, \\
\left[\Delta^{-1}_0\right]^{\mu \nu}_{ab}(P) & = & \delta_{ab}\, 
\left(P^2 g^{\mu \nu} -P^\mu P^\nu + \frac{1}{\xi_C}
\tilde{P}^\mu \tilde{P}^\nu \right)\;,
\eea
\end{subequations}
where $\xi_C$ is the Coulomb gauge parameter and $\tilde{P}^\mu \equiv
(0, {\bf p})$.
The vertex functions are
\begin{subequations}
\bea
\lefteqn{ {\cal V}_{\a \b \g}^{abc} (P_1,P_2,P_3) \equiv 
\frac{i}{T}\, f^{abc} \, \left[ (P_1-P_2)_{\g}\, g_{\a \b}+
(P_2-P_3)_{\a}\, g_{\b \g}+ (P_3-P_1)_{\b}\, g_{\a \g} \right]\;,} \\
{\cal V}^{abcd}_{\a \b \g \d} & \equiv & 
 f^{abe} f^{ecd} \left( g_{\a\g} g_{\b \d} -
g_{\a \d} g_{\b \g} \right) 
+ f^{ace} f^{ebd}  \left( g_{\a\b} g_{\g \d} -
g_{\a \d} g_{\b \g} \right) 
+ f^{ade} f^{ebc}  \left( g_{\a\b} g_{\g \d} -
g_{\a \g} g_{\b \d} \right) \;. 
\eea
\end{subequations}
The last term in Eq.\ (\ref{Sgluon}) is the trace of the
logarithm of the Faddeev-Popov
determinant, with the full inverse ghost propagator ${\cal W}^{-1}$.
The trace runs over ghost 4-momenta and adjoint color indices.

Similar to the treatment of fermions in Sec.\ \ref{Intquarks}
we now define projectors ${\cal Q}_1,\, {\cal Q}_2$ for
soft and hard gluon modes, respectively,
\be
A_1  \equiv {\cal Q}_1 \, A \;\; , \;\;\;\; A_2 \equiv {\cal Q}_2 \,
A\;,
\ee
where
\begin{subequations} \label{Q12}
\bea
{\cal Q}_1(P_1,P_2) & \equiv & \Theta(\Lambda_g -p_1) \, \delta^{(4)}_{P_1,P_2}
\; , \\
{\cal Q}_2(P_1,P_2) & \equiv & \Theta(p_1-\Lambda_g) \, \delta^{(4)}_{P_1,P_2}
\; .
\eea
\end{subequations}
The gluon cut-off momentum $\Lambda_g$ defines which gluons are
considered to be soft or hard, respectively.

\begin{figure}[ht]
\includegraphics[width=12cm]{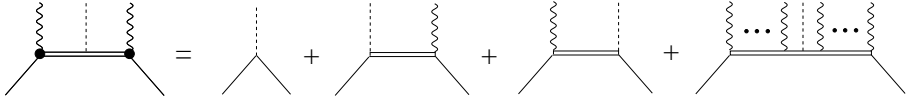}
\caption{The term $A_2 {\cal J}_{{\cal B}}$. The hard gluon field
is denoted by a dashed line, the soft gluon fields by wavy lines.}
\label{YYa}
\end{figure}

\begin{figure}[ht]
\includegraphics[width=12cm]{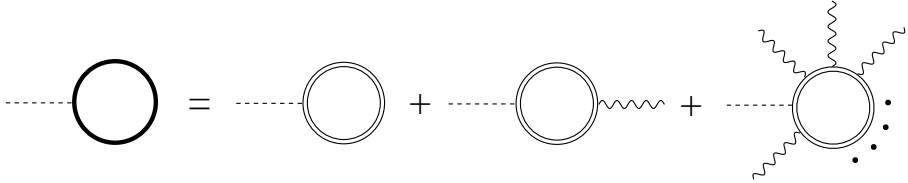}
\caption{The fermionic contribution to the
term $A_2 {\cal J}_{\rm loop}$. There is an additional contribution
from ghosts with similar topology.}
\label{YYb}
\end{figure}

\begin{figure}[ht]
\includegraphics[width=3cm]{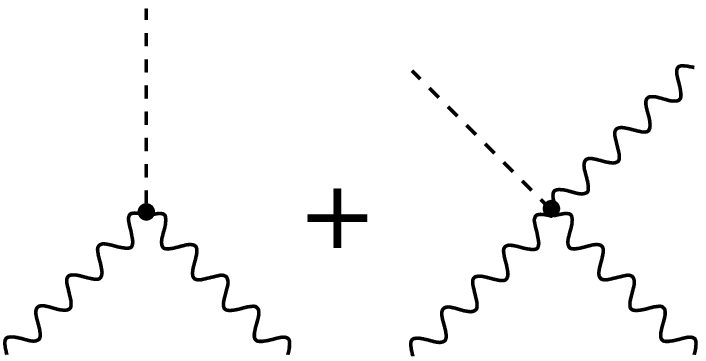}
\caption{The term $A_2 {\cal J}_{{\cal V}}$.}
\label{YYc}
\end{figure}

We now insert $A \equiv A_1 + A_2$ into 
Eq.\ (\ref{ZQCD2}). The integration measure simply
factorizes, ${\cal D} A \equiv {\cal D} A_1\, {\cal D} A_2$.
The action $S[A,\bar{\Psi}_1,\Psi_1]$ can be sorted with respect to
powers of the hard gluon field,
\be \label{expansion}
S[A,\bar{\Psi}_1,\Psi_1] = S[A_1,\bar{\Psi}_1,\Psi_1]
+ A_2 {\cal J} [A_1,\bar{\Psi}_1,\Psi_1] - \frac{1}{2}\,
A_2 \, \Delta^{-1}_{22}[A_1, \bar{\Psi}_1,\Psi_1]\, A_2 
+ S_I[A_1,A_2,\bar{\Psi}_1,\Psi_1] \;.
\ee 
The first term in this expansion, containing no hard gluon fields
at all, is simply the action (\ref{S}), with $A$ replaced by
the relevant gluon field $A_1$.
The second term, $A_2 {\cal J}$,
contains a single power of the hard gluon field, where
\be \label{J}
{\cal J} [A_1,\bar{\Psi}_1,\Psi_1] \equiv 
\left. \frac{\delta S[A, \bar{\Psi}_1,\Psi_1]}{\delta A_2}
\right|_{A_2 =0} \equiv {\cal J}_{{\cal B}}[A_1, \bar{\Psi}_1, \Psi_1]
+ {\cal J}_{\rm loop}[A_1] + {\cal J}_{{\cal V}}[A_1]\,\, .
\ee
The first contribution, 
\be \label{J_B}
{\cal J}_{{\cal B}}[A_1,\bar{\Psi}_1,\Psi_1] 
= \frac{1}{2} \, \bar{\Psi}_1 \, \left( g \, \frac{\delta
{\cal B}}{\delta A_2}\right)_{A_2=0} \Psi_1 \;,
\ee 
arises from the coupling of the relevant fermions to the
``modified'' gluon field ${\cal B}$, i.e., from the second term in 
Eq.\ (\ref{S}). With the notation of Fig.\ \ref{XXb},
all diagrams corresponding to $A_2 {\cal J}_{{\cal B}}$ 
can be summarized into a
single one, cf.\ Fig.\ \ref{YYa}. It contains precisely two relevant
fermion fields, $\bar{\Psi}_1$ and $\Psi_1$.
The second contribution, ${\cal J}_{\rm loop}$, 
arises from the terms ${\rm Tr}_q \ln {\cal
G}_{22}^{-1}$ and ${\rm Tr}_{gh} \ln {\cal W}^{-1}$ in 
Eqs.\ (\ref{S}), (\ref{Sgluon}). The loop consisting of
irrelevant quark modes as internal lines, coupled to a single hard
and arbitrarily many soft gluons, is shown in Fig.\ \ref{YYb}.
Finally, the third contribution, ${\cal J}_{{\cal V}}$, 
arises from the non-Abelian vertices, cf.\ Fig.\ \ref{YYc}.

\begin{figure}[ht]
\includegraphics[width=12cm]{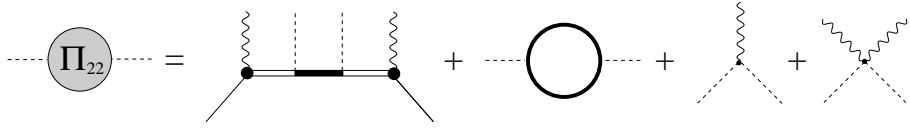}
\caption{The term $A_2 \Pi_{22} A_2$ according to Eq.\ (\ref{Pi}). 
The first diagram on the right-hand side corresponds to the term 
$A_2 \Pi_{\cal B}A_2 $. The second diagram is the fermion-loop contribution
to $A_2 \Pi_{\rm loop} A_2$; there is an analogous one from a
ghost loop. The last two diagrams correspond
to $A_2 \Pi_{\cal V} A_2$.}
\label{FigPi}
\end{figure}

\begin{figure}[ht]
\includegraphics[width=13cm]{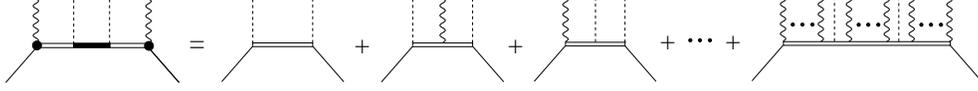}
\caption{The term $A_2 \Pi_{{\cal B}}A_2$.}
\label{ZZa}
\end{figure}

\begin{figure}[ht]
\includegraphics[width=12cm]{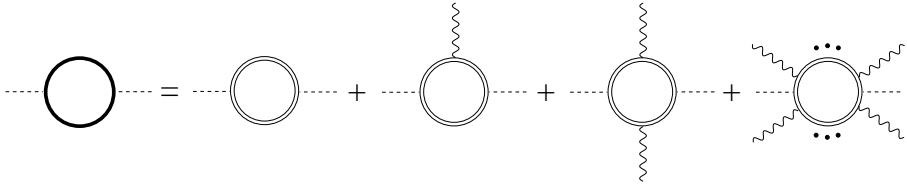}
\caption{The fermionic contribution to the term $A_2 \Pi_{\rm loop}
A_2$.}
\label{ZZb}
\end{figure}

\begin{figure}[ht]
\includegraphics[width=3cm]{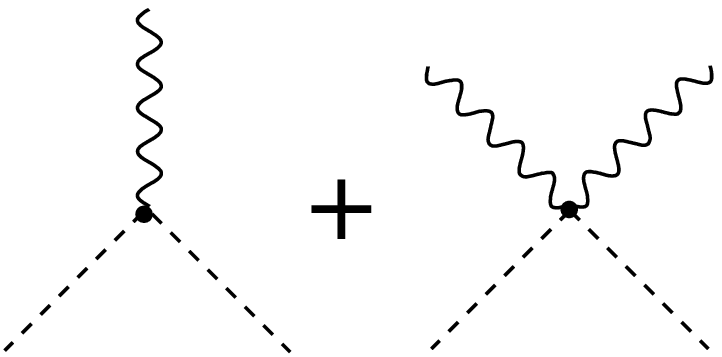}
\caption{The term $A_2 \Pi_{{\cal V}} A_2$.}
\label{ZZc}
\end{figure}

The third term in Eq.\ (\ref{expansion}) is quadratic in $A_2$, 
where
\be
\Delta_{22}^{-1}[A_1,\bar{\Psi}_1,\Psi_1]\equiv 
- \left. \frac{\delta^2 S[A, \bar{\Psi}_1,\Psi_1]}{\delta A_2\,\delta A_2}
\right|_{A_2 =0} \equiv \Delta_{0,22}^{-1} + 
\Pi_{22}[A_1, \bar{\Psi}_1, \Psi_1]\,\, .
\ee
Here, $\Delta_{0,22}^{-1}$ is the free inverse propagator 
for hard gluons.
Similar to the ``current'' ${\cal J}$, cf.\ Eq.\ (\ref{J}),
the ``self-energy'' $\Pi_{22}$ of hard gluons consists of three
different contributions,
\be \label{Pi}
\Pi_{22}[A_1, \bar{\Psi}_1, \Psi_1] =
\Pi_{{\cal B}}[A_1, \bar{\Psi}_1, \Psi_1] +
\Pi_{\rm loop}[A_1] +
\Pi_{{\cal V}}[A_1] \;,
\ee
which has a diagrammatic representation as shown in Fig.\ \ref{FigPi}.
The first two contributions on the right-hand side of Eq.\ (\ref{Pi})
can be expanded as shown in Figs.\ \ref{ZZa} and \ref{ZZb}. 
Figure \ref{ZZc} depicts the three- and four-gluon vertices 
contained in the last term in Eq.\ (\ref{Pi}).
For further use, we explicitly give the first term,
\be
\Pi_{{\cal B}}[A_1, \bar{\Psi}_1, \Psi_1] =
- \frac{1}{2} \, \bar{\Psi}_1 \, \left( g \, \frac{\delta^2
{\cal B}}{\delta A_2 \delta A_2}\right)_{A_2=0} \Psi_1 \;.
\ee

Finally, we collect all terms with more than
two hard gluon fields $A_2$ in Eq.\ (\ref{expansion}) in the
``interaction action'' for hard gluons, 
$S_I[A_1,A_2,\bar{\Psi}_1,\Psi_1]$.
We then perform the functional integration over the hard gluon fields $A_2$.
Since functional integrals must be of Gaussian type in order to be 
exactly solvable, we resort to a method well-known from perturbation theory. 
We add the source term $A_2 J_2$ to the action (\ref{S})
and may then replace the fields $A_2$ in $S_I$
by functional differentiation with respect to $J_2$, at
$J_2=0$. We then move
the factor $\exp\{ S_I[A_1, \delta/\delta J_2, \bar{\Psi}_1, \Psi_1 ]\}$
in front of the functional $A_2$-integral. Then, this functional
integral is Gaussian and can be readily
performed (after a suitable shift of $A_2$), with the result
\bea
{\cal Z} & = & \int {\cal D} \bar{\Psi}_1 \, {\cal D} \Psi_1 {\cal D} A_1
\, \exp\left\{ S[A_1,\bar{\Psi}_1, \Psi_1 ]- \frac{1}{2}\, {\rm Tr}_g \ln
\Delta_{22}^{-1} \right\} \non
&   & \times \left. 
\exp \left\{ S_I\left[A_1, \frac{\delta}{\delta J_2}, 
\bar{\Psi}_1, \Psi_1 \right]\right\}
\,  \exp \left[  \frac{1}{2} \, ({\cal J} + J_2) \,
\Delta_{22}\, ({\cal J} + J_2) \right] \right|_{J_2 = 0}\; . 
\label{Z3}
\eea
The trace over $\ln \Delta_{22}^{-1}$ runs over gluon 4-momenta, as
well as adjoint color and Lorentz indices. We indicate this with
a subscript ``$g$''.
Note that this result is still exact and completely general, since
so far our manipulations of the partition function were independent of
the specific choice (\ref{Q12}) 
for the projection operators ${\cal Q}_{1,2}$.
The next step is to derive the tree-level action
for the effective theory of relevant quark modes and soft gluons.

\subsection{Tree-level effective action} \label{IId}

In order to derive the tree-level effective action, we shall employ
two approximations. The first is based on the
principle assumption in the construction of any
effective theory, namely that soft and hard modes are well separated 
in momentum space. Consequently, momentum conservation
does not allow a hard gluon to
couple to any (finite) number of soft gluons. Under this assumption, the
diagrams generated by $A_2 ({\cal J}_{\rm loop} + {\cal J}_{{\cal V}})$,
cf.\ Fig.\ \ref{YYb}, \ref{YYc}, will not occur in the effective theory.
In the following, we shall therefore omit these terms, so that
${\cal J} \equiv {\cal J}_{{\cal B}}$. 
Note that similar arguments cannot be applied to
the diagrams generated by $A_2 (\Pi_{\rm loop} + \Pi_{{\cal V}}) A_2$,  
cf.\ Fig.\ \ref{ZZb}, \ref{ZZc}, 
since now there are two hard gluon legs which take care of 
momentum conservation.

Our second approximation is that in the ``perturbative'' expansion
of the partition function (\ref{Z3}) with respect to powers
of the interaction action $S_I$, we only take the first term, i.e.,
we approximate $e^{S_I} \simeq 1$. This is analogous to the
derivation of the exact renormalization group in Ref.\ \cite{Wegner},
where it was shown that the corresponding diagrams are of higher order 
and can be neglected.
In our case, diagrams generated by $e^{S_I}$ are those with more
than one {\em resummed\/} hard gluon line.
Even with the approximation $e^{S_I} \simeq 1$, Eq.\ (\ref{Z3}) 
still contains diagrams with
arbitrarily many {\em bare\/} hard gluon lines, arising 
from the expansion of 
\be \label{expansion2}
\ln \Delta_{22}^{-1} = \ln \Delta_{0,22}^{-1}
- \sum_{n=1}^{\infty} \frac{(-1)^n}{n} \, \left(\Delta_{0,22}\, \Pi_{22}
\right)^n \;,
\ee
and from the term ${\cal J}_{{\cal B}} \Delta_{22} {\cal J}_{{\cal B}}$ in
Eq.\ (\ref{Z3}), when expanding
\be \label{expansion3}
\Delta_{22} = \Delta_{0,22} \sum_{n=0}^{\infty} (-1)^n \left( \Pi_{22} \,
\Delta_{0,22} \right)^n\;.
\ee

With these approximations, the partition function reads
\be \label{Z4}
{\cal Z} =  \int {\cal D} \bar{\Psi}_1 \, {\cal D} \Psi_1 {\cal D} A_1\,
\exp\{S_{\rm eff} [A_1,\bar{\Psi}_1, \Psi_1 ] \}\;,
\ee
where the effective action is defined as
\bea 
S_{\rm eff} [A_1,\bar{\Psi}_1, \Psi_1 ] & \equiv & 
S_A[A_1] + \frac{1}{2} \, \bar{\Psi}_1 
\left\{ {\cal G}_{0,11}^{-1} +  g {\cal B}[A_1] \right\} \Psi_1
+ \frac{1}{2}\, {\rm Tr}_q \ln {\cal G}_{22}^{-1}[A_1] 
- \frac{1}{2}\,  {\rm Tr}_g \ln \Delta_{22}^{-1}[A_1,\bar{\Psi}_1,\Psi_1] \non
&  & + \; \frac{1}{2} \, {\cal J}_{{\cal B}}[A_1,\bar{\Psi}_1,\Psi_1]   \,
\Delta_{22}[A_1,\bar{\Psi}_1,\Psi_1] 
\, {\cal J}_{{\cal B}}[A_1,\bar{\Psi}_1,\Psi_1]  \; . 
\label{Seff}
\eea
This is the desired action for the effective theory describing the
interaction of relevant quark modes, $\bar{\Psi}_1, \Psi_1$, and
soft gluons, $A_1$. 
The functional dependence of the various terms on the right-hand
side on the fields $A_1,\bar{\Psi}_1, \Psi_1$
has been restored in order to facilitate the
following discussion of all possible interaction vertices occurring in this
effective theory.

\begin{figure}[ht]
\includegraphics[width=3cm]{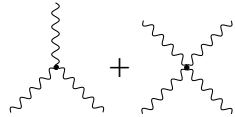}
\caption{The three- and four-gluon vertices in $S_A[A_1]$, describing
the self-interaction of soft gluons in Eq.\ (\ref{Seff}).}
\label{AAa}
\end{figure}

\begin{figure}[ht]
\includegraphics[width=14cm]{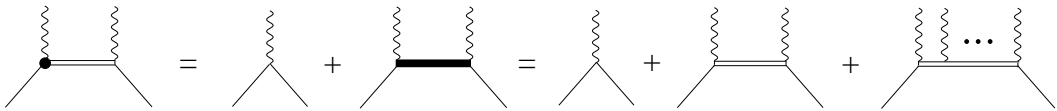}
\caption{The term $\bar{\Psi}_1 \, g {\cal B}[A_1] \, \Psi_1$ in the
effective action (\ref{Seff}).}
\label{AAb}
\end{figure}

\begin{figure}[ht]
\includegraphics[width=12cm]{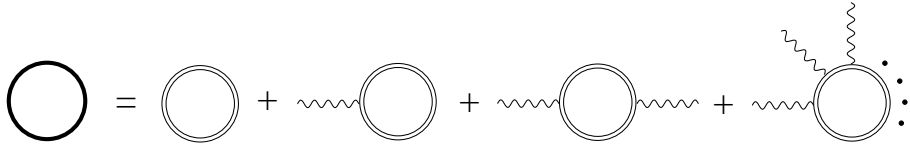}
\caption{The term ${\rm Tr}_q \ln {\cal G}_{22}^{-1}[A_1]$ in
the effective action (\ref{Seff}).}
\label{AAc}
\end{figure}

\begin{figure}[ht]
\includegraphics[width=14cm]{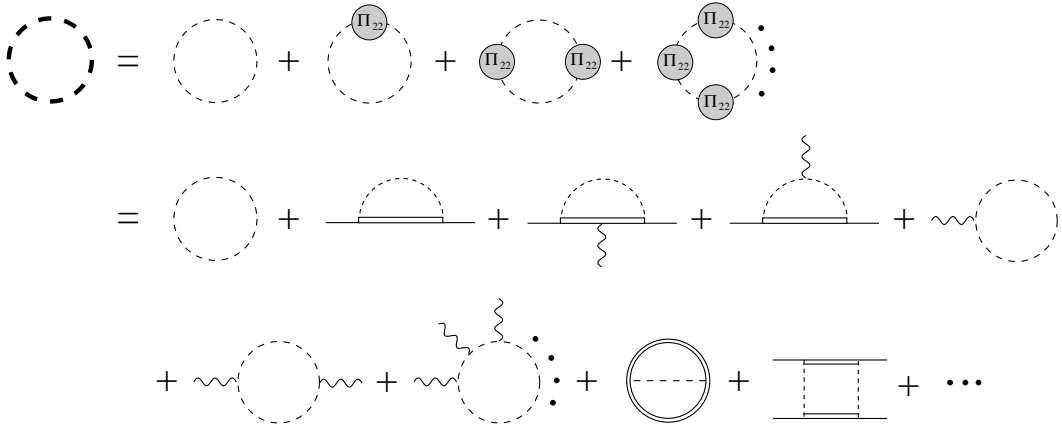}
\caption{The term ${\rm Tr}_g \ln
\Delta_{22}^{-1}[A_1,\bar{\Psi}_1,\Psi_1]$ 
in the effective action (\ref{Seff}). The first line corresponds to
the generic expansion (\ref{expansion2}), with ``self-energy'' insertions
$\Pi_{22}$, as shown in Fig.\ \ref{FigPi}. The second line contains
some examples
for diagrams generated when explicitly inserting the expression for 
$\Pi_{22}$.}
\label{AAd}
\end{figure}

\begin{figure}[ht]
\includegraphics[width=12cm]{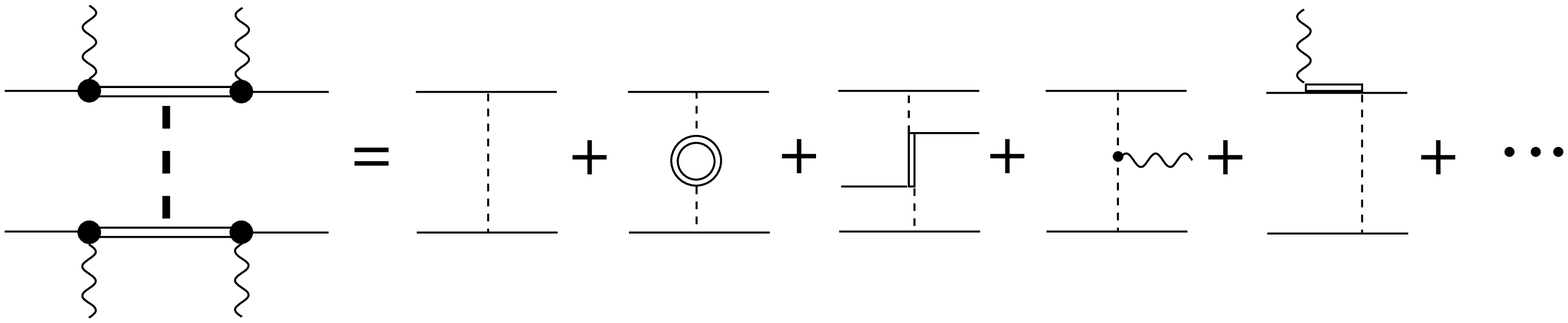}
\caption{The term ${\cal J}_{\cal B} \Delta_{22} {\cal J}_{\cal B}$ in
the effective action (\ref{Seff}). The thick dashed line is a full
hard gluon propagator, i.e., it has the expansion (\ref{expansion2}).
The first diagram on the right-hand side of this figure 
results from the $n=0$ term of this expansion, while the next
three diagrams originate from the $n=1$ term. Even a single insertion
of a hard gluon ``self-energy'' $\Pi_{22}$ gives rise to a variety of
diagrams. Here, we only show the contributions corresponding to
the first diagrams in Figs.\ \ref{ZZa}, \ref{ZZb}, and the three-gluon vertex.
The last diagram arises from the second term
of the expansion shown in Fig.\ \ref{XXb}. }
\label{AAe}
\end{figure}

The diagrams corresponding to these vertices are
shown in Figs.\ \ref{AAa}-\ref{AAe}. The three- and four-gluon 
vertices contained in
$S_A[A_1]$ are displayed in Fig.\ \ref{AAa}. In addition, $S_A[A_1]$ 
contains ghost loops with an arbitrary number of
attached soft gluon legs. The topology is equivalent to that
of the quark loops in Fig.\ \ref{AAc} and is therefore not shown
explicitly.
The interaction between two relevant quarks and the ``modified'' soft
gluon field, corresponding to $\bar{\Psi}_1 \, g {\cal B}[A_1]\, \Psi_1$, is
depicted in Fig.\ \ref{AAb}. 
This is similar to Fig.\ \ref{XXa}, except that now all gluon legs are soft.
Diagrams where an 
arbitrary number of soft gluon legs is attached
to an irrelevant quark loop are generated
by ${\rm Tr}_q \ln {\cal G}_{22}^{-1}$, cf.\ Fig.\ \ref{AAc}.
This is similar to Fig.\ \ref{XXc}, but now only soft gluon
legs are attached to the fermion loop.
The diagrams generated by the loop of a full hard gluon
propagator, ${\rm Tr}_g \ln \Delta_{22}^{-1}$,
are shown in Fig.\ \ref{AAd}. The first line in this figure
features the generic expansion of this term according to 
Eq.\ (\ref{expansion2}), where the hard gluon
``self-energy'' insertion $\Pi_{22}$, cf.\ Eq.\ (\ref{Pi}), is shown
in Fig.\ \ref{FigPi}.
The second line shows examples of diagrams generated by explicitly inserting
$\Pi_{22}$ in the generic expansion. Besides an arbitrary number of 
soft gluon legs, these diagrams also feature an arbitrary number
of relevant quark legs. If there are only two relevant 
quark legs, but
no soft gluon leg, one obtains the one-loop self-energy for relevant
quarks, cf.\ the second diagram in the second line of Fig.\ \ref{AAd}.
The next two diagrams are obtained by adding a soft gluon leg, 
resulting in vertex corrections for the bare vertex between
relevant quarks and soft gluons.
The first of these two diagrams arises from the 
$n=1$ term in Eq.\ (\ref{expansion2}), while the second originates
from the $n=2$ term.
Four relevant quark legs and no soft gluon leg
give rise to the scattering of two relevant quarks via
exchange of two hard gluons, contained in the $n=2$ term
in Eq.\ (\ref{expansion2}), cf.\ the last diagram in
Fig.\ \ref{AAd}. This diagram was also
discussed in the context of the effective theory presented
in Refs.\ \cite{hong,hong2}, cf.\ discussion in Sec.\ \ref{IIIB}.
Finally, the ``current-current'' interaction mediated by a full
hard gluon propagator, ${\cal J}_{{\cal B}} \,\Delta_{22} 
\, {\cal J}_{{\cal B}}$, Fig.\ \ref{AAe}, contains also a
multitude of quark-gluon vertices. The simplest one is the first on
the right-hand side in Fig.\ \ref{AAe}, corresponding to scattering of
two relevant fermions via exchange of a single hard gluon.

The effective action (\ref{Seff}) is formally of the form
(\ref{Seffstandard}).
The difference is that Eq.\ (\ref{Seff}) contains more than one
relevant field: besides relevant quarks there are also soft gluons.
It is obvious that in this case there are many more possibilities
to construct operators ${\cal O}_i$ which occur in 
the expansion (\ref{Seffstandard}).
As pointed out in the introduction, it is therefore advantageous
to derive the effective action (\ref{Seff}) by explicitly integrating
out irrelevant quark and hard gluon modes, 
and not by simply guessing the form of the operators ${\cal O}_i$,
since then one is certain
that one has constructed {\em all\/} possible operators occurring in
the expansion (\ref{Seffstandard}).

As mentioned in the introduction, the standard approach
to derive an effective theory, namely guessing the form of the
operators ${\cal O}_i$ and performing a naive dimensional scaling analysis
to estimate their order of magnitude, fails precisely
when (a) there are non-local operators, or when (b) there is more than one
momentum scale. Both (a) and (b) apply here. As we shall
show below, the HTL/HDL effective action is one limiting
case of Eq.\ (\ref{Seff}), and it is well known that this action
is non-local. Moreover, as is obvious from the above derivation, there
are indeed several momentum scales occurring in Eq.\ (\ref{Seff}).
Let us focus on the case of zero temperature, $T=0$, and, for the
sake of simplicity, assume massless quarks, $m=0$, $\mu = k_F$.
To be explicit, we employ the choice (\ref{P12}) for the projectors
${\cal P}_{1,2}$.
In this case, the first momentum scale is defined by the Fermi energy
$\mu$. The propagator of antiquarks is $\sim 1/(k_0 + \mu + k)$.
If $\Lambda_q, \Lambda_g \alt \mu$, the exchange of an antiquark 
can be approximated by a contact interaction with strength $\sim 1/\mu$,
on the scale of the relevant quarks, $L_q \gg 1/\Lambda_q \agt 1/ \mu$, 
or of the soft gluons, $L_g \gg 1/\Lambda_g \agt 1/ \mu$. 

The second momentum scale is defined by the quark cut-off momentum
$\Lambda_q$. The propagator of irrelevant quark modes 
is $\sim 1/(k_0+\mu - k)$.
On the scale $L_q$ of the relevant quarks, not only
the exchange of an antiquark, but also that of an irrelevant quark
with momentum ${\bf k}$ satisfying $|k- \mu| \geq \Lambda_q$ 
is local, with strength $\sim 1/\Lambda_q$.
However, suppose that the quark cut-off scale happens to be much smaller
than the chemical potential, $\Lambda_q \ll \mu$. In this case,
antiquark exchange is ``much more
localized'' than the exchange of an irrelevant quark, 
$1/\mu \ll 1/ \Lambda_q$.

The third momentum scale is defined by the gluon cut-off momentum
$\Lambda_g$. The propagator of a hard gluon is $\sim 1/P^2$.
On the scale $L_g$ of a soft gluon, the exchange of a hard gluon
with momentum $p \geq \Lambda_g$ can be considered local, with
strength $\sim 1/\Lambda_g^2$.
As we shall show below, in order to derive the value of the QCD gap
parameter in weak coupling and to subleading order, 
the ordering of the scales turns out to be $\Lambda_q \alt g \mu
\ll \Lambda_g \alt \mu$.
Thus, antiquark exchange happens on a length scale of the same order as
hard gluon exchange, which in turn happens on a much smaller length scale
than the exchange of an irrelevant quark,
$1/\mu \alt 1/ \Lambda_g \ll 1/\Lambda_q$.

\section{Examples of effective theories} \label{III}

In this section we show that, for particular choices
of the projectors ${\cal P}_{1,2}$ in Eq.\ (\ref{project}),
several well-known, at first sight unrelated effective theories 
for hot and/or dense quark matter, are in fact nothing but
special cases of the general effective theory defined by
the action (\ref{Seff}).
These are the HTL/HDL effective action for quarks and gluons, and
the high-density effective theory for cold, dense quark matter.

\subsection{HTL/HDL effective action} \label{IIIA}

Let us first focus on the HTL/HDL effective action.
This action defines an effective theory for massless
quarks and gluons with small momenta in a system
at high temperature $T$ (HTL), or large chemical potential $\mu$ (HDL). 
Consequently, the projectors ${\cal P}_{1,2}$ for quarks are given by
\begin{subequations} \label{PHTL}
\bea
{\cal P}_1 (K,Q) & = & \Theta(\Lambda_q -k)\,\delta^{(4)}_{K,Q}\;, \\
{\cal P}_2 (K,Q) & = & \Theta(k - \Lambda_q)\,\delta^{(4)}_{K,Q}\;,
\eea
\end{subequations}
while the projectors for gluons are given by Eq.\ (\ref{Q12}).
(We note that, strictly speaking, the quarks and gluons
of the HTL/HDL effective action should also have small energies
in real time. Since our effective action is defined in imaginary time,
one should constrain the energy only at the end
of a calculation, after analytically continuing the result 
to Minkowski space.)

The essential assumption to derive the HTL/HDL effective action is
that there is a single momentum scale, 
$\Lambda_q = \Lambda_g \equiv \Lambda$,
which separates hard modes with momenta $\sim T$, or $\sim
\mu$, from soft modes with momenta $\sim gT$, or $\sim g\mu$.
In the presence of an additional energy scale $T$, or $\mu$,
naive perturbation theory in terms of powers of the coupling constant
fails. It was shown by Braaten and Pisarski \cite{braatenpisarski} that,
for the $n$-gluon scattering amplitude
the one-loop term, where $n$ soft gluon legs are attached to a
quark or gluon loop, is as important as the tree-level diagram.
The same holds for the scattering of $n-2$ gluons and 2 quarks.
At high $T$ and small $\mu$, the momenta of the quarks and gluons 
in the loop are of the order of the hard scale, $\sim T$. 
This gives rise to the name 
``Hard Thermal Loop'' effective action, and allows to simplify the 
calculation of the respective diagrams.
At large $\mu$ and small $T$, i.e., for the HDL effective action,
the situation is somewhat more involved.
As gluons do not have a Fermi surface, the only
physical scale which determines the order of magnitude of a loop
consisting exclusively of gluon propagators is the temperature. 
Therefore, at small $T$ and large $\mu$, such
pure gluon loops are negligible. On the other hand, 
the momenta of quarks in the loop are $\sim \mu$. Thus, only loops
with at least one quark line need to be considered in the HDL effective action.

In order to show that the HTL/HDL effective action is contained in the
effective action (\ref{Seff}), we first note that a soft particle
cannot become hard by interacting with another soft particle.
This has the consequence that a soft quark cannot 
turn into a hard one by soft-gluon scattering. Therefore,
\be
g {\cal B}[A_1] \equiv g {\cal A}_{11}\;.
\ee
Another consequence is that the last term in Eq.\ (\ref{Seff}),
${\cal J}_{{\cal B}} \Delta_{22} {\cal J}_{{\cal B}}$, vanishes
since ${\cal J}_{{\cal B}}$
is identical to a vertex between a soft quark and a hard gluon, which
is kinematically forbidden. The resulting action then reads
\be 
S_{\mbox{\scriptsize large}\,T/\mu} 
[A_1,\bar{\Psi}_1, \Psi_1 ] \equiv
S_A[A_1] + \frac{1}{2} \, \bar{\Psi}_1 
\left( {\cal G}_{0,11}^{-1} +  g {\cal A}_{11} \right) \Psi_1
+ \frac{1}{2}\, {\rm Tr}_q \ln {\cal G}_{22}^{-1}[A_1] 
- \frac{1}{2}\,  {\rm Tr}_g \ln
\Delta_{22}^{-1}[A_1,\bar{\Psi}_1,\Psi_1] 
 \; . 
\label{SHTL}
\ee
Using the expansion (\ref{explnquark}) we realize that the 
term ${\rm Tr}_q \ln {\cal G}_{22}^{-1}$ generates
all one-loop diagrams, where $n$ soft gluon legs 
are attached to a hard quark loop. This is precisely the quark-loop 
contribution to the HTL/HDL effective action.

For hard gluons with momentum $\sim T$ or $\sim \mu$,
the free inverse gluon propagator is $\Delta_{0,22}^{-1} \sim
T^2$ or $\sim \mu^2$, while the contribution $\Pi_{\rm loop}$ to the hard 
gluon ``self-energy'' (\ref{Pi}) is at most of the order $\sim g^2
T^2$ or $\sim g^2 \mu^2$. Consequently, $\Pi_{\rm loop}$ can be neglected and
$\Pi_{22}$ only contains tree-level diagrams, $\Pi_{22} \equiv \Pi_{{\cal B}}
+ \Pi_{{\cal V}}$.
Using the expansion (\ref{expansion2}) of
${\rm Tr}_g \ln \Delta_{22}^{-1}$, the terms which contain
only insertions of $\Pi_{{\cal V}}$ correspond to one-loop
diagrams where $n$ soft gluon legs
are attached to a hard gluon loop. As was shown in 
Ref.\ \cite{braatenpisarski}, with the exception of the two-gluon
amplitude, the loops with four-gluon vertices are suppressed.
Neglecting these, we are precisely left with 
the pure gluon loop contribution to the HTL effective action. 
As discussed above, for the HDL effective action, this contribution 
is negligible.

The ``self-energy'' $\Pi_{{\cal B}}$ contains only two soft
quark legs attached to a hard quark propagator (via emission and
absorption of hard gluons). Consequently, in 
the expansion (\ref{expansion2}) of ${\rm Tr}_g \ln \Delta_{22}^{-1}$, 
the terms which contain insertions of $\Pi_{{\cal V}}$ and $\Pi_{{\cal B}}$
correspond to one-loop diagrams where an arbitrary number of 
soft quark and gluon legs is attached to the loop. It was shown in Ref.\
\cite{braatenpisarski} that of these diagrams, only the ones with
two soft quark legs and no four-gluon vertices are kinematically
important and thus contribute to the HTL/HDL effective action.
We have thus shown that this effective action, $S_{\rm HTL/HDL}$,
is contained in the effective action (\ref{SHTL}), and constitutes its
leading contribution,
\be
S_{\mbox{\scriptsize large}\, T/\mu} = S_{\rm HTL/HDL} + \,
\mbox{higher orders}\;.
\ee
For the sake of completeness, let us briefly comment on possible
ghost contributions. Ghost loops arise from the term 
${\rm Tr}_{gh} \ln {\cal W}^{-1}$ in
$S_A[A_1]$. Their topology and consequently their properties
are completely analogous to those of the pure gluon loops discussed above.

We conclude with a remark regarding the HDL effective action.
According to Eq.\ (\ref{PHTL}), 
at zero temperature and large chemical potential,
a soft quark or antiquark has a momentum $k \sim g \mu$, i.e.,
it lies at the bottom of the Fermi sea, or at the top of the
Dirac sea, respectively.
These modes are, however, not that important in degenerate Fermi
systems, because it requires a large amount of energy $k_0 \sim \mu$
to excite them. The truly relevant modes are quark modes with large
momenta, $k \sim \mu$, close to the Fermi surface, because it costs
little energy to excite them.
A physically reasonable effective theory for cold, dense quark matter
should therefore feature no antiquark modes at all, and only
quark modes near the Fermi surface. Such a theory will be discussed
in the following.

\subsection{High-density effective theory} \label{IIIB}

An effective theory
for high-density quark matter was first proposed
by Hong \cite{hong} and was further refined by Sch\"afer and others
\cite{HLSLHDET,schaferefftheory,NFL,others}. 
In the construction of this effective theory, one first proceeds
similar to our discussion in Sec.\ \ref{II} and integrates
out antiquark modes. (From a technical point of view, this is
not done as in Sec.\ \ref{II} by functional integration, but by
employing the equations of motion for antiquarks. The result is
equivalent.) On the other hand, at first all quark modes in the Fermi
sea are considered as relevant. 
Consequently, in the notation of Sec.\ \ref{II},
the choice for the projectors ${\cal P}_{1,2}$ would be
\begin{subequations} \label{P12HDET}
\bea
{\cal P}_1(K,Q) & = & \left( \begin{array}{cc}
\Lambda_{\bf k}^+ & 0 \\
0 & \Lambda_{\bf k}^- \end{array} \right) \, \delta^{(4)}_{K,Q} \;, \\
{\cal P}_2(K,Q) & = &  \left( \begin{array}{cc}
\Lambda_{\bf k}^- & 0 \\
0 & \Lambda_{\bf k}^+ \end{array} \right)\, \delta^{(4)}_{K,Q} \;.
\eea
\end{subequations}
Also, at first gluons are not separated into soft and hard modes
either.
After this step, the partition function of the theory assumes the
form (\ref{ZQCD}) with ${\cal Z}_q$ given by Eq.\ (\ref{Zq4}).

In the next step, one departs from the rigorous approach of
integrating out modes, as done in Sec.\ \ref{II}, and follows
the standard way of constructing an effective theory, as
explained in the introduction.
One focusses exclusively on quark modes
close to the Fermi surface as well as on soft gluons. 
However, since quark modes far from the Fermi surface and hard gluons 
are not explicitly integrated out, the effective action does
not automatically contain the terms which
reflect the influence of these modes on the relevant quark and soft gluon
degrees of freedom. Instead, the corresponding terms
have to be written down ``by hand'' and the effective vertices
have to be determined via matching to the underlying microscopic
theory, i.e., QCD.

In order to further organize the terms occurring in the effective
action, one covers the Fermi surface with
``patches''. Each patch is labelled according to
the local Fermi velocity, ${\bf v}_F \equiv \hat{\bf k}\, 
k_F/\mu$ at its center.
A patch is supposed to have a typical size $\Lambda_\parallel$ in
radial ($\hat{\bf k}$) direction, and a size $\Lambda_\perp$
tangential to the Fermi surface.
The momentum of quark modes inside a patch is decomposed into a
large component in the direction of ${\bf v}_F$, the particular
Fermi velocity labelling the patch under consideration, and a small 
residual component, ${\bf l}$, residing exclusively inside the
patch,
\be \label{decomp}
{\bf k} = \mu \, {\bf v}_F + {\bf l}\;.
\ee
The residual component is further decomposed into a component 
pointing in radial direction, ${\bf l}_\parallel
\equiv {\bf v}_F ({\bf v}_F \cdot {\bf l})$, and the orthogonal one,
tangential to the Fermi surface, ${\bf l}_\perp \equiv {\bf l} - {\bf
l}_\parallel$.
The actual covering of the Fermi surface with such patches is not 
unique. One should, however, make sure that neighbouring
patches do not overlap, in order to avoid
double-counting of modes near the Fermi surface.
In this case, the total number of patches on the Fermi surface is 
$\sim \mu^2/\Lambda_\perp^2$. 

\begin{figure}[ht]
\includegraphics[width=10cm]{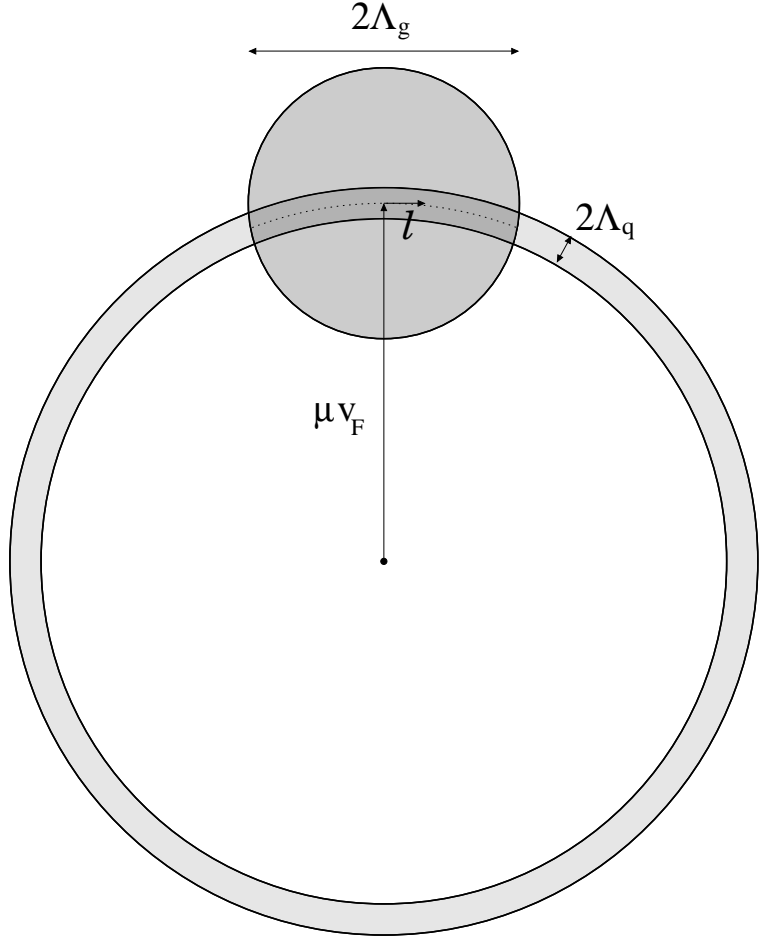}
\caption{A particular patch covering the Fermi surface.
The tangential dimension, $\Lambda_\perp$, is given by the
maximum momentum transferred via a soft gluon, $\Lambda_g$, while
the radial dimension, $\Lambda_\parallel$ is defined by the maximum 
distance of relevant quark modes from the Fermi surface, $\Lambda_q$.
Also shown is a typical momentum transfer ${\bf l}$ via a soft gluon.}
\label{Sphere}
\end{figure}

In the following, we shall show that
the action of the high-density effective theory as discussed in Refs.\
\cite{hong,hong2,HLSLHDET,schaferefftheory,NFL,others} 
is contained in our effective action
(\ref{Seff}). To this end, however, we shall employ 
the choice (\ref{P12}) and (\ref{Q12}) for
the projectors for quark and gluon modes, and not Eq.\ (\ref{P12HDET})
for the quark projectors.
As in Refs.\ \cite{hong,hong2,HLSLHDET,schaferefftheory,NFL,others}, 
the quark mass will be set to zero, $m=0$. We also have to clarify how
the patches covering the Fermi surfaces introduced in Refs.\ 
\cite{hong,hong2,HLSLHDET,schaferefftheory,NFL,others} 
arise within our effective
theory. It is obvious that the radial dimension $\Lambda_\parallel$
of a patch is related to the quark cut-off $\Lambda_q$. We simply
choose $ \Lambda_\parallel \equiv \Lambda_q$. Similarly, since
soft-gluon exchange is not supposed to move a fermion from 
a particular patch to another, the dimension $\Lambda_\perp$ tangential
to the Fermi surface must be related to the gluon cut-off $\Lambda_g$.
Again, we adhere to the most simple choice $\Lambda_\perp \equiv 
\Lambda_g$. Since $\Lambda_g \alt \mu$, this is consistent 
with the matching procedure discussed in Ref.\ \cite{HLSLHDET}, 
where the matching scale is chosen as $\Lambda_\perp = \sqrt{2} \mu$
(which is only slightly larger than $\mu$).
The different scales $\Lambda_q, \, \Lambda_g$, and $\mu$ are 
illustrated in Fig.\ \ref{Sphere}.
The modulus of the residual momentum ${\bf l}$ in Eq.\
(\ref{decomp}) is constrained to $l \leq {\rm max} \, 
(\Lambda_q,\, \Lambda_g)$.

In Nambu-Gor'kov space, the leading, kinetic term in the Lagrangian
of the high-density effective theory reads
\be \label{kintermHDET}
{\cal L}_{\rm kin} = \frac{1}{2} \, \sum_{{\bf v}_F}
\bar{\Psi}_1(X,{\bf v}_F) \gamma_0 \, 
\left( \begin{array}{cc}
 i V \cdot D & 0 \\
 0 & i \bar{V} \cdot D_C \end{array} \right)\,
\Psi_1(X,{\bf v}_F)\; ,
\ee
cf.\ for instance Eq.\ (1) of Ref.\ \cite{schaferefftheory}.
Here, we have introduced the 4-vectors
\be
V^\mu \equiv (1, {\bf v}_F)\;\;\;\; , \;\;\;\;\; 
\bar{V}^\mu \equiv (1, - {\bf v}_F)\;.
\ee
The covariant derivative for charge-conjugate fields is
defined as $D^\mu_{C} \equiv \partial^\mu + ig A^\mu_a T_a^T$.
The contribution (\ref{kintermHDET}) arises from
the term $\bar{\Psi}_1 \, \left( {\cal G}_{0,11}^{-1}\, + g {\cal
A}_{11} \right)\, \Psi_1$ in Eq.\ (\ref{Seff}). In order to see this,
use ${\cal P}_1^2 \equiv {\cal P}_1$ to write
\bea
\frac{1}{2} \, \bar{\Psi}_1 \, {\cal G}_{0,11}^{-1} \, \Psi_1 & = &
\frac{1}{2}\, \bar{\Psi}_1 \,\gamma_0 {\cal P}_1 \gamma_0 \, 
{\cal G}_{0,11}^{-1} \, {\cal P}_1\, \Psi_1 \non
& = & \frac{1}{2} \sum_{K,Q} \bar{\Psi}_1(K) \gamma_0 \, \frac{1}{T}\,
\left( \begin{array}{cc}
         k_0 + \mu -k & 0 \\
         0 & k_0 - \mu +k \end{array} \right) \, \delta^{(4)}_{K,Q}
\, \Psi_1(Q) \non
& \simeq & \frac{1}{2} \sum_{{\bf v}_F,L} \bar{\Psi}_1(L,{\bf v}_F) 
\gamma_0 \, \frac{1}{T}\,
\left( \begin{array}{cc}
         V \cdot L & 0 \\
         0 & \bar{V} \cdot L \end{array} \right) \, \Psi_1(L,{\bf
v}_F)\;.
\label{freekintermHDET}
\eea
In the last step, we have approximated $k \simeq \mu + {\bf v}_F
\cdot {\bf l}$, which holds up to terms of order $O(l^2/\mu)$, cf.\ 
Eq.\ (\ref{decomp}). This is a good approximation if the modulus of
a typical residual quark momentum in the effective theory
is $l \ll {\rm max}\, (\Lambda_q, \, \Lambda_g) \alt \mu$.
We have also introduced the 4-vector $L^\mu \equiv (k_0, {\bf l})$
and, applying the decomposition (\ref{decomp}), we have written
the sum over ${\bf k}$ as a double sum over
${\bf v}_F$ and ${\bf l}$. The latter sum runs over all residual
momenta ${\bf l}$ inside a given patch, while the former runs over
all patches. With this decomposition, the spinors 
$\bar{\Psi}_1$, $\Psi_1$ are defined locally on a given patch
(labelled by the Fermi velocity ${\bf v}_F$), and depend on the
4-momentum $L$.
Note that a Fourier transformation to coordinate space
converts $V \cdot L \rightarrow i V \cdot \partial$.

Now consider the term $\bar{\Psi}_1 g {\cal A}_{11} \Psi_1$.
Since ${\cal A}_{11}$ is not diagonal in
momentum space, cf.\ Eq.\ (\ref{calA}), in principle the two
quark spinors $\bar{\Psi}_1$, $\Psi_1$ can belong to different
patches. However, we have chosen the tangential dimension of
a patch such that a (typical) soft gluon can by definition never move
a fermion across the border of a particular patch, 
$|{\bf k} - {\bf q}| \ll \Lambda_g$. Therefore,
both spinors reside in the same patch and, to leading order,
$\hat{\bf k} \simeq \hat{\bf q} \simeq {\bf v}_F$. With these
assumptions we may write
$\Lambda_{\bf k}^- \diracslash{A}_a(K-Q) \Lambda_{\bf q}^+
\simeq V \cdot A_a (K-Q) \, \gamma_0 \,\Lambda^+_{\bf k}$,
$\Lambda_{\bf k}^+ \diracslash{A}_a(K-Q) \Lambda_{\bf q}^-
\simeq \bar{V} \cdot A_a (K-Q) \, \gamma_0 \, \Lambda^-_{\bf k}$.
Then, introducing the residual momentum ${\bf l}'$
corresponding to the quark 3-momentum ${\bf q}$ and
defining ${L'}^\mu \equiv (q_0, {\bf l}')$, 
the respective term in the effective action becomes
\be 
\frac{1}{2}\, \bar{\Psi}_1 \, g {\cal A}_{11} \,\Psi_1
\simeq \frac{1}{2}\, \frac{g}{\sqrt{VT^3}} 
\sum_{{\bf v}_F, L, L'} 
\bar{\Psi}_1(L,{\bf v}_F) \gamma_0 \, 
\left( \begin{array}{cc}
 V \cdot A_a(L-L') \, T_a & 0 \\
 0 & - \bar{V} \cdot A_a(L-L') \, T_a^T \end{array} \right)\,
\Psi_1(L',{\bf v}_F) \;.
\label{gaugeHDET}
\ee
In coordinate space, the sum of Eqs.\ (\ref{freekintermHDET}) 
and (\ref{gaugeHDET}) becomes Eq.\ (\ref{kintermHDET}).

Subleading terms of order $O(1/\mu)$ in the high-density effective
theory are of the form
\be
{\cal L}_{O(1/\mu)} = - \frac{1}{2} \sum_{{\bf v}_F} \bar{\Psi}_1(X,{\bf
v}_F) \gamma_0 \, \frac{1}{2\mu} \left( \begin{array}{cc}
D_\perp^2 -\frac{g}{2} \, \sigma^{\mu \nu} F_{\perp \mu \nu}^aT_a & 0 \\
0 & - D_{C\perp}^2 - \frac{g}{2}\, \sigma^{\mu \nu} F_{\perp \mu
\nu}^a T_a^T \end{array} \right)\Psi_1(X,{\bf v}_F) \;,
\label{subleadingHDET}
\ee
cf.\ Eq.\ (2) of Ref.\ \cite{schaferefftheory}.
Here, $D_\perp^\mu \equiv 
\{0, ({\bf 1} - {\bf v}_F {\bf v}_F) \cdot {\bf D}\}$,
and similarly for $D_{C \perp}^\mu$. The commutator of two gamma matrices
is defined as usual,
$\sigma^{\mu \nu} \equiv (i/2) [\gamma^\mu , \gamma^\nu]$, and
$F_{\perp \mu \nu}^a T_a \equiv (i/g) [D_{\perp \mu}, D_{\perp \nu}]$.
As we shall see in the following, this contribution arises from the term
$- g^2 \, \bar{\Psi}_1 \,{\cal A}_{12} \,{\cal G}_{22} \,{\cal
A}_{21}\, \Psi_1$ in Eq.\ (\ref{Seff}).

First, note that, with the projectors (\ref{P12}), the irrelevant quark
propagator ${\cal G}_{22}$ contains quark as well as antiquark modes.
In order to derive Eq.\ (\ref{subleadingHDET}), however, we have to
discard the quark and keep only the antiquark modes. In essence,
this is a consequence of the simpler choice (\ref{P12HDET}) for the 
projectors ${\cal P}_{1,2}$ in the high-density effective theory
of Refs.\ \cite{hong,hong2,HLSLHDET,schaferefftheory,NFL,others}.
In this case, the propagator ${\cal G}_{22}$ may be simplified.
A calculation quite similar to that of Eqs.\
(\ref{freekintermHDET}) and (\ref{gaugeHDET}) now leads (in 
coordinate space) to
\be
{\cal G}_{22}^{-1} \equiv {\cal G}_{0,22}^{-1} + g{\cal A}_{22}
\simeq \gamma_0 \, \tau_3 \left( \begin{array}{cc}
2 \mu + i \bar{V} \cdot D & 0 \\
0 &  2\mu - i V \cdot D_C \end{array} \right) \; ,
\ee
where $\tau_3$ acts in Nambu-Gor'kov space.
This result may be readily inverted to yield
\be \label{G22}
{\cal G}_{22} \simeq \gamma_0 \, \tau_3\, \frac{1}{2 \mu}
\sum_{n=0}^\infty \frac{1}{(2 \mu)^n} \left( \begin{array}{cc}
- i \bar{V} \cdot D & 0 \\
0 & i V \cdot D_C \end{array} \right)^n\;.
\ee
Utilizing the projectors (\ref{P12HDET}),
one may also derive a simpler form for $g{\cal A}_{12}$ and
$g{\cal A}_{21}$. Consider, for instance, the term 
$\bar{\Psi}_1 \, g {\cal A}_{12}\, \Psi_2$. We follow the same steps
that led to Eqs.\ (\ref{freekintermHDET}), i.e., we assume that the spinors 
$\bar{\Psi}_1$ and $\Psi_2$ reside in the same patch, such that
$\hat{\bf k} \simeq \hat{\bf q} \simeq {\bf v}_F$. This allows to derive
the identity $\Lambda_{\bf k}^\mp \, \diracslash{A}^a(K-Q) \Lambda_{\bf
q}^\mp \simeq \Lambda_{\bf k}^\mp \diracslash{A}^a_\perp(K-Q)$,
where $A_\perp^{\mu a} \equiv  
\{ 0, ({\bf 1} - {\bf v}_F {\bf v}_F) \cdot {\bf A}^a \}$. 
Now introduce the 4-vectors $L^\mu$, ${L'}^\mu$, as in Eq.\
(\ref{gaugeHDET}), which leads to
\be
\frac{1}{2} \, \bar{\Psi}_1 \, g {\cal A}_{12}\, \Psi_2
 \simeq  \frac{1}{2} \,\frac{g}{\sqrt{VT^3}}  \sum_{{\bf v}_F,L,L'} 
\bar{\Psi}_1 (L,{\bf v}_F)
\left( \begin{array}{cc}
\diracslash{A}^a_\perp(L-L')\, T_a & 0 \\
0 & - \diracslash{A}^a_\perp(L-L') \, T_a^T
\end{array} \right) \Psi_2(L',{\bf v}_F) \;.
\ee
We may add a term
$\diracslash{L}_\perp$ to the diagonal Nambu-Gor'kov
components, which trivially vanishes between spinors $\bar{\Psi}_1$ and
$\Psi_2$. This has the advantage that, in coordinate space,
\be \label{A12}
g {\cal A}_{12} \simeq \left( \begin{array}{cc}
i \Diracslash{D}_\perp & 0 \\
0 & i \Diracslash{D}_{C \perp} \end{array} \right)\;,
\ee
i.e., this term transforms covariantly under gauge transformations,
and no longer as a gauge field.
A similar calculation for $g{\cal A}_{21}$ gives the result
$g{\cal A}_{21} \equiv g {\cal A}_{12}$.
Combining Eqs.\ (\ref{G22}) and (\ref{A12}), the term
$- g^2 \, \bar{\Psi}_1 \,{\cal A}_{12} \,{\cal G}_{22} \,{\cal
A}_{21}\, \Psi_1$ corresponds to the following contribution in the
Lagrangian,
\be \label{subleadingHDET2}
- \frac{1}{2} \sum_{{\bf v}_F}
\bar{\Psi}_1(X,{\bf v}_F) \gamma_0  \left( \begin{array}{cc}
\Diracslash{D}_\perp & 0 \\
0 & - \Diracslash{D}_{C \perp} \end{array} \right)
\frac{1}{2 \mu}
\sum_{n=0}^\infty \frac{1}{(2 \mu)^n} \left( \begin{array}{cc}
- i \bar{V} \cdot D & 0 \\
0 & i V \cdot D_C \end{array} \right)^n 
\left( \begin{array}{cc}
\Diracslash{D}_\perp & 0 \\
0 &  \Diracslash{D}_{C \perp} \end{array} \right) \Psi_1(X,{\bf v}_F)
\;. 
\ee
Taking only the $n=0$ term, and utilizing $\gamma^\mu \gamma^\nu
\equiv g^{\mu \nu} - i \sigma^{\mu\nu}$, one arrives at Eq.\ 
(\ref{subleadingHDET}). Note that our definition for transverse
quantities, e.g.\ $A_\perp^\mu \equiv 
\{ 0, ({\bf 1} - {\bf v}_F {\bf v}_F) \cdot {\bf A} \}$, 
slightly differs from that of Refs.\ \cite{hong,hong2},
where $A_\perp^\mu \equiv A^\mu - V^\mu V \cdot A$. However, both 
definitions agree when sandwiched between spinors 
$\bar{\Psi}_{1,2}$ and $\Psi_{2,1}$.

At order $O(1/\mu^2)$, besides the $n=1$ term in Eq.\ 
(\ref{subleadingHDET2}),
there are also four-fermion interaction terms,
cf.\ Eqs.\ (3-5) of Ref.\ \cite{schaferefftheory}. 
In the effective
action (\ref{Seff}), these contributions arise from 
the term ${\cal J}_{{\cal B}} \Delta_{22} {\cal J}_{{\cal B}}$
which originates from integrating out hard gluons.
(Since this is not done explicitly in the construction of the
high-density effective theory in Refs.\ 
\cite{hong,hong2,HLSLHDET,schaferefftheory,NFL,others}, this term
is not automatically generated, but has to be added ``by hand''.)
To leading order, this term corresponds to the exchange of a
hard gluon between two quarks, cf.\ the first diagram on the 
right-hand side of Fig.\ \ref{AAe}. 
If the quarks are close to the Fermi surface, the 
energy in the hard gluon propagator can be neglected, and
$\Delta_{0,22} \alt 1/\Lambda_g^2$. Since $1/\Lambda_g^2 \agt
1/\mu^2$, the contribution from hard-gluon exchange is of order
$O(1/\mu^2)$.
Four-fermion interactions also receive corrections at
one-loop order, cf.\ Fig.\ 5 of Ref.\ \cite{hong2}. In
Eq.\ (\ref{Seff}), they are contained in the
term ${\rm Tr} \ln \Delta_{22}^{-1}$, see the last diagram in Fig.\ \ref{AAd}.

Besides the quark terms in the Lagrangian of the high-density effective
theory \cite{hong,hong2,HLSLHDET,schaferefftheory,NFL,others}, 
there are also contributions from
gluons. The first is 
the standard Yang-Mills Lagrangian $-(1/4) F_{\mu \nu}^a F^{\mu
\nu}_a$, cf.\ Eq.\ (1) of Ref.\ \cite{schaferefftheory}.
This part is contained in the term $S_A[A_1]$ in Eq.\ (\ref{Seff}),
cf.\ Eq.\ (\ref{SA}).
The second contribution is a mass term for magnetic gluons,
\be
{\cal L}_{m_g} = - \frac{m_g^2}{2} \, {\bf A}^a \cdot {\bf A}^a \;,
\ee
cf.\ Eq. (19) of Ref.\ \cite{SonStephanov}, Eq.\ (18) of Ref.\ 
\cite{hong2}, or Eq.\ (27) of Ref.\ \cite{HLSLHDET}, 
where $m_g$ is
the gluon mass parameter (\ref{gluonmass}).
This term has to be added ``by hand'' in order to obtain the correct
value for the HDL gluon polarization tensor within the high-density
effective theory. In Eq.\ (\ref{Seff}) this contribution arises
from the $n=2$ term of the expansion (\ref{explnquark})
of ${\rm Tr} \ln {\cal G}_{22}^{-1}$. The gluon polarization tensor
has contributions
from particle-hole and particle-antiparticle excitations. 
The latter give rise to ${\cal L}_{m_g}$. 
While this term arises naturally within our derivation of the effective theory,
it does not in the high-density effective theory of Refs.\ 
\cite{hong,hong2,HLSLHDET,schaferefftheory,NFL,others}, 
because only antiquarks, but not irrelevant
quark modes, are explicitly integrated out. 
Irrelevant quark modes can then only be taken into account 
by adding the appropriate counter terms.

Sometimes, the full HDL action is added
to the Lagrangian of the high-density effective theory, cf.\ Eq.\ (8)
of Ref.\ \cite{schaferefftheory}.
This procedure requires a word of caution. For instance,
an important contribution to the HDL polarization tensor arises
from particle-hole excitations around the Fermi surface.
Such excitations are still relevant degrees of freedom in the
effective theory. However, in order for them to appear in the
gluon polarization tensor they would first have
to be integrated out. Therefore, strictly speaking 
such contributions cannot occur in the tree-level effective action. 
Of course, in an effective
theory one is free to add whatever contributions one deems necessary.
However, one has to be careful to avoid double counting.
As will be shown in Sec.\ \ref{IV}, the full HDL polarization tensor
will appear quite naturally in an approximate solution to the
Schwinger-Dyson equation for the gluon propagator, however, not at
tree-, but only at (one-)loop level.

It was claimed in Refs.\ \cite{hong2,HLSLHDET,schaferefftheory}
that a consistent power-counting scheme within the high-density
effective theory requires $\Lambda_\perp = \Lambda_\parallel$.
In contrast, we shall show in Sec.\ \ref{IV} 
that a computation of the gap parameter 
to subleading order requires $\Lambda_q \equiv \Lambda_\parallel
\ll \Lambda_\perp \equiv \Lambda_g$.
This means that irrelevant quark modes become local on a scale
$l_q \gg 1/ \Lambda_q$, while antiquark modes become local already on
a much smaller scale, $l_{\bar{q}} \gg 1/\mu$, cf.\ discussion at the
end of Sec.\ \ref{II}. As mentioned in the introduction,
for two different scales power counting of terms in
the effective action becomes a non-trivial problem.
While the high-density effective
theory of Refs.\
\cite{hong,hong2,HLSLHDET,schaferefftheory,NFL,others} 
contains effects
from integrating out antiquarks, i.e., from the scale $1/\mu$, the
effective action (\ref{Seff}) in addition keeps track of
the influence of irrelevant quark modes, i.e., from physics on the scale
$1/\Lambda_q \gg 1/\mu$. Since all terms in the effective action
(\ref{Seff}) are kept, one can be certain not to miss any important
contribution just because the naive dimensional power-counting scheme
is invalidated by the occurrence of two vastly different length
scales.

\section{Calculation of the QCD gap parameter} \label{IV}

In this section, we demonstrate how the effective theory derived in
Sec.\ \ref{II} can be applied to compute the gap parameter of
color-superconducting quark matter to subleading order.
For the sake of definiteness, we shall consider a spin-zero, two-flavor
color superconductor.

\subsection{CJT formalism for the effective theory} \label{IVa}

The gap parameter in superconducting systems is not accessible
by means of perturbation theory; one has to apply non-perturbative,
self-consistent, many-body resummation techniques to calculate it.
For this purpose, it is convenient to employ the CJT formalism
\cite{CJT}. The first step is to add source terms 
to the effective action (\ref{Seff}),
\be \label{Seffsources}
S_{\rm eff}[A_1,\bar{\Psi}_1, \Psi_1]\; \longrightarrow\;
S_{\rm eff}[A_1,\bar{\Psi}_1, \Psi_1] + J_1 A_1 + 
\frac{1}{2}\, A_1 K_1 A_1  + \frac{1}{2}
\left( \bar{\Psi}_1 H_1 + \bar{H}_1 \Psi_1 + 
\bar{\Psi}_1 {\cal K }_1 \Psi_1 \right) \; ,
\ee
where we employed the compact matrix notation defined in Eq.\ (\ref{compact}).
$J_1$, $\bar{H}_1$, and $H_1$ 
are local source terms for the soft gluon and relevant quark fields, 
respectively, while $K_1$ and ${\cal K}_1$ are bilocal source terms.
The bilocal source ${\cal K}_1$ for quarks is also a matrix in
Nambu-Gor'kov space.
Its diagonal components are source terms which couple quarks to
antiquarks, while its off-diagonal components couple quarks
to quarks. The latter have to be introduced for systems which can become
superconducting, i.e., where the ground state has
a non-vanishing diquark expectation value,
$\langle \psi_1 \psi_1 \rangle \neq 0$.

One then performs a Legendre transformation with respect
to all sources and arrives at the CJT effective action \cite{CJT,kleinert}
\bea
\Gamma\left[A,\bar{\Psi},\Psi,\Delta, {\cal G}\right]
& = & S_{\rm eff} \left[A,\bar{\Psi},\Psi\right]
  -\frac{1}{2}\, {\rm Tr}_g \ln \Delta^{-1}
  -\frac{1}{2}\, {\rm Tr}_g \left( D^{-1} \Delta - 1 \right) \nonumber\\
& + & \frac{1}{2}\, {\rm Tr}_q\ln {\cal G}^{-1}
  +\frac{1}{2}\, {\rm Tr}_q\left(G^{-1}{\cal G}-1\right)
  +\Gamma_2\left[A,\bar{\Psi},\Psi,\Delta,{\cal G}\right]\; .   \label{Gamma}
\eea
Here, $S_{\rm eff}[A, \bar{\Psi}, \Psi]$ is the tree-level
action defined in Eq.\ (\ref{Seff}), which now depends on the
{\em expectation values\/} $A \equiv \langle A_1 \rangle$, $
\bar{\Psi} \equiv \langle \bar{\Psi}_1 \rangle$, and $\Psi \equiv
\langle \Psi_1 \rangle$ 
for the one-point functions of soft gluon and relevant quark fields.
In a slight abuse of notation, we use the same symbols for the
expectation values as for the original fields, prior to integrating
out modes. This should not lead to confusion, as the original fields
no longer occur in any of the following expressions.

The quantities $D^{-1}$ and $G^{-1}$ in Eq.\ (\ref{Gamma}) are
the inverse {\em tree-level\/} propagators for soft gluons and
relevant quarks, respectively, which are determined from 
the effective action $S_{\rm eff}$, see below.
The quantities $\Delta$ and ${\cal G}$ are the expectation values
for the two-point functions, i.e., the {\em full\/} propagators, of 
soft gluons and relevant quarks.
The functional $\Gamma_2$ is the sum of all two-particle irreducible
(2PI) diagrams. These diagrams are vacuum diagrams, i.e., they
have no external legs. They are constructed from the vertices
defined by the interaction part of $S_{\rm eff}$,
linked by full propagators $\Delta$, ${\cal G}$. 
The expectation values for the one- and two-point functions of
the theory are determined from the stationarity conditions
\be \label{statcond}
0 = \frac{\delta \Gamma}{\delta A} = \frac{\delta \Gamma}{\delta
\bar{\Psi}} = \frac{\delta \Gamma}{\delta \Psi} = \frac{\delta \Gamma}{\delta
\Delta } = \frac{\delta \Gamma}{\delta {\cal G}} \; .
\ee
The first condition yields the Yang-Mills equation for the 
expectation value $A$ of the soft gluon field. The second and third
condition correspond to the Dirac equation for $\Psi$ and
$\bar{\Psi}$, respectively. The effective action (\ref{Seff})
contains a multitude of terms which depend on $A, \bar{\Psi}, \Psi$,
and thus the Yang-Mills and Dirac equations are
rather complex, wherefore we refrain from explicitly presenting them here.
Nevertheless, for the Dirac equation the solution is trivial,
since $\bar{\Psi}_1,\, \Psi_1$ are Grassmann-valued fields, and
their expectation values must vanish identically,
$\bar{\Psi} = \langle \bar{\Psi}_1 \rangle = \Psi = \langle \Psi_1
\rangle \equiv 0$. On the other hand, for the Yang-Mills equation,
the solution $A$ is in general non-zero but, at least for
the two-flavor color superconductor considered here, it was shown 
\cite{Gerhold,DirkDennis} to be parametrically small, $A \sim \phi^2/(g^2
\mu)$, where $\phi$ is the color-superconducting gap parameter. 
Therefore, to subleading order in the gap equation it can be neglected.

The fourth and fifth condition (\ref{statcond}) are Dyson-Schwinger
equations for the soft gluon and relevant quark propagator,
respectively,
\begin{subequations} \label{DSE}
\bea 
\Delta^{-1} & = & D^{-1} + \Pi\;, \label{DSEgluon}\\
{\cal G}^{-1} & = & G^{-1} + \Sigma \;, \label{DSEquark}
\eea
\end{subequations}
where
\begin{subequations} \label{selfenergy}
\bea 
\Pi & \equiv & - 2 \, \frac{\delta \Gamma_2}{\delta \Delta^T}\; ,
\label{selfenergygluon}\\
\Sigma & \equiv & 2 \, \frac{\delta \Gamma_2}{\delta {\cal G}^T}
\label{selfenergyquark}
\eea
\end{subequations}
are the gluon and quark self-energies, respectively.
The Dyson-Schwinger equation for the relevant quark propagator is
a $2 \times 2$ matrix equation in Nambu-Gor'kov space,
\be \label{NGinvquark}
{\cal G}^{-1} =   \left( \begin{array}{cc}
                            [ G^+]^{-1} &  0 \\
                     0 & [ G^-]^{-1} \end{array} \right)
                + \left( \begin{array}{cc}
                            \Sigma^+ &  \Phi^- \\
                     \Phi^+ & \Sigma^- \end{array} \right) \;,
\ee
where $\Sigma^+$ is the regular self-energy for quarks and
$\Sigma^-$ the corresponding one for charge-conjugate quarks.
The off-diagonal self-energies $\Phi^\pm$, the so-called {\em gap
matrices}, connect regular with 
charge-conjugate quark degrees of freedom. A non-zero $\Phi^\pm$
corresponds to the condensation of quark Cooper pairs.
Only two of the four components of this matrix equation are
independent, say $[G^+]^{-1} + \Sigma^+$ and $\Phi^+$, 
the other two can be obtained via 
$[G^-]^{-1} + \Sigma^-= C \{[G^+]^{-1} + \Sigma^+\}^T C^{-1}$,
$\Phi^- \equiv \gamma_0 [\Phi^+]^\dagger \gamma_0$.
Equation (\ref{NGinvquark}) can be formally solved for ${\cal G}$
\cite{manuel2},
\be \label{NGquarkprop}
{\cal G} \equiv \left( \begin{array}{cc}
                            {\cal G}^+ &  \Xi^- \\
               \Xi^+ & {\cal G}^- \end{array} \right) \; ,
\ee
where
\be
{\cal G}^\pm \equiv \left\{ [G^\pm]^{-1} + \Sigma^\pm - 
\Phi^\mp \left( [G^\mp]^{-1} + \Sigma^\mp \right)^{-1} \Phi^\pm \right\}^{-1}
\ee
is the propagator describing normal propagation of quasiparticles
and their charge-conjugate counterpart, while
\be \label{Xi}
\Xi^\pm \equiv - \left( [G^\mp]^{-1} + \Sigma^\mp \right)^{-1}
\Phi^\pm {\cal G}^\pm
\ee
describes anomalous propagation of quasiparticles, which is possible
if the ground state is a color-superconducting quark-quark condensate,
for details, see Ref.\ \cite{DHRreview}.

The tree-level gluon propagator is defined as
\be
D^{-1} \equiv - \frac{\delta^2 S_{\rm eff} [A,\bar{\Psi}, \Psi]}{\delta 
A\,\delta A} \;.
\ee
Since we ultimately evaluate the tree-level propagator at
the stationary point of $\Gamma$, Eq.\ (\ref{statcond}), where
$\bar{\Psi} = \Psi =0$, we may omit all terms in $S_{\rm eff}$, 
Eq.\ (\ref{Seff}), which are proportional to the quark fields.
The only terms which contribute to the tree-level gluon propagator
are therefore 
\be
D^{-1} \equiv 
- \frac{\delta^2 }{\delta A\,\delta A}\left(
S_A+ \frac{1}{2}\, {\rm Tr}_q\ln {\cal G}_{22}^{-1}
- \frac{1}{2}\, {\rm Tr}_g\ln \Delta_{22}^{-1} \right) \;.
\ee
Using the expansions (\ref{expquark}),
(\ref{explnquark}), (\ref{expansion2}), and
(\ref{expansion3}), and exploiting the cyclic property of the trace, 
one finds
\be
D^{-1} = - \frac{\delta^2 S_A}{\delta A \, \delta A} 
- \frac{g}{2}\, 
{\rm Tr}_q\left( \frac{\delta {\cal G}_{22}}{\delta A} \, 
\frac{\delta {\cal A}_{22}}{\delta A} \right) 
+\frac{1}{2}{\rm Tr}_g \left(\frac{\delta \Delta_{22}}{ \delta A } \,
\frac{\delta \Pi_{22}}{\delta A} +\Delta_{22} \, 
\frac{\delta^2 \Pi_{22}}{\delta A\,\delta A}
\right)
\;.\label{D0noA}
\ee
In order to proceed, note
that the Dyson-Schwinger equations (\ref{DSE}) are evaluated
at the stationary point of the effective action, where 
$\bar{\Psi} = \Psi = 0, \, A \simeq 0$.
For $A=0$, the first term yields the free inverse propagator for
soft gluons, $\Delta_{0,11}^{-1}$, cf.\ Eq.\ (\ref{Sgluon}),
plus a contribution from the
Faddeev-Popov determinant, 
$(\delta^2 {\rm Tr}_{gh} \ln {\cal W}^{-1} / \delta A
\delta A)_{A=0}$. The contributions from the three- and four-gluon
vertex vanish for $A=0$. 
Furthermore, according to Eq.\ (\ref{calA}), 
\be \label{Gammatilde}
\frac{\delta {\cal A}_{22}(K,Q)}{\delta A(P)} =
\frac{1}{\sqrt{VT^3}} \, \hat{\Gamma} \, \delta^{(4)}_{K,Q+P}
\equiv \tilde{\Gamma}(K,Q;P)\; .
\ee
This is a matrix in fundamental color, flavor, and Nambu-Gor'kov space,
as well as in the space of quark 4-momenta $K,Q$. It is a vector in
Minkowski and adjoint color space ($\hat{\Gamma}$ carries a 
Lorentz-vector and a gluon color index), as well as in the space
of gluon 4-momenta $P$.
We evaluate $(\delta {\cal G}_{22}/\delta A)_{A=0}$ using the expansion
(\ref{expquark}). Only the term for $n=1$ survives when taking $A=0$.
For $\bar{\Psi} = \Psi = 0$, we have $\Pi_{\cal B}=0$, cf.\ Fig.\ \ref{ZZa},
and we only need to consider $\Pi_{22} = \Pi_{\rm loop} + \Pi_{{\cal V}}$.
Then, the term ${\cal V}^{(3)} \equiv 
(\delta \Pi_{{\cal V}} / \delta A )_{A=0}$ corresponds
to a triple-gluon vertex, cf.\ Fig.\ \ref{ZZc},
where two hard gluons couple to one soft gluon.
The term $(\delta \Pi_{\rm loop}/\delta A )_{A=0}$ is a correction to
this vertex: it couples two hard gluons to a soft one through an
(irrelevant) quark loop, cf.\ Fig.\ \ref{ZZb}.
According to arguments well-known from the HTL/HDL effective theory,
this vertex correction can never be of the same order as
the tree-level vertex ${\cal V}^{(3)}$,
since the two incoming gluons are hard. We therefore neglect 
$(\delta \Pi_{\rm loop}/\delta A )_{A=0}$ in the following. 
Similarly, ${\cal V}^{(4)} \equiv (\delta^2 \Pi_{{\cal V}} / \delta A \delta
A )_{A=0}$ is a four-gluon vertex, cf.\ Fig.\ \ref{ZZc},
where two hard gluons couple to
two soft ones, and $(\delta^2 \Pi_{\rm loop} / \delta A \delta A )_{A=0}$ 
is the one-(quark-)loop correction to this vertex, cf.\ Fig.\
\ref{ZZb}. Applying
the same arguments as above, we only keep ${\cal V}^{(4)}$.
Arguments from the HTL/HDL effective theory also tell us that to leading order 
we may approximate $\Delta_{22} \simeq \Delta_{0,22}$.
Finally, utilizing the same arguments we approximate
$\delta \Delta_{22}/ \delta A  \simeq - \Delta_{0,22} {\cal V}^{(3)}
\Delta_{0,22}$.
Then, the inverse tree-level gluon propagator of Eq.\ (\ref{D0noA}) becomes
\be
D^{-1} = \Delta_{0,11}^{-1} +\frac{g^2}{2}\, 
{\rm Tr}_q \left( {\cal G}_{0,22}\, \tilde{\Gamma} \, 
{\cal G}_{0,22}\, \tilde{\Gamma} \right) 
- \frac{1}{2}\, {\rm Tr}_g \left( \Delta_{0,22} \, {\cal V}^{(3)} \,
\Delta_{0,22}\,{\cal V}^{(3)} \right) + 
\frac{1}{2}\, {\rm Tr}_g \left(\Delta_{0,22}\,  {\cal V}^{(4)} 
\right) - \left. \frac{\delta^2 {\rm Tr}_{gh} \ln {\cal W}^{-1} }{\delta A \, 
\delta A} \right|_{A = 0}
\;.\label{D0noA2}
\ee
The second term represents an (irrelevant) quark-loop, while
the third term is a hard gluon loop. The fourth term is a hard
gluon tadpole. Finally, the last term in Eq.\ (\ref{D0noA2}) 
corresponds to a ghost loop necessary to cancel loop contributions
from unphysical gluon degrees of freedom.
Note that, in the effective theory, loop
contributions involving irrelevant quarks and hard gluons 
occur already in the tree-level action (\ref{Seff}). Therefore,
such loops also arise in the inverse tree-level propagator 
(\ref{D0noA2}) for the soft gluons of the effective theory. 
For the projection operators (\ref{Q12}) and (\ref{PHTL})
the inverse tree-level propagator (\ref{D0noA2}) is precisely the
HTL/HDL-resummed inverse gluon propagator.
For small temperatures, $T \ll \mu$, 
the contribution from the gluon and ghost
loops is negligible as compared to that from the quark loop,
\be
D^{-1} \simeq  \Delta_{0,11}^{-1} +\frac{g^2}{2}\, 
{\rm Tr}_q \left( {\cal G}_{0,22}\, \tilde{\Gamma} \, 
{\cal G}_{0,22}\, \tilde{\Gamma} \right) 
\;.\label{D0noA3}
\ee

The inverse tree-level quark propagator is defined as
\be \label{S0}
G^{-1} \equiv
- 2 \,\frac{\delta^2 S_{\rm eff} [A,\bar{\Psi}, \Psi ]}{
\delta\bar{\Psi}\,\delta{\Psi}}\;.
\ee
For $\bar{\Psi} =\Psi =0$, the last term in Eq.\ (\ref{Seff}) does not
contribute to $G^{-1}$, because it has at least four external quark legs,
and the two functional derivatives $\delta/ \delta \bar{\Psi}$, 
$\delta/ \delta \Psi$ amputate only two of them. The first and the
third term in Eq.\ (\ref{Seff}) do not depend on $\bar{\Psi}, \Psi$ at
all, therefore
\be
G^{-1}  =   {\cal G}_{0,11}^{-1} +  g {\cal B}[A]
+ \frac{\delta^2 {\rm Tr}_g \ln \Delta_{22}^{-1} }{\delta\bar{\Psi}\,
\delta \Psi } \;.
\ee
Using the expansion formulae (\ref{expansion2}) and (\ref{expansion3})
and the fact that $\Pi_{22}$ depends
on $\bar{\Psi}, \Psi$ only through $\Pi_{\cal B}$, we obtain
\be
\frac{\delta^2{\rm Tr}_g \ln \Delta_{22}^{-1} }{\delta\bar{\Psi}\,
\delta{\Psi}}
= {\rm Tr}_g \left( \Delta_{22} 
\frac{\delta^2 \Pi_{{\cal B}}}{\delta\bar{\Psi}\,\delta{\Psi}}\right)\;.
\ee
We have exploited the fact that this expression is evaluated
at $\bar{\Psi} = \Psi = 0$, i.e., terms with external quark legs
will eventually vanish. The trace runs only over adjoint colors, Lorentz
indices, and (hard) gluon 4-momenta. 
Since $\Delta_{22}$ is a hard gluon propagator,
the contribution from $\Pi_{22}$ to $\Delta_{22}$ may be neglected
to the order we are computing, and we may set $\Delta_{22} 
\simeq \Delta_{0,22}$. Furthermore, 
$(\delta^2 \Pi_{{\cal B}}/ \delta \bar{\Psi} \delta \Psi)_{A=0}
\equiv - g^2 \,\tilde{\Gamma}\, {\cal G}_{0,22}\, \tilde{\Gamma}$, 
cf.\ Fig.\ \ref{ZZa}. At $\bar{\Psi}= \Psi = 0,\, A \simeq 0$ we are left with
\be \label{S0noA}
G^{-1}  =  {\cal G}_{0,11}^{-1}
- g^2 \, {\rm Tr}_g \left( \Delta_{0,22} \, \tilde{\Gamma} \,
{\cal G}_{0,22} \, \tilde{\Gamma} \right)\;.
\ee 
As was the case for the tree-level gluon propagator, also 
the tree-level quark propagator receives a loop contribution; here it
arises from a loop involving an irrelevant quark and a hard gluon
line. The term
$\tilde{\Gamma} {\cal G}_{0,22} \tilde{\Gamma}$ under the gluon trace remains
a matrix in the quark indices, i.e., fundamental color, flavor,
Dirac, and quark 4-momenta.

We now proceed to solve the Dyson-Schwinger equations (\ref{DSE}) for the
soft gluon and relevant quark propagator. To this end, we have
to determine $\Gamma_2$. Of course, it is not feasible to consider
{\em all\/} possible 2PI diagrams. The advantage of the CJT formalism is
that {\em any\/} truncation of $\Gamma_2$ defines a meaningful,
self-consistent many-body approximation for which one can solve
the Dyson-Schwinger equations (\ref{DSE}). In our truncation of
$\Gamma_2$ we only take into account 2-loop diagrams 
which are 2PI with respect to the soft gluon and
relevant quark propagators $\Delta, \, {\cal G}$,
\be 
\label{EqGamma2}
\Gamma_2 = - \frac{g^2}{4}\,{\rm Tr}_{q,g} \left( {\cal G} \,
\tilde{\Gamma} \, {\cal G} \, \tilde{\Gamma} \, \Delta \right)
- \frac{g^2}{2} \, {\rm Tr}_{q,g} \left( {\cal G} \, \tilde{\Gamma} \,
{\cal G}_{0,22} \, \tilde{\Gamma} \, \Delta \right)
- \frac{g^2}{4}\, {\rm Tr}_{q,g} \left( {\cal G}\, \tilde{\Gamma}\, {\cal
G}\, \tilde{\Gamma}\, \Delta_{0,22} \right) \; .
\ee
The traces now run over quark as well as over gluon indices. Consider,
for instance, the term ${\cal G} \, \tilde{\Gamma}\,  {\cal G}\,
\tilde{\Gamma}$. It is a matrix in the space of 
fundamental color, flavor, Dirac and quark 4-momenta, of which the
trace is taken through ${\rm Tr}_q$. In addition,
due to the two factors $\tilde{\Gamma}$ it carries two
Lorentz-vector, adjoint-color, and gluon-4-momenta indices.
The trace ${\rm Tr}_g$ contracts these indices with the corresponding
ones from the gluon propagator $\Delta$.

\begin{figure}[ht]
\includegraphics[width=8cm]{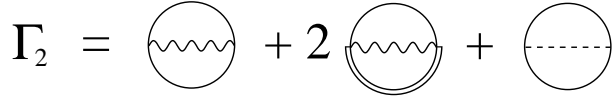}
\caption{Diagrammatic representation of $\Gamma_2$, Eq.\
(\ref{EqGamma2}).}
\label{Gamma2eff}
\end{figure}

The diagrams corresponding to Eq.\ (\ref{EqGamma2}) are shown
in Fig.\ \ref{Gamma2eff}. 
The first two terms are constructed from the quark-gluon
coupling $\sim \bar{\Psi}\, g {\cal B}\, \Psi$. Using Eq.\ (\ref{B}),
one may either obtain an ordinary quark-gluon vertex 
$\sim g\, \bar{\Psi}\, {\cal A} \,\Psi$, 
involving one soft gluon and two relevant quark
legs, or a vertex $\sim g^2 \, \bar{\Psi}\, {\cal A} \, 
{\cal G}_{22}\, {\cal A}\, \Psi$, with (at least) two soft 
gluon legs and two relevant quark
legs. To lowest order, we approximate ${\cal G}_{22} \simeq {\cal
G}_{0,22}$, which neglects vertices with more than two soft gluon
legs. Taking two ordinary quark-gluon vertices and tying 
them together to obtain a 2PI 2-loop diagram, we arrive at the first
term in Eq.\ (\ref{EqGamma2}), or the first 
diagram in Fig.\ \ref{Gamma2eff}. Taking one of the two-gluon-two-quark
vertices and tying the legs together, one obtains the second term
in Eq.\ (\ref{EqGamma2}), or the second
diagram in Fig.\ \ref{Gamma2eff}, respectively. Finally, the third term/diagram
arises from the last term in Eq.\ (\ref{Seff}). To lowest order,
this corresponds to a four-quark vertex 
$\sim g^2 \, \bar{\Psi} \, \tilde{\Gamma}\, \Psi
\, \Delta_{0,22} \bar{\Psi} \, \tilde{\Gamma}\, \Psi$. Tying the quark
legs together to form a 2PI diagram, one obtains the corresponding
term/diagram in Eq.\ (\ref{EqGamma2})/Fig.\ \ref{Gamma2eff}. 

The combinatorial factors in front of the various terms in Eq.\ 
(\ref{EqGamma2}) are explained as follows. In the first diagram, there
are two ordinary quark-gluon vertices. According to Eq.\ (\ref{Seff}),
each comes with a factor 1/2. Moreover, since there are two vertices,
the diagram is, in the perturbative sense, a diagram of second order,
which causes an additional factor 1/2 \cite{FTFT}. Finally, there are
two possibilities to connect the quark lines between the two
vertices. In total, we then have a prefactor 
$ -(1/2)^2 \times 1/2 \times 2 = -1/4$, where the minus sign arises
from the fermion loop. The second diagram arises from the
two-quark-two-gluon vertex, which already comes with a prefactor $-1/2$ in
Eq.\ (\ref{Seff}). It is perturbatively of first order, and 
there is only one possibility to tie the quark and gluon lines
together, so there is no additional combinatorial factor (and no
additional minus sign) for this diagram. Finally, the third diagram
arises from the four-quark vertex, $(1/2) {\cal J}_{{\cal B}}
\Delta_{0,22} {\cal J}_{{\cal B}}$, in Eq.\ (\ref{Seff}). 
This vertex comes with a factor 
1/2 and is perturbatively of first order. 
However, there are two additional factors 1/2 residing in ${\cal J}_{{\cal
B}}$, since ${\cal J}_{{\cal B}} \sim (1/2) \bar{\Psi} \, \hat{\Gamma}\,
\Psi$, cf.\ Eq.\ (\ref{J_B}). 
Again, there are two possibilities to tie the quark lines
together, so that, in total, we have a prefactor 
$-1/2 \times (1/2)^2 \times 2 = - 1/4$, where the minus sign 
again stands for the quark loop.

\begin{figure}[ht]
\includegraphics[width=8cm]{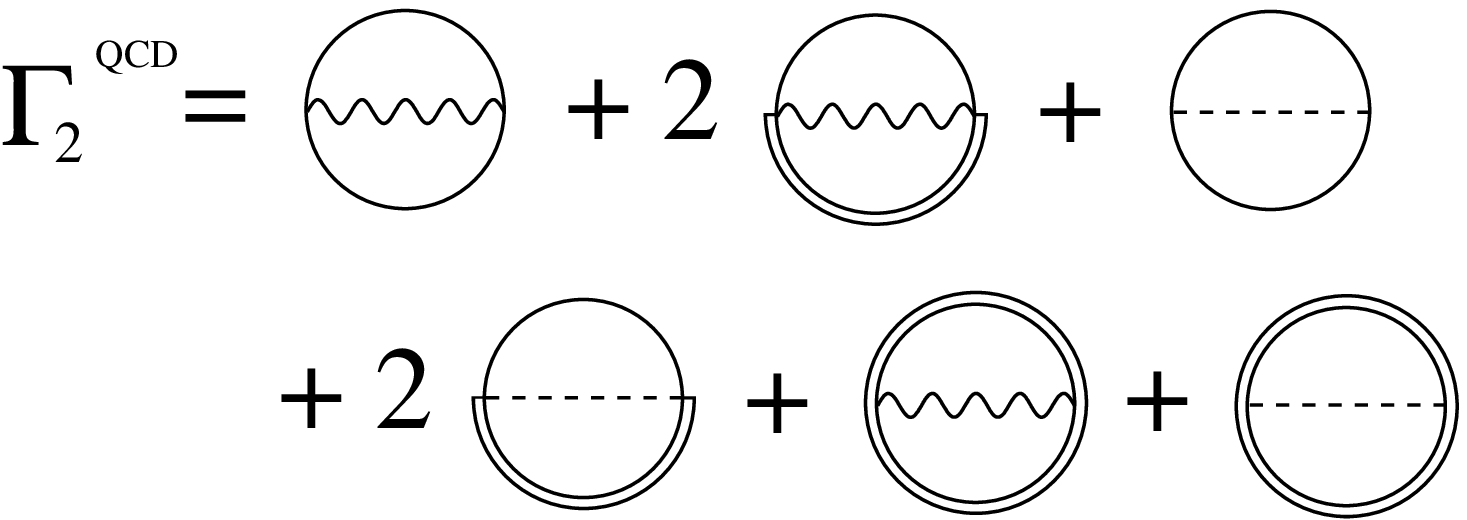}
\caption{Diagrammatic representation of $\Gamma_2^{\rm QCD}$, after
decomposing quark lines into relevant and
irrelevant, as well as gluon propagators into soft and hard
contributions.}
\label{Gamma2QCD}
\end{figure} 

At this point, it is instructive to compare $\Gamma_2$, Eq.\
(\ref{EqGamma2}), in the effective theory with 
$\Gamma_2^{\rm QCD}$ which one would have written
down in QCD at the same loop level.
$\Gamma_2^{\rm QCD}$ would be equivalent to the first diagram
of Fig.\ \ref{Gamma2eff}, but now the quark and gluon lines represent the
full propagators for {\em all\/} momentum modes, relevant {\em and\/}
irrelevant as well as soft {\em and\/} hard. In order to compare
with $\Gamma_2$ of the effective theory, we decompose the quark
propagators into relevant and irrelevant modes, and the gluon
propagator into soft and hard modes. One obtains the six diagrams
shown in Fig.\ \ref{Gamma2QCD}. The first three are precisely the
same that occur in $\Gamma_2$ of the effective theory, including
the combinatorial prefactors. The last three diagrams do not occur
in $\Gamma_2$ of the effective theory, because they are not 2PI
with respect to the relevant quark propagator ${\cal G}$ and
the soft gluon propagator $\Delta$. Nevertheless, they are
still included in the CJT effective action of the effective theory,
Eq.\ (\ref{Gamma}): opening the
relevant quark line of the fourth diagram, we recognize the loop
contribution to the tree-level quark propagator $G^{-1}$, cf.\
Eq.\ (\ref{S0noA}). Now consider the fifth term in Eq.\ (\ref{Gamma}):
here, this loop contribution to $G^{-1}$
is multiplied with ${\cal G}$ and traced over, which yields
the fourth diagram in $\Gamma_2^{\rm QCD}$.
Similarly, opening the soft gluon line
of the fifth diagram, we identify this diagram as 
the irrelevant quark-loop contribution
to the tree-level gluon propagator $D^{-1}$, cf.\ Eq.\ (\ref{D0noA3}).
The third term in Eq.\ (\ref{Gamma}), where this contribution is
multiplied by $\Delta$ and traced over, then yields the fifth diagram
of $\Gamma_2^{\rm QCD}$.
Finally, the sixth diagram resides in the term $\sim {\rm Tr}_g \ln 
\Delta_{22}^{-1}$ of the tree-level effective action $S_{\rm eff}$,
cf.\ Fig.\ \ref{AAd}.
Therefore, in principle, the CJT effective action (\ref{Gamma}) for
the effective theory contains the same information as the
corresponding one for QCD. However, while in QCD self-consistency
is maintained for {\em all\/} momentum modes via the solution of the
stationarity condition (\ref{statcond}), in the effective theory
self-consistency is only required for the {\em relevant\/} quark 
and {\em soft\/} gluon modes. In this sense, the effective theory 
provides a simplification of the full problem. 

\subsection{Dyson-Schwinger equations for relevant quarks and soft
gluons} \label{IVb}

After having specified $\Gamma_2$ in Eq.\ (\ref{EqGamma2}), 
we are now in the position to write down the Dyson-Schwinger equations
(\ref{DSE}) explicitly. For the full inverse propagator of soft gluons
we obtain with Eqs.\ (\ref{DSEgluon}), (\ref{selfenergygluon}), 
(\ref{D0noA3}), and (\ref{EqGamma2})
\be \label{DSEg}
\Delta^{-1} = \Delta_{0,11}^{-1} +\frac{g^2}{2}\,\left[ 
{\rm Tr}_q\left( {\cal G}_{0,22}\, \tilde{\Gamma} \, 
{\cal G}_{0,22}\, \tilde{\Gamma} \right) + 2\, 
{\rm Tr}_q\left( {\cal G}\, \tilde{\Gamma} \, 
{\cal G}_{0,22}\, \tilde{\Gamma} \right)+ 
{\rm Tr}_q\left( {\cal G}\, \tilde{\Gamma} \, 
{\cal G}\, \tilde{\Gamma} \right) \right] \;.
\ee
The first term in square brackets takes into account the effect of
quark-antiquark excitations as well as quark-hole excitations far
from the Fermi surface. The second term is the contribution from
excitations where one quark is close to the Fermi surface (a relevant
quark) while the second is far from the Fermi surface or an
antiquark (an irrelevant quark). The relevant quark propagator
${\cal G}$ can have diagonal elements in Nambu-Gor'kov space,
corresponding to normal propagation of quasiparticles, as well as
off-diagonal elements, corresponding to anomalous propagation of
quasiparticles, cf.\ Eq.\ (\ref{NGquarkprop}).
However, in the second term in square
brackets the latter contribution is absent, because ${\cal
G}_{0,22}$ is purely diagonal in Nambu-Gor'kov space, cf.\ Eq.\ (\ref{G0FT}).
This is different for the last term in square brackets, which
corresponds to quark-hole excitations close to the Fermi surface.
Both quark propagators have to be determined self-consistently and may
have off-diagonal elements in Nambu-Gor'kov space. Consequently, the
trace over Nambu-Gor'kov space gives two contributions, a loop where
both quarks propagate normally, and another one where they propagate
anomalously. Diagrams of this type have been evaluated in Ref.\
\cite{Meissner2f3f} and lead to the Meissner effect for gluons
in a color superconductor.

For the full inverse propagator of relevant quarks we obtain
with Eqs.\ (\ref{DSEquark}), (\ref{selfenergyquark}), (\ref{S0noA}),
and (\ref{EqGamma2})
\be \label{DSEq}
{\cal G}^{-1} = {\cal G}_{0,11}^{-1}
- g^2 \, \left[ {\rm Tr}_g\left( \Delta_{0,22} \, \tilde{\Gamma} \,
{\cal G}_{0,22} \, \tilde{\Gamma} \right)
+{\rm Tr}_g\left( \Delta \, \tilde{\Gamma} \,
{\cal G}_{0,22} \, \tilde{\Gamma} \right)
+  {\rm Tr}_g\left( \Delta_{0,22} \, \tilde{\Gamma} \,
{\cal G} \, \tilde{\Gamma} \right)
+  {\rm Tr}_g\left( \Delta \, \tilde{\Gamma} \,
{\cal G} \, \tilde{\Gamma} \right) \right] \; .
\ee
The first two terms in square brackets do not have off-diagonal
components in Nambu-Gor'kov space. They contribute only to the
regular quark self-energy. The other two terms in square brackets
have both diagonal and off-diagonal components in Nambu-Gor'kov
space. The diagonal components contribute to the regular quark self-energy,
in particular, the fourth term leads to the quark
wave-function renormalization factor computed first in Ref.\ 
\cite{manuel}. It gives rise to non-Fermi liquid behavior 
\cite{rockefeller}. The off-diagonal components enter the
gap equation for the color-superconducting gap parameter.

The system of Eqs.\ (\ref{DSEg}) and (\ref{DSEq}) has to be solved
self-consistently for the full propagators of quarks and gluons.
However, as was shown in Ref.\ \cite{dirkselfenergy}, 
in order to extract the color-superconducting gap parameter to
subleading order it is sufficient to consider the gluon propagator 
in HDL approximation;
corrections arising from the color-superconducting gap in the
quasiparticle spectrum are of sub-subleading order in the gap
equation. For our purpose this means that it is not necessary to 
self-consistently solve Eq.\ (\ref{DSEg}) together with Eq.\
(\ref{DSEq}); we may approximate ${\cal G}$ on the right-hand
side of Eq.\ (\ref{DSEg}) by ${\cal G}_{0,11}$. 
In essence, this is equivalent to
considering only the first term
on the right-hand side of Eq.\ (\ref{DSEq}) when solving
Eq.\ (\ref{DSEg}). Of course, under this approximation
the effect of the regular quark self-energy (leading to wave-function
renormalization) and of the anomalous quark self-energy
(which accounts for the gap in the quasiparticle excitation spectrum)
are neglected.

With this approximation, and using ${\cal G}_0 \equiv {\cal G}_{0,11}
\oplus {\cal G}_{0,22}$, we may combine the terms in Eq.\ (\ref{DSEg})
to give
\be \label{DeltaHDL}
\Delta^{-1} \simeq \Delta_{0,11}^{-1} + \frac{g^2}{2}\,
{\rm Tr}_q\left( {\cal G}_0\, \tilde{\Gamma} \, 
{\cal G}_0\, \tilde{\Gamma} \right) \;.
\ee
Taking the gluon cut-off scale $\Lambda_g$ to fulfill 
$g \mu \ll \Lambda_g \lesssim \mu$,
soft gluons are defined to have momenta of order $g \mu$. 
We compute the fermion loop in Eq.\ (\ref{DeltaHDL}) under this
assumption (taking the soft gluon energy to be of the same
order of magnitude as the gluon momentum). We then realize
that the soft gluon propagator determined by Eq.\ (\ref{DeltaHDL}) 
is just the gluon propagator in HDL approximation. We indicate this
fact in the following by a subscript, $\Delta \equiv \Delta_{\rm HDL}$.
Armed with this (approximate) solution of the Dyson-Schwinger equation
(\ref{DSEg}) we now proceed to solve Eq.\ (\ref{DSEq}).
We consider the two independent components $[G^+]^{-1} + \Sigma^+$
and $\Phi^+$ in Nambu-Gor'kov space separately. Due to translational
invariance, it is convenient to define
$[G^+]^{-1}(K,Q) \equiv (1/T) [G^+]^{-1}(K)\,
\delta^{(4)}_{K,Q}$, $\Sigma^+(K,Q) \equiv (1/T) \Sigma^+(K)\,
\delta^{(4)}_{K,Q}$, 
and using Eqs.\ (\ref{NGvertex}),
(\ref{G0FT}), (\ref{D_0}), 
(\ref{Gammatilde}), we obtain 
the Dyson-Schwinger equation for $[G^+]^{-1} + \Sigma^+$,  
\bea 
[G^+]^{-1}(K) + \Sigma^+(K) & = & [ G_{0,11}^+]^{-1}(K) \non
& - & g^2 \, \frac{T}{V} \sum_Q \left\{ \frac{}{} 
\left[ \Delta_{0,22}\right]^{\mu \nu}_{ab}(K-Q) +  
\left[ \Delta_{\rm HDL}\right]^{\mu \nu}_{ab}(K-Q) \right\} 
\gamma_\mu T^a \, G_{0,22}(Q) \, \gamma_\nu T^b  \non 
& - &  g^2 \, \frac{T}{V} \sum_Q \left\{ \frac{}{}
\left[ \Delta_{0,22}\right]^{\mu \nu}_{ab}(K-Q) +  
\left[ \Delta_{\rm HDL}\right]^{\mu \nu}_{ab}(K-Q) \right\}
\gamma_\mu T^a \, {\cal G}^+(Q) \, \gamma_\nu T^b \;.
\label{DSEq2}
\eea
Note that the first sum over $Q$ runs over irrelevant quark momenta,
$0 \leq q < \mu - \Lambda_q$ and $\mu+ \Lambda_q < q < \infty$, while
the second sum runs over relevant quark momenta,
$\mu - \Lambda_q \leq q \leq \mu + \Lambda_q$.
There is no double counting of gluon exchange contributions, since
the hard gluon propagator $\Delta_{0,22}$ has support only for gluon momenta
$|{\bf k} - {\bf q}| > \Lambda_g$, while the HDL propagator 
is restricted to gluon momenta $|{\bf k} - {\bf q}| \leq \Lambda_g$.
To subleading order in the gap equation, we do not have to solve
this Dyson-Schwinger equation self-consistently. It is sufficient
to use the approximation ${\cal G}^+ \simeq G_{0,11}^+$ on the
right-hand side of Eq.\ (\ref{DSEq2}) and to keep only the last term
which, as discussed above, is responsible for non-Fermi liquid
behavior in cold, dense quark matter. The net result is then simply
a wave-function renormalization for the free quark propagator
$G_{0,11}^+$ \cite{manuel},
\be \label{fullinversequarkprop}
[G^+]^{-1}(K) + \Sigma^+(K) \simeq [ G_{0,11}^+]^{-1}(K)
+ \bar{g}^2  \, k_0 \, \gamma_0 \, \ln \frac{M^2}{k_0^2} 
\equiv \left[ Z^{-1}(k_0)\, k_0 + \mu \right] \,
\gamma_0 - \vg \cdot {\bf k}\; ,
\ee
where $\bar{g} \equiv g/ (3 \sqrt{2} \pi)$ and $M^2 = (3 \pi/4) m_g^2$,
with the gluon mass parameter $m_g$ defined in Eq.\ (\ref{gluonmass}). 
Neglecting effects from the finite life-time of
quasi-particles \cite{manuel2}, which are of sub-subleading
order in the gap equation, the wave-function renormalization factor is
\be \label{wavefunc}
Z(k_0) = \left( 1 + \bar{g}^2  \,\ln \frac{M^2}{k_0^2} \right)^{-1}\;.
\ee
Due to translational invariance, it is convenient to define
$\Phi^+(K,Q) \equiv (1/T) \, 
\Phi^+(K) \, \delta^{(4)}_{K,Q}$ and $\Xi^+(K,Q) \equiv T \, \Xi^+(K)\,
\delta^{(4)}_{K,Q}$, and the Dyson-Schwinger equation for $\Phi^+(K)$ becomes
\be
\Phi^+ (K)  =  g^2 \, \frac{T}{V} \sum_Q \left\{ \frac{}{}
\left[ \Delta_{0,22}\right]^{\mu \nu}_{ab}(K-Q) 
+  \left[ \Delta_{\rm HDL}\right]^{\mu \nu}_{ab}(K-Q)\frac{}{}
\right\} \, \gamma_\mu (T^a)^T \, \Xi^+(Q) \, \gamma_\nu T^b   \;.
\label{Phi}
\ee
Here, the sum runs only over relevant quark momenta, $\mu - \Lambda_q 
\leq q \leq \mu + \Lambda_q$. This is the gap equation for the 
color-superconducting gap parameter within our effective theory.
There is no contribution from irrelevant fermions, since their
propagator is diagonal in Nambu-Gor'kov space.

While the gluon cut-off was taken to be $\Lambda_g \lesssim \mu$, so
that soft gluons have typical momenta of order $g \mu$,
so far we have not specified the magnitude of $\Lambda_q$.
In weak coupling, the color-superconducting gap function is strongly
peaked around the Fermi surface \cite{son,rdpdhr,schaferwilczek}. 
For a subleading-order calculation of the gap parameter, it is
therefore sufficient to consider as relevant quark modes those within
a thin layer of width $2 \Lambda_q$ around the Fermi surface.
For the following, our principal assumption is
$\Lambda_q \lesssim g \mu \ll \Lambda_g \lesssim \mu$.
As we shall see below, this assumption is crucial to 
identify sub-subleading corrections to the gap equation (\ref{Phi}), which
arise, for instance, from the pole of the gluon propagator.
Note that this assumption is different from that of
Refs.\ \cite{hong2,schaferefftheory}, where it is assumed that 
$\Lambda_q \simeq \Lambda_g$.

For a two-flavor color superconductor, the color-flavor-spin
structure of the gap matrix is \cite{DHRreview}
\be \label{gapmatrix}
\Phi^+(K) = J_3 \tau_2 \gamma_5\, \Lambda_{\bf k}^+ \, 
\Theta(\Lambda_q -|k-\mu|)\, \phi(K)\;, 
\ee
where $(J_3)_{ij} \equiv -i \epsilon_{ij3}$ and $(\tau_2)_{fg} \equiv
-i \epsilon_{fg}$ represent the fact
that quark pairs condense in the color-antitriplet, flavor-singlet channel.
The Dirac matrix $\gamma_5$ restricts quark pairing to the even-parity
channel (which is the preferred one due to the $U(1)_A$ anomaly of
QCD). In the effective action (\ref{Seff}), antiquark and irrelevant
quark degrees of freedom are integrated out. The condensation of
antiquark or irrelevant quark pairs, while in principle possible,  
is thus not taken into account;
the bilocal source terms in Eq.\ (\ref{Seffsources}) only allow for
the condensation of relevant quark degrees of freedom.
The condensation of antiquarks or irrelevant quarks
could also be accounted for, if one introduces bilocal
source terms already in Eq.\ (\ref{quarkaction}), 
i.e., {\em prior\/} to integrating out any of the quark degrees of freedom.
While there is in principle no obstacle in following this course of
action, it is, however, not really necessary if one is interested in 
a calculation of the color-superconducting gap parameter to subleading
order in weak coupling: antiquarks contribute to the gap
equation beyond subleading order \cite{aqgap}, and the gap function for quarks
falls off rapidly away from the Fermi surface, i.e., in the region
of irrelevant quark modes, and thus also contributes at most to
sub-subleading order to the gap equation.
Consequently, the Dirac structure of the gap matrix (\ref{Phi})
contains only the projector $\Lambda_{\bf k}^+$ onto positive energy 
states. The theta function accounts for the fact that
the gap function $\phi(K)$ pertains only to
relevant quark modes.

Inserting Eq.\ (\ref{fullinversequarkprop}) and the corresponding
one for $[G^-]^{-1} +\Sigma^-$, as well as Eq.\ (\ref{gapmatrix}),
into the definition (\ref{Xi}) for the
anomalous quark propagator, one obtains
\be
\Xi^+(Q) = J_3 \tau_2 \gamma_5 \, \Lambda_{\bf q}^- \,
\Theta(\Lambda_q - |q-\mu|) \, \frac{\phi(Q)}{[q_0/Z(q_0)]^2 - \epsilon_q^2}\;.
\ee
One now plugs this expression into the gap equation (\ref{Phi}), 
multiplies both sides with $J_3 \tau_2 \gamma_5 \Lambda_{\bf k}^+$,
and traces over color, flavor, and Dirac degrees of freedom. These
traces simplify considerably since both hard 
and HDL gluon propagators are diagonal in adjoint color space,
$[\Delta_{0,22}]^{\mu \nu}_{ab} \equiv \delta_{ab}\, \Delta_{0,22}^{\mu
\nu}$, $[\Delta_{\rm HDL}]^{\mu \nu}_{ab} \equiv \delta_{ab}\, 
\Delta_{\rm HDL}^{\mu \nu}$. The
result is an integral equation for the gap function $\phi(K)$,
\be \label{gapequation}
\phi(K) = \frac{g^2}{3} \, \frac{T}{V} \sum_Q \left[ \frac{}{}
\Delta_{0,22}^{\mu \nu}(K-Q) 
+  \Delta_{\rm HDL}^{\mu \nu}(K-Q)\frac{}{}
\right] \, {\rm Tr}_s \left( \Lambda_{\bf k}^+ \gamma_\mu 
\Lambda_{\bf q}^- \gamma_\nu \right) \, 
\frac{\phi(Q)}{[q_0/Z(q_0)]^2 - \epsilon_q^2}\;.
\ee
The sum over $Q$ runs only over relevant quark momenta, 
$|q- \mu| \leq \Lambda_q$. Also, the 3-momentum ${\bf k}$ is relevant,
$|k-\mu| \leq \Lambda_q$.

\subsection{Solution of the gap equation} \label{IVc}

In pure Coulomb gauge, both the hard gluon and the HDL propagators
have the form
\be \label{propgen}
\Delta^{00}(P) = \Delta^{\ell}(P) \;\;\;\; , \;\;\;\;\;
\Delta^{0i}(P) = 0 \;\;\;\;, \;\;\;\;\;
\Delta^{ij}(P) =  (\delta^{ij} - \hat{p}^i \hat{p}^j) \,
\Delta^{t}(P)\;,
\ee
where $\Delta^{\ell,t}$ are the propagators for
longitudinal and transverse gluon degrees of freedom.
For hard gluons 
\begin{subequations} \label{prop022}
\bea
\Delta^{\ell}_{0,22}(P) & = & -\frac{1}{p^2}\;,  \\
\Delta^{t}_{0,22}(P) & = &  - \frac{1}{P^2}\; ,
\eea
\end{subequations}
while for soft, HDL-resummed gluons
\begin{subequations} \label{propHDL}
\bea 
\Delta^{\ell}_{\rm HDL}(P) & = & - \frac{1}{p^2 - \Pi^{\ell}_{\rm HDL}(P)} 
\; , \\
\Delta^t_{\rm HDL}(P) & = & - \frac{1}{P^2 - \Pi^t_{\rm HDL}(P)} \;,
\eea
\end{subequations}
with the HDL self-energies \cite{LeBellac}
\begin{subequations} \label{HDLselfenergies}
\bea
\Pi^{\ell}_{\rm HDL} (p_0,p) & = & - 3 \, m_g^2 \left[ 1 - \frac{p_0}{2p}\,
\ln \left( \frac{p_0 + p}{p_0-p} \right) \right]\; , \\
\Pi^t_{\rm HDL} (p_0,p) & = & \frac{3}{2}\, m_g^2 \left[
\frac{p_0^2}{p^2}  + \left( 1 - \frac{p_0^2}{p^2} \right)
\, \frac{p_0}{2p} \,\ln \left( \frac{p_0 + p}{p_0-p} \right) \right]\;.
\eea
\end{subequations}
The HDL propagators (\ref{propHDL}) have quasiparticle poles at
$p_0 = \pm \omega_{\ell,t}(p)$, and a cut between
$p_0 = - p$ and $p_0 = p$ \cite{LeBellac}. 
The gluon energy on the quasiparticle mass-shell is always
larger than the gluon mass parameter, $\omega_{\ell,t}(p) \geq m_g$,
where the equality holds for zero momentum, $p=0$.

\begin{figure}[ht]
\includegraphics[width=12cm]{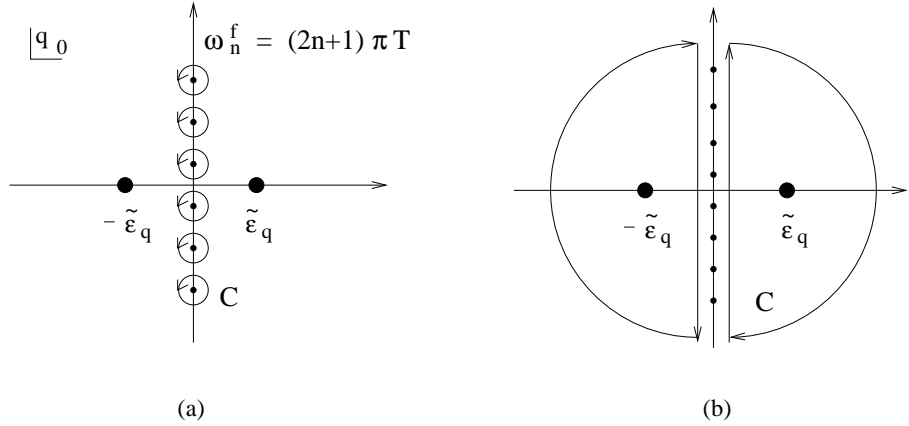}
\caption{(a) The contour ${\cal C}$ in Eq.\ (\ref{genMsum})
encloses the poles of $\tanh [q_0/(2T)]$ on the
imaginary $q_0$ axis. (b) Deforming the contour ${\cal C}$ and
adding semicircles at infinity to enclose the poles
of the quark propagator on the real $q_0$ axis.}
\label{contour1}
\end{figure} 

We first perform the Matsubara sum, using the method of contour
integration in the complex $q_0$ plane \cite{LeBellac,FTFT}, 
\be \label{genMsum}
T \sum_n f(q_0) \equiv \frac{1}{2 \pi i } \oint_{\cal C} dq_0 \, \frac{1}{2} \,
\tanh\left( \frac{q_0}{2T} \right) \, f(q_0)\;
\ee
where the contour ${\cal C}$ consists of circles running around
the poles $\omega_n^{\rm f}= (2 n+1) \pi T $ of 
$\tanh [q_0/((2T)]$ on the imaginary $q_0$ axis, cf.\ Fig.\
\ref{contour1} (a).
Inserting the propagators
(\ref{prop022}) and (\ref{propHDL}) into Eq.\ (\ref{gapequation}), 
we have to compute four distinct
terms. The first one arises from the exchange of static electric hard gluons.
Since $\Delta^{\ell}_{0,22}(P)$ does not depend on $p_0=k_0 - q_0$, 
only the quark propagator gives rise to a pole of $f(q_0)$, cf.\ 
Fig.\ \ref{contour1} (b).
After deforming the contour and closing it at infinity
as shown in Fig.\ \ref{contour1} (b),
one employs the residue theorem to pick up the
poles of the quark propagator, 
\be \label{appEHGE}
T \sum_n \Delta_{0,22}^{\ell}(P) \, 
\frac{\phi(Q)}{[q_0/Z(q_0)]^2 - \epsilon_q^2} = 
\frac{1}{p^2}\, \tanh \left( \frac{\tilde{\epsilon}_q}{2T} \right) \,
\frac{Z^2(\tilde{\epsilon}_q)}{4 \, \tilde{\epsilon}_q} \,
\left[ \phi(\tilde{\epsilon}_q,{\bf q}) + \phi(-\tilde{\epsilon}_q,{\bf
q})\right]\;,
\ee
with $\tilde{\epsilon}_q \equiv \epsilon_q \, Z(\tilde{\epsilon}_q)$.
Here, we have used the fact that
the quark wave-function renormalization factor is an even function
of its argument, $Z(q_0) \equiv Z(-q_0)$, cf.\ Eq.\ (\ref{wavefunc}).
An essential assumption in order to derive Eq.\ (\ref{appEHGE}) 
is that the gap function $\phi(Q)$ is analytic in the complex $q_0$
plane. This assumption will also be made in all subsequent
considerations.

\begin{figure}[ht]
\includegraphics[width=12cm]{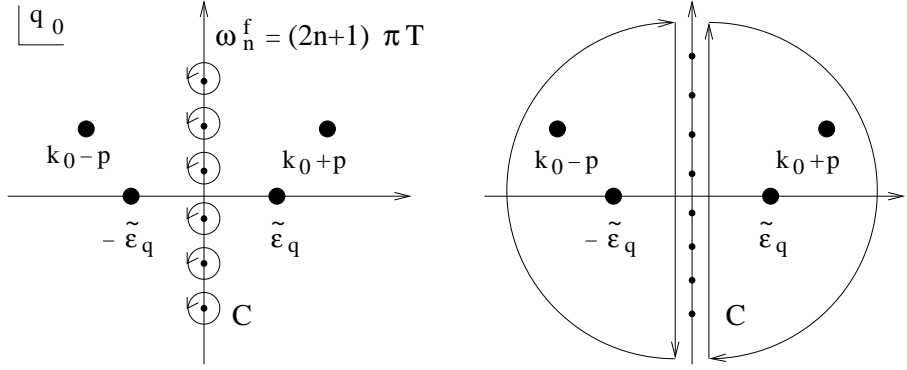}
\caption{Same as in Fig.\ \ref{contour1}, but for magnetic
hard gluon exchange. Now also the gluon propagator has 
poles at $k_0 \pm p$ in the complex $q_0$ plane. 
These are further away from the imaginary axis than the poles
$\tilde{\epsilon}_q$ of the quark propagator, because for
our choice of quark and gluon cut-offs, $\Lambda_q \ll \Lambda_g$,
we have $\tilde{\epsilon}_q \alt \Lambda_q \ll \Lambda_g \leq p$.}
\label{contour2}
\end{figure} 

By the same method one computes the second term in Eq.\
(\ref{gapequation}), corresponding to
magnetic hard gluon exchange. This is slightly more
complicated, since not only the quark propagator but
also $\Delta^{t}_{0,22}(P)$ has poles at $p_0= \pm p$, 
which are located at $q_0 = k_0 \pm p$ in the
complex $q_0$ plane, cf.\ Fig.\ \ref{contour2}. The external
quark energy $k_0$ is fixed and, prior to analytic continuation
$k_0 \rightarrow \tilde{\epsilon}_k + i \eta$
to the quasiparticle mass-shell, is equal to one particular
fermionic Matsubara frequency, cf.\ Fig.\ \ref{contour2}. The residue theorem
now yields four contributions, two from the quark and two from the
gluon poles. Using $\tanh [(k_0 \pm p)/(2T)] \equiv \pm \coth(p/2T)$
and analytically continuing $k_0 \rightarrow \tilde{\epsilon}_k + i
\eta$ we find
\bea
T \sum_n \Delta_{0,22}^{t}(P) \, 
\frac{\phi(Q)}{[q_0/Z(q_0)]^2 - \epsilon_q^2} &  =  &
 \tanh \left( \frac{\tilde{\epsilon}_q}{2T} \right) \,
\frac{Z^2(\tilde{\epsilon}_q)}{4 \, \tilde{\epsilon}_q}\,
\left[\frac{\phi(\tilde{\epsilon}_q,{\bf q})}{(\tilde{\epsilon}_k
- \tilde{\epsilon}_q + i \eta)^2 - p^2}
+ \frac{\phi(-\tilde{\epsilon}_q,{\bf q})}{(\tilde{\epsilon}_k
+ \tilde{\epsilon}_q + i \eta)^2 - p^2}\right]  \non
&   &  \hspace*{-2cm} + \, \coth \left( \frac{p}{2T} \right) \, \frac{1}{4p} \,
\left[ \frac{Z^2(p+\tilde{\epsilon}_k) \,\phi(p+\tilde{\epsilon}_k,{\bf q})}{
(p+\tilde{\epsilon}_k  + i \eta)^2 - \epsilon_q^2
Z^2(p+\tilde{\epsilon}_k)}
+ \frac{Z^2(p- \tilde{\epsilon}_k) \, \phi(\tilde{\epsilon}_k-p,{\bf q})}{
(p-\tilde{\epsilon}_k - i \eta)^2 - \epsilon_q^2
Z^2(p-\tilde{\epsilon}_k)} \right] \;.
\eea
Since the gluon momentum is hard, $p \geq \Lambda_g$, and thus much
larger than the quasiparticle energies $\tilde{\epsilon}_k, \tilde{\epsilon}_q$
which are at most of the order of the quark cut-off $\Lambda_q \ll
\Lambda_g$, to order $O(\Lambda_q/\Lambda_g)$ we may neglect the
terms $(\tilde{\epsilon}_k \pm \tilde{\epsilon}_q + i \eta)^2$ in the
energy denominators of the first term. Furthermore, in the second term
we may approximate $Z(p \pm \tilde{\epsilon}_k) \simeq Z(p) = 1 +
O(g^2)$ and $\phi(p \pm \tilde{\epsilon}_k, {\bf q})
\simeq \phi(p,{\bf q})$. Note that the gap function is far off-shell
for $p \geq \Lambda_g \gg \Lambda_q \geq |\mu -q|$. Then, to order
$O(\Lambda_q/\Lambda_g)$, we may also neglect 
$\tilde{\epsilon}_k, \tilde{\epsilon}_q$ in the energy denominators
of the second term. We obtain
\bea
T \sum_n \Delta_{0,22}^{t}(P) \, 
\frac{\phi(Q)}{[q_0/Z(q_0)]^2 - \epsilon_q^2} & = &
- \frac{1}{p^2} \, \tanh \left( \frac{\tilde{\epsilon}_q}{2T} \right) \,
\frac{Z^2(\tilde{\epsilon}_q)}{4 \, \tilde{\epsilon}_q}\,
\left[ \phi(\tilde{\epsilon}_q,{\bf q}) + \phi(-\tilde{\epsilon}_q,{\bf
q})\right]\, \left[ 1 + O\left( \frac{\Lambda_q^2}{\Lambda_g^2}\right) 
\right] \non
&   &  + \coth \left( \frac{p}{2T} \right) \,
\frac{\phi(p,{\bf q})}{2 \, p^3}\,
\left[ 1 + O\left( \frac{\Lambda_q^2}{\Lambda_g^2} \right)\right]\,
\;. \label{appMHGE}
\eea
Let us estimate to which order the two remaining terms contribute to
the gap equation (\ref{gapequation}).
At $T=0$, we may set the hyperbolic functions to one. 
We shall also ignore the difference
between the on-shell and off-shell gap functions, and take
$\phi(p,{\bf q}) \simeq \phi(\pm \tilde{\epsilon}_q,{\bf q}) \equiv \phi
 = const.$. For the purpose of power counting, we may restrict
ourselves to the leading contribution of the Dirac traces in Eq.\
(\ref{gapequation}), which is of order one, cf.\ Eqs.\ (\ref{traces})
below. In order to obtain the leading contribution of the first term in
Eq.\ (\ref{appMHGE}), we may also set $Z^2(\tilde{\epsilon}_q) \simeq
1$. The integral over the absolute magnitude of the quark
momentum is $\int dq \, q^2$, while the angular integration is 
$\int d \cos \theta \equiv \int d p \, p / (kq)$.
Then, the first term in Eq.\ (\ref{appMHGE}) leads to the
following contribution in the gap equation
\be
 g^2\, \frac{\phi}{k} \int_{\mu-\Lambda_q}^{\mu+ \Lambda_q} 
dq \, \frac{q}{\epsilon_q} \int_{\Lambda_g}^{k+q} \frac{dp}{p}
\simeq g^2 \, \phi \, \ln \left( \frac{2\Lambda_q}{\phi} 
\right) \, \ln \left( \frac{ 2\mu}{\Lambda_g} \right) 
\sim g^2 \, \phi \, \frac{1}{g} = g\, \phi\;,
\ee
where we approximated $k \simeq q \simeq \mu$ and
employed the weak-coupling solution (\ref{gapsol})
to estimate $\ln ( 2 \Lambda_q/\phi ) \sim 1/g$. Furthermore, for
$\Lambda_g \alt \mu$, the angular logarithm is 
$\ln( 2\mu/\Lambda_g) \sim O(1)$. According to the discussion
presented in the introduction, the contribution
from hard magnetic gluon exchange is thus of subleading order 
in the gap equation. Note that the term arising from hard electric
gluon exchange, Eq.\ (\ref{appEHGE}), is of the same order
as the first term in Eq.\ (\ref{appMHGE}), and thus also contributes
to subleading order. The way we estimated the first term
on the right-hand side of Eq.\ (\ref{appMHGE}) is equivalent to just
taking the hard magnetic gluon propagator in the static limit,
$\Delta^t_{0,22}(P) \simeq  1/p^2$, which is correct
up to terms of order $O(\Lambda_q^2/\Lambda_g^2)$. To this order,
the propagator for hard magnetic gluons is thus (up to a sign)
identical to the one
for hard electric gluons. Since the ratio $\Lambda_q/\Lambda_g
\simeq g \mu/ \mu \equiv g$, this approximation introduces corrections
at order $O(g^3 \phi)$ in the gap equation, which is {\em beyond\/}
sub-subleading order, $O(g^2 \phi)$. 

Similarly, we estimate the contribution of the second term
in Eq.\ (\ref{appMHGE}) to the gap equation (\ref{gapequation}),
\be
 g^2\, \frac{\phi}{k} \int_{\mu-\Lambda_q}^{\mu+ \Lambda_q} 
dq \, q \int_{\Lambda_g}^{k+q} \frac{dp}{p^2}
\sim g^2 \, \phi \,  \frac{\Lambda_q}{\Lambda_g} 
\sim g^3 \phi\;,
\ee
i.e., for our choice $\Lambda_q/\Lambda_g \sim g$,
this term contributes beyond sub-subleading order.
Note that this estimate is conservative, as we assumed the
off-shell gap function to be of the same
order as the gap at the Fermi surface, $\phi(p,{\bf q}) \sim \phi$.
However, we know \cite{rdpdhr} that, for energies far from the
Fermi surface, $\tilde{\epsilon}_q \sim \Lambda_q \alt g \mu$,
even the on-shell gap function
is suppressed by one power of $g$ compared to the value
of the gap at the Fermi surface, $\phi(\Lambda_q,{\bf q}) \sim g \phi$.
The off-shell gap function at $q_0 = p \agt \Lambda_g \gg \Lambda_q$ 
may be even smaller. In order to decide this issue, one would have
to perform a computation of the gap function for arbitrary
values of the energy $q_0$, and not just on the
quasiparticle mass-shell, $q_0 \equiv \tilde{\epsilon}_q$.
We note that for the choice $\Lambda_q \simeq \Lambda_g$ for the cut-offs 
\cite{hong2,schaferefftheory}, the ratio
$\Lambda_q/\Lambda_g$ is of order one and cannot be used as a 
parameter to sort the various contributions according to
their order of magnitude. The expansion of the denominators
in powers of $\Lambda_q/\Lambda_g$ as seen on the right-hand side
of Eq.\ (\ref{appMHGE}) is then inapplicable. 

\begin{figure}[ht]
\includegraphics[width=12cm]{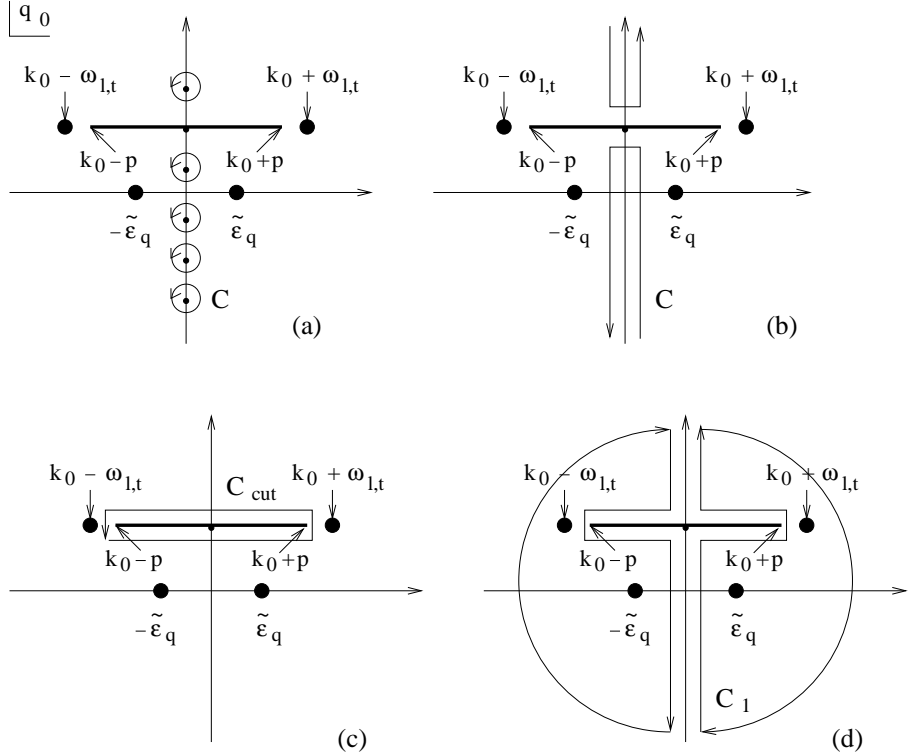}
\caption{Evaluating the Matsubara sum for HDL-resummed gluon
propagators. (a) The original contour ${\cal C}$ in Eq.\
(\ref{genMsum}). There is no circle around the point 
$k_0 = q_0$, where the corresponding term in the Matsubara sum 
has a cut arising from the HDL gluon propagator.
(b) Deforming the contour ${\cal C}$.
(c) The contour ${\cal C}_{\rm cut}$ running
around the cut. (d) The contour ${\cal C}_1 = {\cal C} + {\cal C}_{\rm
cut}$ which is closed at infinity.}
\label{contour3}
\end{figure} 

The third and fourth terms in the gap equation (\ref{gapequation})
arise from soft, HDL-resummed electric and magnetic gluon exchange.
Evaluating the Matsubara sum via contour integration in the complex
$q_0$ plane is considerably more difficult than in the
previous cases, because the HDL gluon propagators
$\Delta^{\ell,t}_{\rm HDL}$ do not only have poles but also cuts.
The analytic structure is shown in Fig.\ \ref{contour3} (a).
Besides the poles of the quark propagator at $q_0 = \pm \tilde{\epsilon}_q$,
there are also those from the gluon propagator at 
$q_0 = k_0 \pm \omega_{\ell,t}(p)$. The cut of the gluon propagator
between $- p \leq p_0 \leq p$ translates into a cut
between $k_0 -p \leq q_0 \leq k_0 + p$. Prior to analytic
continuation, the gluon poles and the cut are shifted away from
the real axis and located at the 
(imaginary) external Matsubara frequency $k_0$.

The Matsubara sum over $q_0$ is evaluated in the standard way, 
cf.\ Eq.\ (\ref{genMsum}), with
the caveat that the contribution at $q_0 = k_0$,
where the cut of the gluon propagator is located, has to be omitted. 
This is similar to the zero-temperature case where the Matsubara sum
becomes a continuous integral along the imaginary $q_0$ axis and
where one has to avoid integrating over the cut.
Alternatively, the term $q_0 = k_0$ can be included in the
Matsubara sum if one shifts the cut by some small amount
$\pm i \epsilon$ along the imaginary $q_0$ axis.
The final result will be the same, as one still has to circumvent
the cut by a proper choice of the integration contour.

We now deform the contour as shown in Fig.\ \ref{contour3} (b), and
add and subtract a contour integral running around the cut, Fig.\
\ref{contour3} (c). The integral over the contour ${\cal C} + {\cal
C}_{\rm cut}$ can be closed at infinity, yielding the contour
${\cal C}_1$ shown in Fig.\ \ref{contour3} (d). One obtains 
\be \label{gapequation2}
T \sum_n \Delta_{\rm HDL}^{\ell, t}(P) \, 
\frac{\phi(Q)}{[q_0/Z(q_0)]^2 - \epsilon_q^2} = 
\frac{1}{2 \pi i} \left[ \oint_{{\cal C}_1} - \oint_{{\cal C}_{\rm
cut}} \right] d q_0\, \frac{1}{2} \,
\tanh\left( \frac{q_0}{2T} \right) \, \Delta_{\rm HDL}^{\ell,t}(P)\,
\frac{\phi(Q)}{[q_0/Z(q_0)]^2 - \epsilon_q^2}\;.
\ee
Evaluating the integral over ${\cal C}_1$ is rather similar
to the case of hard gluon exchange: one just picks up the
poles of the quark and gluon propagators inside the contour ${\cal
C}_1$. 
After analytic continuation $k_0 \rightarrow \tilde{\epsilon}_k + i \eta$
one obtains
\bea
\lefteqn{\frac{1}{2 \pi i} \oint_{{\cal C}_1} d q_0\, \frac{1}{2} \,
\tanh\left( \frac{q_0}{2T} \right) \, \Delta_{\rm HDL}^{\ell,t}(P)\,
\frac{\phi(Q)}{[q_0/Z(q_0)]^2 - \epsilon_q^2} \simeq } \non
& \simeq & -  \tanh \left( \frac{\tilde{\epsilon}_q}{2T} \right) \,
\frac{Z^2(\tilde{\epsilon}_q)}{4 \, \tilde{\epsilon}_q}\,
\left[ \Delta_{\rm HDL}^{\ell,t}(\tilde{\epsilon}_k - \tilde{\epsilon}_q
+ i \eta, {\bf p})\, \phi(\tilde{\epsilon}_q,{\bf q}) 
+ \Delta_{\rm HDL}^{\ell,t}(\tilde{\epsilon}_k + \tilde{\epsilon}_q
+ i \eta, {\bf p})\, \phi(-\tilde{\epsilon}_q,{\bf q})\right] \non
&   &  + \coth \left( \frac{\omega_{\ell,t}}{2T} \right) \,
\frac{1}{2  \, \omega_{\ell,t}^2}\,
\left[ \phi(\omega_{\ell,t} + \tilde{\epsilon}_k,{\bf q})\,
Z_{\ell,t}(- \omega_{\ell,t},p)
- \phi(\tilde{\epsilon}_k-\omega_{\ell,t} ,{\bf q})\,
Z_{\ell,t}( \omega_{\ell,t},p) \right]\,
\left[ 1 + O\left( \frac{\epsilon_q^2}{\omega_{\ell,t}^2} 
\right)\right]
\;. \label{C_1}
\eea
Here, we approximated the quark wave-function renormalization
factor $Z(\omega_{\ell,t}\pm \tilde{\epsilon}_k) \simeq 1 + O(g^2)$.
We also expanded the denominators of the quark propagator 
$(\tilde{\epsilon}_k \pm \omega_{\ell,t} + i \eta)^2 - \epsilon_q^2
\simeq \omega_{\ell,t}^2 \, [ 1 +
O(\epsilon_q^2/\omega_{\ell,t}^2)]$.
For our choice of the cut-off $\Lambda_q \alt g \mu \sim m_g$, we may estimate
$\omega_{\ell,t} \geq m_g \agt \Lambda_q \geq \epsilon_q$, i.e.,
the corrections of order 
$O(\epsilon_q^2/\omega_{\ell,t}^2)$ are small everywhere
except for a small region of phase space where $p \simeq 0$ and
$\epsilon_q \simeq \Lambda_q$. 
(In principle, in the expansion of the denominators 
there are also linear terms, $\sim \pm \tilde{\epsilon}_k/
\omega_{\ell,t}$, but these are very small everywhere for
external momenta close to the Fermi surface, $k \simeq \mu$.)
Note that the gap function
is again off-shell at the gluon pole, although not as far as
in the case of hard gluon exchange, cf.\ Eq.\ (\ref{appMHGE}).
The residues of the HDL gluon propagators at the respective poles
are \cite{LeBellac}
\begin{subequations}
\bea
Z_{\ell}(\omega_\ell,p) & =& - \frac{\omega_\ell (\omega_\ell^2 - p^2)}{
p^2 (p^2 + 3 m_g^2 - \omega_\ell^2)} \; , \\
Z_t (\omega_t,p) & = & -\frac{\omega_t (\omega_t^2 - p^2)}{
3 m_g^2 \omega_t^2 - (\omega_t^2 - p^2)^2}\;.
\eea
\end{subequations}
To very good approximation, one finds that
$Z_t(\omega_t,p) \simeq - 1/(2 \omega_t)$ for all momenta $p$. 
In the longitudinal case, the residue is very well approximated
by $Z_l(\omega_l,p) \simeq - \omega_l/(2\, p^2)$ for small momenta
$p \alt m_g$, while for large momenta, $m_g \ll
p$, $Z_l(\omega_l,p) \sim  \exp [-2 p^2/(3 m_g^2)]/p$, 
i.e., it is exponentially suppressed \cite{RDPphysicaA}.

These approximate forms allow for a simple power counting of the
gluon-pole contribution in Eq.\ (\ref{C_1}) to the gap equation 
(\ref{gapequation}). To this end, we approximate the gap function
by its value at the Fermi surface, 
$\phi(\pm \omega_{\ell,t} + \tilde{\epsilon}_k, {\bf
q}) \simeq \phi$, and consider the limiting case $T=0$ where
$\coth[\omega_{\ell,t}/(2T)] = 1$.
Then, the contribution from the longitudinal gluon pole is 
\be \label{appESGE}
 g^2\, \frac{\phi}{k} \int_{\mu-\Lambda_q}^{\mu+ \Lambda_q} 
dq \, q \left[ \int_{|k-q|}^{m_g} 
\frac{dp}{2 \, p\, \omega_\ell} + \int_{m_g}^{\Lambda_g}
\frac{dp}{ \omega_\ell^2}\, 
\exp \left(-\frac{2 p^2}{3 m_g^2}\right) \right]
\sim g^2 \, \phi \, \frac{\Lambda_q}{m_g} 
\sim g^2 \phi \;.
\ee
In the first $p$ integral,
which only runs up to the scale $m_g$,
one may approximate $\omega_\ell \simeq m_g$, while
in the second $p$ integral, which runs from $m_g$ to
$\Lambda_g \alt \mu$, one may take $\omega_\ell \simeq p$.
To obtain the right-hand side of Eq.\ (\ref{appESGE}) we have 
set $k \simeq q \simeq \mu$, and we have employed
our choice $\Lambda_q \alt g \mu$ for the quark cut-off. This also
allowed us to approximate logarithms of $\Lambda_q/m_g$ by numbers
of order $O(1)$. With this choice for the quark cut-off,
the contribution (\ref{appESGE}) is of
sub-subleading order, $\sim O(g^2 \phi)$, to the gap
equation. 

With a more careful evaluation of 
the integrals, one could extract the precise numerical prefactor
of the sub-subleading contribution (\ref{appESGE}). 
Note, however, that further suppression factors
may arise from the off-shellness of the gap function at 
$\phi(\pm \omega_{\ell,t} + \tilde{\epsilon}_k, {\bf q})$,
which consequently would render this
contribution beyond sub-subleading order. As noted previously,
this issue can only be decided if $\phi(q_0,{\bf q})$ is known
also off the quasiparticle mass-shell, and not only on-shell.
We also note that the $1/p^2$
factor in the residue $Z_\ell$ is an artifact of the Coulomb gauge
\cite{RDPphysicaA}, and does not appear in e.g.\ covariant gauge.
One would have to collect all other terms
of sub-subleading order to make sure that the complete 
sub-subleading contribution is gauge invariant and the term
(\ref{appESGE}) not cancelled by some other terms.

Similarly, we estimate the contribution from the transverse gluon pole,
\be \label{appMSGE}
 g^2\, \frac{\phi}{k} \int_{\mu-\Lambda_q}^{\mu+ \Lambda_q} 
dq \, q \int_{|k-q|}^{\Lambda_g} \frac{dp\, p}{2 \, \omega_t^3}
\sim g^2 \, \phi \int_0^{\Lambda_q} d \xi\,
\int_{m_g}^{\Lambda_g} \frac{d \omega_t}{\omega_t^2}
\sim g^2 \, \phi \, \frac{\Lambda_q}{m_g} \sim g^2 \, \phi\;,
\ee
where we defined $\xi \equiv q - \mu$. We
approximated $dp \, p \simeq d \omega_t \, \omega_t$ since,
for the purpose of power counting, to very good approximation one
may take the dispersion relation of the transverse gluon 
equal to that of a relativistic particle with mass
$m_g$, $\omega_t(p) \simeq (p^2 + m_g^2)^{1/2}$. 
We also used  $\Lambda_q \alt m_g \ll \Lambda_g $ and $k \simeq q \simeq \mu$.
In conclusion, also the
transverse gluon pole possibly contributes to sub-subleading order in the
gap equation, with the same caveats concerning the off-shellness of
the gap function as mentioned previously. 

Let us now focus on the integral around the cut of the gluon
propagator in Eq.\ (\ref{gapequation2}). 
We substitute $q_0$ by $p_0 = k_0 - q_0 \equiv \omega$ and
use the fact that $\tanh[q_0/(2T)] \equiv - \coth [\omega/(2T)]$.
Since the gluon propagator is the only part of the integrand 
which is discontinuous across the cut, we obtain after analytic continuation 
$k_0 \rightarrow \tilde{\epsilon}_k + i \eta$
\bea
\lefteqn{- \frac{1}{2 \pi i} \oint_{{\cal C}_{\rm cut}} d q_0\, \frac{1}{2} 
\tanh\left( \frac{q_0}{2T} \right)  \Delta_{\rm HDL}^{\ell,t}(P)\,
\frac{\phi(Q)}{[q_0/Z(q_0)]^2 - \epsilon_q^2}} \non
 & = & 
\int_{-p}^p d\omega\, \frac{1}{2} \coth \left( \frac{\omega}{2T}
\right) \, \frac{Z^2(\tilde{\epsilon}_k - \omega)\, 
\phi(\tilde{\epsilon}_k - \omega,{\bf q})}{(\tilde{\epsilon}_k -
\omega + i \eta)^2 - [Z(\tilde{\epsilon}_k - \omega)\, \epsilon_q]^2} \, 
\rho^{\ell,t}_{\rm cut}(\omega,{\bf p})\;,
\label{C_cut}
\eea
where $\rho^{\ell,t}_{\rm cut}(\omega,p) \equiv {\rm Im} \Delta^{\ell,t}_{\rm
HDL} (\omega+i \eta,p)/ \pi$ is the spectral density of the
HDL propagator arising from the cut. Explicitly,
\begin{subequations}
\bea
\rho^{\ell}_{\rm cut}(\omega,{\bf p}) & = & 
\frac{2 M^2}{\pi}\, \frac{\omega}{p}\,
\left\{ \left[ p^2 + 3\, m_g^2 \left(
1 - \frac{\omega}{2p}\, \ln
\left| \frac{ p+ \omega}{p-\omega} \right| \right) \right]^2
+ \left( 2 M^2\, \frac{\omega}{p} \right)^2
\right\}^{-1} \;, \\
\rho^t_{\rm cut} (\omega,{\bf p}) & = & 
\frac{M^2}{\pi}\, \frac{\omega}{p}\,
\frac{p^2}{p^2-\omega^2} \, 
\left\{ \left[ p^2 + \frac{3}{2}\, m_g^2 \left(
\frac{\omega^2}{p^2-\omega^2} + \frac{\omega}{2p}\, \ln
\left| \frac{ p+ \omega}{p-\omega} \right| \right) \right]^2
+ \left( M^2\, \frac{\omega}{p} \right)^2
\right\}^{-1}
\;. \label{rhotcut}
\eea
\end{subequations}
In order to power count the contribution from the cut of
$\Delta^{\ell}_{\rm HDL}$ to the gap equation, it is sufficient
to approximate the spectral density by \cite{rdpdhr}
\be
\rho^{\ell}_{\rm cut}(\omega,{\bf p}) \simeq
\frac{2 M^2}{\pi}\, \frac{\omega}{p}\,\frac{1}{
( p^2 + 3\, m_g^2 )^2} \;.
\ee
This form reproduces the correct behavior for $\omega \ll p$.
For $\omega \alt p$, it overestimates the spectral density when
$p\alt m_g$, while it slightly underestimates 
it for $p \agt m_g$. For the gap equation, however,
this region is unimportant, since the respective
contribution is suppressed by the
large energy denominator $(\tilde{\epsilon}_k -
\omega + i \eta)^2 - [Z(\tilde{\epsilon}_k - \omega)\, \epsilon_q]^2
\simeq p^2$ in Eq.\ (\ref{C_cut}). To leading order,
we may set $Z(\tilde{\epsilon}_k - \omega) \simeq 1$. We also
approximate $\phi(\tilde{\epsilon}_k - \omega,{\bf
q}) \simeq \phi$. 
Then, the $\omega$ integral can be performed analytically. (One may compute
this integral with the principal value prescription; the
contribution from the complex pole contributes to the imaginary part
of the gap function, which we neglect throughout this computation.)
This produces at most logarithmic singularities, which
are integrable. We therefore simply approximate
the $\omega$ integral by a number of order $O(1)$.
Consequently, the contribution from Eq.\ (\ref{C_cut}) to
the gap equation is of order
\be \label{appESGEcut}
g^2 \, \frac{\phi}{k} \int_{\mu-\Lambda_q}^{\mu+ \Lambda_q} 
dq \, q \int_{|k-q|}^{\Lambda_g} dp\, \frac{m_g^2}{(p^2 + 3\, m_g^2)^2}
\sim g^2 \, \phi \int_0^{\Lambda_q} d \xi \left( \int_\xi^{m_g}
\frac{dp}{m_g^2} +  m_g^2 \int_{m_g}^{\Lambda_g}
\frac{dp}{p^4} \right) \sim g^2 \, \phi\, \frac{\Lambda_q}{m_g}
\sim g^2 \, \phi\;,
\ee
where we approximated the $p$ integral by a method similar 
to the one employed in Eq.\ (\ref{appESGE}). For our choice
$\Lambda_q \alt g \mu$, Eq.\ (\ref{appESGEcut}) constitutes another 
(potential) contribution of sub-subleading order to the gap equation.

Finally, we estimate the contribution from the cut of the transverse
gluon propagator. For all momenta $p$ and energies $-p \leq \omega
\leq p$, a very good approximation for the spectral density
(\ref{rhotcut}) is given by the formula
\be
\rho_{\rm cut}^t (\omega, {\bf p}) \simeq \frac{M^2}{\pi} \,
\frac{\omega \, p}{p^6 + (M^2\, \omega)^2}\;.
\ee
This approximate result constitutes an upper bound for
the full result (\ref{rhotcut}). The advantage of using this
approximate form is that, interchanging the order of the
$p$ and $\omega$ integration in the gap equation, 
the former may immediately be performed.
Approximating $Z(\tilde{\epsilon}_k - \omega) \simeq 1$,
neglecting the dependence of the gap function on the
direction of ${\bf q}$, and defining $\lambda \equiv {\rm max}(
|k-q|, \omega)$, at $T=0$ the contribution to the gap equation is
\bea \label{appMSGEcut}
\lefteqn{g^2 \int_{\mu-\Lambda_q}^{\mu + \Lambda_q}
d q \, \frac{q}{k} \int_0^{\Lambda_g} d \omega
\, \left( \frac{\phi(\tilde{\epsilon}_k- \omega,q)}{
(\tilde{\epsilon}_k -\omega)^2 - \epsilon_q^2} 
+ \frac{\phi(\tilde{\epsilon}_k+ \omega,q)}{
(\tilde{\epsilon}_k +\omega)^2 - \epsilon_q^2} \right)
\left[ {\rm arctan} \left( \frac{\Lambda_g^3}{\omega M^2} \right)
- {\rm arctan} \left(\frac{\lambda^3}{\omega M^2} \right) \right]} \\
& \sim & g^2 \int_0^{\Lambda_q} \frac{d \xi }{\epsilon_q}
\int_0^M d \omega \left[ \phi(\tilde{\epsilon}_k- \omega,q)
\left( \frac{1}{\tilde{\epsilon}_k - \omega -  \epsilon_q} -
\frac{1}{\tilde{\epsilon}_k - \omega +  \epsilon_q} \right)
+ \phi(\tilde{\epsilon}_k +\omega,q)
\left( \frac{1}{\tilde{\epsilon}_k + \omega -  \epsilon_q} -
\frac{1}{\tilde{\epsilon}_k + \omega +  \epsilon_q} \right)\right]\;.
\nonumber
\eea
Here, we have used the fact that the particular combination
of arctan's in the first line effectively cuts off the $\omega$
integral at the scale $\omega \sim M$. As usual, we have 
set $k \simeq q \simeq \mu$.
If we simply neglect the off-shell behavior of the gap function
and approximate $\phi(\tilde{\epsilon}_k \pm \omega,q) \simeq
\phi$, this contribution would (at least) be of subleading order.
Note that the corresponding contribution in previous treatments of the
QCD gap equation, cf.\ for instance Eq.\ (67) of Ref.\ \cite{rdpdhr},
was discarded as being of higher order. 
At this point, we refrain from a more careful evaluation
of the contribution (\ref{appMSGEcut}), because this 
requires a calculation of the gap function off the quasiparticle
mass-shell. Since the purpose of the present work is to
show that our method reproduces previous results, we follow
Ref.\ \cite{rdpdhr} and also
discard the contribution (\ref{appMSGEcut}) in the following.

The remaining term from the evaluation of the Matsubara sum
in Eq.\ (\ref{gapequation2}) is the contribution from
the quark pole, i.e., the first line of Eq.\ (\ref{C_1}).
This has to be combined with the subleading-order terms 
from hard-gluon exchange, i.e., from Eq.\ (\ref{appEHGE}) and
from the first line of Eq.\ (\ref{appMHGE}), in order to
obtain the gap equation which contains all contributions of
leading and subleading order. Before doing so, however, we
also evaluate the Dirac traces in Eq.\ (\ref{gapequation}).
In pure Coulomb gauge, we only require
\begin{subequations} \label{traces}
\bea
{\rm Tr}_s \left( \Lambda_{\bf k}^+ \gamma_0 
\Lambda_{\bf q}^- \gamma_0 \right) & = & \frac{(k+q)^2 - p^2}{2\, k\,q} \;
, \\
(\delta^{ij} - \hat{p}^i \hat{p}^j)\, 
{\rm Tr}_s \left( \Lambda_{\bf k}^+ \gamma_i \Lambda_{\bf q}^-
\gamma_j \right) & = & -2 - \frac{p^2}{2\, k\,q} + 
\frac{(k^2-q^2)^2}{2\, k\, q\, p^2}\;,
\eea
\end{subequations}
where we used $p^2 \equiv ({\bf k}-{\bf q})^2 = k^2 + q^2 - 2\, k\, q \,
\hat{\bf k} \cdot \hat{\bf q}$ to eliminate $\hat{\bf k} \cdot
\hat{\bf q}$ in favor of $p^2$. Let us estimate the order of magnitude
of the terms arising from the traces at the Fermi surface, $k \equiv
\mu$. Setting $q \equiv \mu + \xi$, where 
$-\Lambda_q \leq \xi \leq \Lambda_q$, one obtains
\begin{subequations} \label{traces2}
\bea
{\rm Tr}_s \left( \Lambda_{\bf k}^+ \gamma_0 
\Lambda_{\bf q}^- \gamma_0 \right) & = & 
2 - \frac{p^2}{2\, k \, q} + O\left( \frac{\xi^2}{\mu^2} \right) \;, 
\label{traces2a} \\
(\delta^{ij} - \hat{p}^i \hat{p}^j)\, 
{\rm Tr}_s \left( \Lambda_{\bf k}^+ \gamma_i \Lambda_{\bf q}^-
\gamma_j \right) & = & -2 - \frac{p^2}{2\, k \, q} 
+ O\left( \frac{\xi^2}{\mu^2} \right)  \;. \label{traces2b}
\eea
\end{subequations}
As shown above, the contribution from hard-gluon exchange is at most
of subleading order. Thus, for this contribution
it is sufficient to keep only the leading terms in Eq.\
(\ref{traces2}), i.e., one may safely neglect terms
of order $O(\xi^2/\mu^2) \alt O(\Lambda_q^2/\Lambda_g^2) \sim O(g^2)$ 
or higher.
Note that, since for hard gluon exchange $p \sim \mu \agt \Lambda_g$,
the terms $p^2/(2 k q)$ cannot be omitted. However, since
the magnetic gluon propagator is effectively $\sim 1/p^2$, cf.\
Eq.\ (\ref{appMHGE}), i.e., (up to a sign) identical to the electric
propagator, these terms will ultimately cancel
between the electric and the magnetic contribution.
This cancellation is well-known, see for instance Ref.\ \cite{asqwdhr}, and 
is special to the spin-zero case. It does not occur in spin-one
color superconductors where there is an additional exponential
prefactor which suppresses the magnitude of the spin-one gap relative
to the spin-zero case \cite{asqwdhr}.

As is well-known, electric soft-gluon exchange also
contributes to subleading order in the gap equation.
Thus, as in the case of hard-gluon exchange, we may drop the terms
of order $O(\xi^2/\mu^2)$ in Eq.\ (\ref{traces2a}).
On the other hand, magnetic soft-gluon exchange constitutes the
leading order contribution to the gap equation. We therefore would have to
keep all terms up to {\em sub\/}leading order, i.e., $\sim
O(\xi/\mu)$. Fortunately, the corrections to the result
(\ref{traces2b}) are of order $O(\xi^2/\mu^2) \sim O(g^2)$, i.e., they
are of {\em sub-sub\/}leading order and thus can also be omitted.

We combine Eqs.\ (\ref{appEHGE}), (\ref{appMHGE}), and the first
line of Eq.\ (\ref{C_1}), and assume that the gap function is
even in its energy argument, $\phi(-\tilde{\epsilon}_q, {\bf q})
= \phi(\tilde{\epsilon}_q, {\bf q})$, and isotropic,
$\phi(\tilde{\epsilon}_q, {\bf q}) \equiv \phi(\tilde{\epsilon}_q,q)
\equiv \phi_q$.
Then, on the quasiparticle mass-shell
$k_0 = \tilde{\epsilon}_k$ the gap equation (\ref{gapequation})
becomes
\bea
\phi_k & = & \frac{g^2}{24 \pi^2} \, 
\int_{\mu - \Lambda_q}^{\mu + \Lambda_q}  dq\, 
\frac{q}{k}\,\frac{Z^2(\tilde{\epsilon}_q)}{\tilde{\epsilon}_q}\,
\tanh \left( \frac{\tilde{\epsilon}_q}{2T} \right) \, \phi_q
\int_{|k-q|}^{k+q} dp\, p \, 
\left\{ \Theta(p - \Lambda_g) \, \frac{4}{p^2} + 
\Theta(\Lambda_g - p)\right. \non
&   & \times \left. \sum_{s = \pm} \left[
 \Delta_{\rm HDL}^{\ell}(\tilde{\epsilon}_k - s \tilde{\epsilon}_q
+ i \eta, p) \, \left( - 1 +  \frac{p^2}{4\,k\,q} \right) 
+  \Delta_{\rm HDL}^{t}(\tilde{\epsilon}_k - s \tilde{\epsilon}_q
+ i \eta, p) \, \left(  1 +  \frac{p^2}{4\,k\,q} \right) \right]
\right\}\;.
\label{gapequation3}
\eea
The next step is to divide the integration region in the $p-q$ plane
into two parts, separated by the gluon ``light cone'' 
$|\tilde{\epsilon}_k - s \tilde{\epsilon}_q| = p$. For our choice
$\Lambda_q \ll \Lambda_g$ the region,
where $|\tilde{\epsilon}_k - s \tilde{\epsilon}_q| < p$, is very large,
while its complement is rather small.
In order to estimate the contribution from the latter 
to the gap equation,  we may approximate the HDL gluon propagators by their
limiting forms for large gluon energies, cf.\ Eqs.\ (\ref{propHDL}), 
(\ref{HDLselfenergies}),
\be
p_0 \gg p \; : \;\;\;\;\;\;\;
\Delta^\ell_{\rm HDL} (P) \simeq \frac{p_0^2}{m_g^2\,p^2} \;\;\;\; ,
\;\;\;\;\;
\Delta^t_{\rm HDL} (P) \simeq \frac{1}{m_g^2} \;.
\ee
Following the power-counting scheme employed previously, 
the contribution from the electric sector is of order
\be
g^2 \, \frac{\phi}{k} \int_{\mu-\Lambda_q}^{\mu+\Lambda_q}
dq\, \frac{q}{\epsilon_q} \, 
\frac{(\tilde{\epsilon}_k - s \tilde{\epsilon}_q)^2}{m_g^2}
\int_{|k-q|}^{|\tilde{\epsilon}_k - s \tilde{\epsilon}_q|}
\frac{dp}{p}
\sim g^2 \, \frac{\phi}{m_g^2} \int_0^{\Lambda_q} d \xi \, \epsilon_q
\sim g^2 \, \phi\, \frac{\Lambda_q^2}{m_g^2} \sim g^2 \, \phi\;.
\ee
This is a contribution of sub-subleading order, as long as one
adheres to the choice $\Lambda_q \alt g \mu$.
Analogously, we estimate the contribution from the magnetic
sector to be
\be
g^2 \, \frac{\phi}{k} \int_{\mu-\Lambda_q}^{\mu+\Lambda_q}
dq\, \frac{q}{\epsilon_q} \, 
\int_{|k-q|}^{|\tilde{\epsilon}_k - s \tilde{\epsilon}_q|}
dp \,p \, \frac{1}{m_g^2}
\sim g^2 \, \frac{\phi}{m_g^2} \int_0^{\Lambda_q} \frac{d
\xi}{\epsilon_q} \, \xi^2
\sim g^2 \, \phi\, \frac{\Lambda_q^2}{m_g^2} \sim g^2 \, \phi\;.
\ee
Consequently, all contributions from the region
$|\tilde{\epsilon}_k - s \tilde{\epsilon}_q| \geq p$ are of
sub-subleading order, and the further analysis can be restricted
to the region $|\tilde{\epsilon}_k - s \tilde{\epsilon}_q| < p$.
In this region, it is permissible to use the low-energy limit 
of the HDL gluon propagator, which follows from Eqs.\ (\ref{propHDL}),
(\ref{HDLselfenergies})
keeping only the leading terms in the gluon energy,
\be \label{lel}
p_0 \ll p \; : \;\;\;\;\;\;\;
\Delta_{\rm HDL}^{\ell} (P) 
 \simeq  - \frac{1}{p^2 + 3\, m_g^2} \;\;\;\;  , \;\;\;\;\;
\Delta_{\rm HDL}^t (P)  \simeq  
\frac{p^4}{p^6 + M^4 \,p_0^2} \; .
\ee
Here, we only retained the real part of the transverse gluon propagator,
since the imaginary part contributes to
the imaginary part of the gap function, which is usually ignored.
(In Ref.\ \cite{rdpdhr} it was argued that, at least close
to the Fermi surface, the contribution of the
imaginary part is of sub-subleading order in the gap equation.)
With the approximation (\ref{lel}), the gap equation 
(\ref{gapequation3}) becomes
\bea
\phi_k & = & \frac{g^2}{24 \pi^2} \, 
\int_{\mu - \Lambda_q}^{\mu + \Lambda_q}  dq\, 
\frac{q}{k}\,\frac{Z^2(\tilde{\epsilon}_q)}{\tilde{\epsilon}_q}\,
\tanh \left( \frac{\tilde{\epsilon}_q}{2T} \right) \, \phi_q
\left\{  4 \, \ln \left( \frac{k+q}{\Lambda_g}\right) \right. \non
&   & + \left.
\sum_{s=\pm}
\int_{|\tilde{\epsilon}_k - s \tilde{\epsilon}_q|}^{\Lambda_g} 
dp\,\left[ \frac{p}{p^2 + 3\, m_g^2} \, 
\left(  1 -  \frac{p^2}{4\,k\,q} \right) 
+ \frac{p^5}{p^6 + M^4 (\tilde{\epsilon}_k - s \tilde{\epsilon}_q)^2}
 \, \left(  1 +  \frac{p^2}{4\,k\,q} \right) \right] \right\}
\;,
\label{gapequation4}
\eea
where we already performed the integration over hard gluon momenta 
$p\geq \Lambda_g$. The integration over soft gluon momenta can 
also be performed analytically. Formally, the terms $\sim p^2/(4kq)$ 
give rise to subleading-order contributions, $\sim \Lambda_g^2/(8kq)$,
but they ultimately cancel, since they come with different signs in
the electric and the magnetic part. Other contributions from these
terms are at most of sub-subleading order. 
Exploiting the symmetry of the integrand around
the Fermi surface and setting $k \simeq \mu$, we arrive at
\be
\phi_k = \frac{g^2}{12 \pi^2} \int_{0}^{\Lambda_q}
d(q-\mu) \, \frac{Z^2(\tilde{\epsilon}_q)}{\tilde{\epsilon}_q} \,
\tanh \left( \frac{\tilde{\epsilon}_q}{2T} \right) \, \phi_q\,
\left[ 2\, \ln \left( \frac{4 \mu^2}{\Lambda_g^2} \right)
+ \ln \left( \frac{\Lambda_g^2}{3\, m_g^2} \right)
+ \frac{1}{3} \, \ln \left( 
\frac{\Lambda_g^{6}}{M^4 |\tilde{\epsilon}_k^2 -
\tilde{\epsilon}_q^2|} \right) \right]\;.
\ee
Here, we have neglected terms $\sim \tilde{\epsilon}_k -
s\tilde{\epsilon}_q$ against $3\, m_g^2$ under the logarithm
arising from soft electric gluons, and terms
$\sim (\tilde{\epsilon}_k - s\tilde{\epsilon}_q)^6$ against 
$M^4 (\tilde{\epsilon}_k - s\tilde{\epsilon}_q)^2$ under the
logarithm from soft magnetic gluons.

Now observe that the gluon cut-off $\Lambda_g$ cancels in the final
result,
\be
\phi_k = \frac{g^2}{18 \pi^2} \int_{0}^{\Lambda_q}
d(q-\mu) \, \frac{Z^2(\tilde{\epsilon}_q)}{\tilde{\epsilon}_q} \,
\tanh \left( \frac{\tilde{\epsilon}_q}{2T} \right) \, \phi_q\,
\frac{1}{2}\, \ln  \left( \frac{\tilde{b}^2 \mu^2}{|\tilde{\epsilon}_k^2 -
\tilde{\epsilon}_q^2|} \right) \;,
\label{gapequation5}
\ee
where $\tilde{b} \equiv 256 \pi^4 [2/(N_f g^2)]^{5/2}$.
This is Eq.\ (19) of Ref.\ \cite{qwdhr}, since
$\bar{g}^2 \equiv g^2/(18 \pi^2)$, with the upper limit
of the $(q-\mu)$ integration, $\delta$, replaced by the
quark cut-off $\Lambda_q$.

The solution of the gap equation (\ref{gapequation5}) is well-known,
and given by Eq.\ (\ref{gapsol}).
As was shown in Ref.\ \cite{rdpdhr}, the dependence on the cut-off
$\Lambda_q$ enters only at sub-subleading order, i.e., it
constitutes an $O(g)$ correction to the prefactor in 
Eq.\ (\ref{gapsol}). Therefore, to subleading order we do not need
a matching calculation to eliminate $\Lambda_q$.

The result (\ref{gapequation5}) shows that the standard gap equation 
of QCD can be obtained from the effective action (\ref{Seff}). The
above, rather elaborate derivation of Eq.\ (\ref{gapequation5}) 
demonstrates that, in order
to obtain this result, it is mandatory to choose $\Lambda_q \ll 
\Lambda_g$. This also enabled us to identify potential 
sub-subleading order contributions. However, we argued that,
at this order, the off-shell behavior of the gap function 
has to be taken into account.

\section{Summary and Outlook} \label{V}

In this paper we have presented a formal derivation of an
effective action for non-Abelian gauge theories, Eq.\ (\ref{Seff}).
We first introduced cut-offs in momentum space for
quarks, $\Lambda_q$, and gluons, $\Lambda_g$. These cut-offs
separate relevant from irrelevant quark modes and
soft from hard gluon modes. We then 
explicitly integrated out irrelevant quark and hard gluon modes.
The effective action (\ref{Seff}) is completely general and, as
shown explicitly in Sec.\ \ref{IIIA}, after appropriately
choosing $\Lambda_q$ and $\Lambda_g$, it comprises
well-known effective actions as special cases, for instance,
the ``Hard Thermal Loop'' (HTL) and
``Hard Dense Loop'' (HDL) effective actions. We also demonstrated,
cf.\ Sec.\ \ref{IIIB}, that the high-density effective theory introduced
by Hong and others \cite{hong,hong2,HLSLHDET,schaferefftheory,NFL,others}
is contained in the effective action (\ref{Seff}).

We then showed how the QCD gap equation can be derived from
the effective action (\ref{Seff}). The gap equation is a
Dyson-Schwinger equation for the anomalous part of the
quark self-energy. It has to be solved
self-consistently, which is feasible only after truncating
the set of all possible diagrams
contributing to the Dyson-Schwinger equation. Such truncations
can be derived in a systematic way within the
Cornwall-Jackiw-Tomboulis (CJT) formalism \cite{CJT}.
Here, we only include diagrams of the sunset-type, cf.\ 
Fig.\ \ref{Gamma2eff}, in the
CJT effective action, which gives rise to one-loop diagrams
(with self-consistently determined quark and gluon propagators)
in the quark and gluon self-energies.

Usually, the advantage of an effective theory is that 
the degree of importance of various operators can be estimated
(via power counting) at the level of the effective action, i.e.,
{\em prior\/} to the actual calculation of a physical quantity.
This tremendously simplifies the computation of quantities which
are accessible within a perturbative framework.
On the other hand, the requirement of self-consistency for the
solution of the Dyson-Schwinger equation invalidates any such
power-counting scheme on the level of the effective action. 
For instance, perturbatively,
the right-hand side of the gap equation (\ref{gapeq}) is proportional
to $g^2$. However, self-consistency generates additional
large logarithms $\sim \ln (\mu/\phi) \sim 1/g$ which
cancel powers of $g$.

Nevertheless, there is still a distinct advantage in using an 
effective action for the derivation and the solution of Dyson-Schwinger 
equations for quantities which have to be determined self-consistently, 
such as the color-superconducting gap function in QCD.
This advantage originates from 
the introduction of the cut-offs which separate various
regions in momentum space. They allow for a {\em rigorous\/} power
counting of different contributions to the Dyson-Schwinger equation.
We explicitly demonstrated this in Sec.\ \ref{IV},
where we reviewed the calculation of the color-superconducting
gap parameter to subleading order. 

In order to obtain the standard result (\ref{gapsol}), it was mandatory to
choose $\Lambda_q  \alt g \mu \ll \Lambda_g \alt \mu$. This is in contrast
to previous statements in the literature 
\cite{hong2,HLSLHDET,schaferefftheory} 
that a consistent power-counting scheme requires $\Lambda_q \sim \Lambda_g$.
In particular, the choice $\Lambda_q \ll \Lambda_g$ has the consequence
that the gluon energy in the QCD gap equation is restricted
to values $p_0 \alt \Lambda_q$, while the gluon momentum can be much larger,
$p \alt \Lambda_g$. This naturally explains why it is permissible to use
the low-energy limit (\ref{lel}) of the HDL gluon propagators
in order to extract the dominant contribution 
to the gap equation (which arises from soft magnetic gluons).
In previous calculations of the gap within the framework
of an effective theory \cite{hong2,HLSLHDET,schaferefftheory}, the
low-energy limit for the HDL propagators was used 
without further justification,
even though for the choice $\Lambda_q \sim \Lambda_g$ 
the gluon energy can be of the same order as the gluon momentum.
The physical picture which arises from the choice $\Lambda_q \alt g
\mu \ll \Lambda_g \alt \mu$ is summarized in Fig.\ \ref{Sphere}.
Relevant quarks are located within a thin layer of width
$\sim \Lambda_q$ around the Fermi surface. Soft gluon exchange
mediates between quarks within a ``patch'' of size $\sim \Lambda_g$
inside this layer. The area of the patch is much larger than its
thickness. Hard gluon exchange mediates between quark states 
inside and outside of the patch.

In the course of the calculation, we were able to
identify various potential contributions of sub-subleading order.
However, we argued that, at this order, a solution
of the gap equation must take into account the {\em off-shell\/}
behavior of the gap function. For a complete sub-subleading
order calculation it also appears to be necessary to include
2PI diagrams beyond those of sunset topology 
in $\Gamma_2$, cf.\ Eq.\ (\ref{EqGamma2}) and Fig.\ \ref{Gamma2eff}.
Besides an improvement of the result for the color-superconducting
gap parameter beyond subleading order, we believe that
our rather general effective action (\ref{Seff}) can serve as a 
convenient starting point to investigate other 
interesting problems pertaining to hot and/or dense 
quark matter.

\section*{Acknowledgement}

The authors would like to thank Jean-Paul Blaizot,
Bengt Friman, Owe Philipsen, Rob Pisarski,
Tony Rebhan, Hai-cang Ren, Thomas Sch\"afer, Andreas Schmitt,
York Schr\"oder, Igor Shovkovy, and Dam Son for interesting
and stimulating discussions. Q.W.\ acknowledges support
by the Virtual Institute VH-VI-041 of the Helmholtz Association
of National Research Centers.


\end{document}